\documentclass[superscriptaddress,preprintnumbers,nofootinbib,11pt]{revtex4}
\usepackage{amsmath,amssymb}
\usepackage[dvipdf,dvips]{graphicx}
\usepackage{color}
\usepackage{hyperref}
\usepackage{url}
\usepackage{slashed}
\usepackage{subfigure}
\usepackage[usenames,dvipsnames]{xcolor}
\usepackage{amsmath}
\usepackage{amsfonts}
\usepackage{float} 
\usepackage{amssymb}
\usepackage{epsfig}
\usepackage{graphics}
\usepackage{euscript}
\usepackage{slashed}
\usepackage{epstopdf}
\usepackage[utf8]{inputenc}
\allowdisplaybreaks

\usepackage{verbatim}
\usepackage{graphicx}
\usepackage{latexsym}

\def \pt{\partial}

\def \STr{\textmd{STr}}
\def \and{\textmd{and}}
\def \nn{\nonumber} 
\def \TT{\textmd{TT}} 
\def \T{\textmd{T}} 

\def \be{\begin{equation}}
\def \ee{\end{equation}}

\def \bea{\begin{eqnarray}}
\def \eea{\end{eqnarray}}

\begin{document}

\title{A link that matters: Towards phenomenological tests of unimodular asymptotic safety}

\author{Gustavo P. de Brito} \email{gpbrito@cbpf.br}
\affiliation{Centro Brasileiro de Pesquisas F\'{i}sicas (CBPF),\\ Rua Dr Xavier Sigaud 150, Urca, Rio de Janeiro, RJ, Brazil, CEP 22290-180}
\affiliation{Institute for Theoretical Physics, University of Heidelberg,\\ 
	Philosophenweg 16, 69120 Heidelberg, Germany}

\author{Astrid Eichhorn} \email{eichhorn@sdu.dk}
\affiliation{CP3-Origins, University of Southern Denmark, Campusvej 55, DK-5230 Odense M, Denmark}

\author{Antonio D. Pereira} \email{adpjunior@id.uff.br}
\affiliation{Instituto de F\'isica, Universidade Federal Fluminense, Campus da Praia Vermelha, Av. Litor\^anea s/n, 24210-346, Niter\'oi, RJ, Brazil}

\begin{abstract}
 Constraining quantum gravity from observations is a challenge. We expand on the idea that the interplay of quantum gravity with matter could be key to meeting this challenge.
 Thus, we set out to confront different potential candidates for quantum gravity -- unimodular asymptotic safety, Weyl-squared gravity and asymptotically safe gravity -- with constraints arising from demanding an ultraviolet complete Standard Model. Specifically, we show that within approximations, demanding that quantum gravity solves the Landau-pole problems in Abelian gauge couplings and Yukawa couplings strongly constrains the viable gravitational parameter space.
 In the case of Weyl-squared gravity with a dimensionless gravitational coupling, we also investigate whether the gravitational contribution to beta functions in the matter sector  calculated from functional Renormalization Group techniques is universal, by studying the dependence on the regulator, metric field parameterization and choice of gauge.
\end{abstract}

\maketitle

\section{Introduction \label{Introduction}}

Observational constraints on quantum gravity are hard to come by. Based on a simple dimensional argument, one typically expects a power-law suppression of quantum-gravity effects\footnote{In the presence of a second ``meso"-scale, as hinted at by some quantum-gravity approaches, e.g., \cite{Dowker:aza}, this situation can change.} with $(E/M_{\rm Pl})^{\#}$, with $E$ being the energy scale relevant for experiments, $M_{\rm Pl}$ being the Planck mass and $\#>0$. Nevertheless, mathematical and internal consistency  are not the only conditions that could allow to constrain candidate quantum-gravity theories while direct probes of Planck-scale physics remain (mostly) out of reach. Observational consistency tests for quantum gravity arise from the interplay of quantum gravity with matter.

A consistent microscopic description of all degrees of freedom of nature must account for both gravity and matter. In a quantum-field theoretic setting this can be achieved by including a metric as well as matter fields in the microscopic description. The key point for observational consistency tests is that the properties of the microscopic description actually determine some properties of the matter sector at energy scales far below the Planck mass, i.e., in the infrared (IR). This allows to restrict the microscopic dynamics by demanding that the resulting low-energy behavior is consistent with observations. 

Such a UV-IR connection appears to contradict the well-established  principle of separation of scales in nature, which loosely speaking states that physics at macroscopic scales decouples from microscopic physics. Yet, effective descriptions for physics at macroscopic scales typically feature finitely many parameters that are sensitive to the microphysics. 
In the Standard Model of particle physics the marginal couplings  are sensitive to the microphysics. They exhibit a logarithmic scale-dependence. Accordingly, changes of $\mathcal{O}(1)$ at the Planck scale lead to changes of   $\mathcal{O}(1)$ at the electroweak scale. This is in contrast to canonically irrelevant couplings, which are power-law suppressed due to their canonical dimension: A large interval of UV values is mapped to a rather small interval of IR values by the RG flow. Accordingly, the information on microscopic physics encoded in those couplings is ``washed out'' by the RG flow, and not accessible in the IR. \\
Therefore, the marginal couplings of the Standard Model are a prime target to set up observational consistency tests for quantum gravity. Specifically, they provide two tests, under the assumption of no new physics between the electroweak scale and the Planck scale\footnote{Analogous considerations hold in settings beyond the Standard Model.}: \\
 i) The first is a microscopic consistency test: The perturbative Landau poles of the Standard Model \footnote{Landau-poles in the Abelian gauge and the Higgs-Yukawa sector \cite{Maiani:1977cg,Cabibbo:1979ay,Dashen:1983ts,Callaway:1983zd,Beg:1983tu,Lindner:1985uk,Kuti:1987nr,Hambye:1996wb} indicate that the non-perturbative triviality problems from QED \cite{GellMann:1954fq,Gockeler:1997dn,Kim:2000rr,Gies:2004hy} and $\phi^4$ theory \cite{Freedman:1981wr,Aizenman:1981du,Frohlich:1982tw} carry over to the full Standard Model.}, most prominently in the Abelian gauge coupling as well as the Yukawa sector, must be resolved by quantum-gravity fluctuations. 
 A fundamental description, i.e., valid up to arbitrarily short distances, of the building blocks of nature in the quantum field theory framework requires theory to be either asymptotically free or safe. Both cases provide a framework for an ultraviolet completion of effective field theories, in which an enhanced  symmetry, quantum scale symmetry \cite{Wetterich:2019qzx}, rules the microscopic dynamics. Scale-symmetry is a consequence of vanishing interactions in the case of asymptotic freedom, or a balance between residual interactions in the case of asymptotic safety, see \cite{Eichhorn:2018yfc} for an overview of potential mechanisms for asymptotic safety.  
 Theories which pass this first test can be subjected to the second.\\
 ii) The second test exploits the finite number of free parameters in an asymptotically safe/free model, and the fact that a UV-IR link can be established  based on marginal couplings.
 The free parameters of an asymptotically free/safe model are the (marginally) relevant couplings, corresponding to those interactions which can trigger a departure from scale invariance in the Renormalization Group (RG) flow to the IR. In the case of asymptotic freedom, they are the powercounting relevant couplings, i.e., couplings with positive mass dimension, and marginally relevant ones, i.e., couplings with vanishing mass dimension and a leading-order  antiscreening quantum contribution. The presence of residual interactions at the interacting fixed point underlying asymptotic safety changes the scaling, and relevant couplings no longer automatically correspond to power-counting relevant ones. 
Conversely, a coupling that is irrelevant at a fixed point must automatically assume one specific value in the IR, since it is not a free parameter, but determined in terms of the finitely many relevant couplings. Intuitively speaking, the powerful symmetry of scale-invariance in the ultraviolet (UV) leaves imprints in the IR akin to any other enhanced symmetry in the UV. The theoretically determined values of irrelevant couplings need not agree with their measured values.
Hence, given a set of fields and symmetries, an asymptotically free/safe fixed point is not automatically phenomenologically viable.
  
Here, we use these ideas to make the first steps to differentiate between quantum-gravity-matter models with regards to their phenomenological viability. We set out to compare Weyl-squared gravity, asymptotically safe gravity based on the Reuter fixed point, and unimodular asymptotically safe gravity and  concentrate on Standard-Model like matter sectors. We focus on these three, as they are all based on a formulation purely in terms of metric degrees of freedom, but differ in the symmetries that are realized. Thus it is of interest to understand the resulting differences in the interplay with matter. For a clear discussion of the relation of various symmetry-restrictions on gravity at the classical level, see \cite{Gielen:2018pvk}. 
 
Before focusing on the interplay of these gravity-models with matter, we review the motivation for them as well as their status as potential candidates for a description of quantum gravity below.

\section{Introduction to (unimodular) asymptotically safe gravity and Weyl-squared gravity}

There are strong indications for the existence \cite{Reuter:2001ag,Lauscher:2001ya,%
  Litim:2003vp,Codello:2006in,Codello:2008vh,%
  Benedetti:2009rx,Niedermaier:2010zz,Manrique:2010am,Manrique:2011jc,%
  Dietz:2012ic,Donkin:2012ud,Codello:2013fpa,Falls:2013bv,Becker:2014qya,%
  Christiansen:2014raa,Demmel:2015oqa,Percacci:2015wwa,Gies:2016con,Biemans:2016rvp,Denz:2016qks,%
  Christiansen:2017bsy,Knorr:2017fus,Knorr:2017mhu,Falls:2017lst,deAlwis:2017ysy,Falls:2018ylp,Pagani:2019vfm} of the asymptotically safe Reuter fixed point \cite{Hawking:1979ig,Reuter:1996cp},  see, e.g., \cite{Percacci:2017fkn,Eichhorn:2017egq,Eichhorn:2018yfc,Reuter:2019byg} for recent reviews, with promising indications for a quantum-gravity induced  UV completion for the Standard Model with an enhanced predictive power \cite{Shaposhnikov:2009pv,Eichhorn:2017ylw,Eichhorn:2017lry,Eichhorn:2018whv} in four dimensions \cite{Eichhorn:2019yzm}. The Reuter fixed point is an interacting fixed point of the Renormalization Group (RG) flow, rendering the model asymptotically safe. The enhanced scale-symmetry controls the infinitely many operators that are expected to be present at the Planck scale from effective field theory arguments. While they are present in an asymptotically safe context, there are infinitely many relations between the couplings that have to be satisfied as a consequence of scale symmetry. Accordingly, the model remains predictive despite the existence of  infinitely many higher-order interactions. In particular, there are indications for a near-perturbative nature of the fixed point \cite{Codello:2006in,Niedermaier:2009zz,Niedermaier:2010zz,Falls:2013bv,Falls:2014tra,Falls:2017lst,Eichhorn:2018akn,Falls:2018ylp,Eichhorn:2018ydy,Eichhorn:2018nda}, providing a basis for systematic approximations of the RG flow. 
Building on strong indications for the Reuter fixed point, challenging questions pertaining to the unitarity of theory \cite{Becker:2017tcx}, singularity-resolution \cite{Adeifeoba:2018ydh,Platania:2019kyx,Bosma:2019aiu,Knorr:2019atm}, background independence \cite{Morris:2016spn,Percacci:2016arh,Ohta:2017dsq,Eichhorn:2018akn,Pagani:2019vfm}, the relation to a Lorentzian setting \cite{Manrique:2011jc,Eichhorn:2019xav}, as well as the relation to other quantum-gravity approaches \cite{Eichhorn:2018phj,deAlwis:2019aud} can now be tackled. Most of these relate to the internal structure and consistency of the model and need to be answered before an asymptotically safe description of nature can be deemed viable. Yet, there is another crucial requirement that a viable model has to satisfy, namely the consistency with observational constraints. Here, it appears to be possible to make progress that could allow to rule out such a model based on phenomenological consistency in the matter sector, e.g., \cite{Shaposhnikov:2009pv,Harst:2011zx,Eichhorn:2011pc,Dona:2013qba,Meibohm:2015twa,Oda:2015sma,Meibohm:2016mkp,Biemans:2017zca,Eichhorn:2017eht,Eichhorn:2017muy,Eichhorn:2017ylw,Eichhorn:2017lry,Eichhorn:2017als,Hamada:2017rvn,Christiansen:2017cxa,Gies:2018jnv,Bonanno:2018gck,Eichhorn:2018whv,Pawlowski:2018ixd,Eichhorn:2019yzm,deBrito:2019epw,Wetterich:2019zdo}.\newline	

Unimodular gravity is attractive due to several reasons: It is actually based on the symmetry group that follows from an analysis of the massless spin-2-representation of the Poincar\'e group \cite{vanderBij:1981ym,Herrero-Valea:2018ilg}, namely the transverse diffeomorphisms (``TDiff", the local-volume preserving diffeomorphisms). Further, it gets rid of unnecessary ``baggage" that classical Einstein gravity carries through the presence of a non-dynamical degree of freedom, namely the conformal factor. It follows the spirit of the Hawking-King-McCarthy-Malament theorem \cite{Hawking:1976fe,Malament:1977} which states that under suitable global causality conditions, the conformal geometry of a spacetime is encoded in the causal relations. This suggests a more minimalistic starting point for quantum gravity which does not allow all components of the metric to fluctuate, but explicitly removes the conformal factor. At the same time, this solves the conformal-factor problem in Euclidean gravity based on the Einstein-Hilbert action and would therefore also constitute an interesting starting point for semi-classical considerations of the gravitational path integral \cite{Feldbrugge:2017fcc}. A further motivation for unimodular gravity as the starting point for the quantization of gravity is present in a canonical setting, where the Hamiltonian is nonvanishing in the unimodular case \cite{Henneaux:1989zc,Smolin:2009ti}.
Moreover, it brings a different perspective to the fine-tuning-questions surrounding the cosmological constant. While the observed value of the cosmological constant appears to be compatible with standard asymptotically safe gravity, its inclusion requires the selection of a very specific RG trajectory. Instead, in a unimodular setting, the cosmological constant appears at the level of the equations of motion \cite{Weinberg:1988cp,Henneaux:1989zc,Finkelstein:2000pg}. These are derived from the full quantum effective action, where all quantum fluctuations have been integrated out. Therefore, there is no longer a ``typical", large energy-scale present in the setting. Instead, the most ``natural" scale is $k=0$, which motivates the conjecture that the most natural choice for a constant of integration in this setting should be close to zero in units of the Planck scale. Indeed, it has been shown explicitly that the cosmological constant is not subject to quantum corrections \cite{Alvarez:2015sba}.
Finally, it has been argued \cite{Torres:2017ygl,Adeifeoba:2018ydh} that singularity resolution in simple asymptotic-safety inspired models of black holes requires a unimodular setting, unless the microscopic fixed-point value for the cosmological constant accidentally happens to vanish. While the equivalence to General Relativity at the classical level is undisputed \cite{Ellis:2010uc,Ellis:2013eqs} (and extensions including higher order operators also have been investigated \cite{Eichhorn:2015bna,Nojiri:2015sfd,Saez-Gomez:2016gum}), the relation of unimodular quantum gravity and ``standard"  quantum gravity (with full diffeomorphism (``Diff") invariance) is under debate see, e.g., \cite{Alvarez:2005iy,Ardon:2017atk}. We point out that a decisive comparison of the quantum theories actually requires some knowledge of the UV completion. Specifically, theory spaces of TDiff versus Diff invariant gravity  differ, as the former does not contain the cosmological constant. Therefore one would actually expect unimodular asymptotic safety \cite{Eichhorn:2013xr,Eichhorn:2015bna,Benedetti:2015zsw} to differ from the Reuter universality class.
Further, there are different variants of unimodular gravity. \newline

Weyl-squared gravity (see, e.g., \cite{Scholz:2017pfo} for a review) is in some sense the exact opposite of unimodular gravity: Whereas the latter features a fixed, non-dynamical conformal factor, the former declares conformal/Weyl transformations (i.e., rescaling of the metric by a  local conformal factor, $g_{\mu\nu} \rightarrow \Omega^2(x)g_{\mu\nu}$) to be gauge transformations.  This enhanced symmetry strongly constrains the viable dynamics. In fact, the only invariant term that can appear in the action is the Weyl-squared invariant $C^2= C^{\mu\nu\alpha\beta}C_{\mu\nu\alpha\beta}$ with 
 \begin{align}
C^2 = R_{\mu\nu\alpha\beta} R^{\mu\nu\alpha\beta} - 2 R_{\mu\nu} R^{\mu\nu} + \frac{1}{3} R^2 \,.
\end{align}
The Weyl-tensor transforms according to $C^{\mu}_{\nu\kappa\lambda} \rightarrow C^{\mu}_{\nu\kappa\lambda}$ under a Weyl transformation. Thus,
the most general, local gravitational action invariant under Weyl transformations which is expressed just in terms of the metric is given by
\begin{align}\label{C2_action}
S_{\textmd{WG}} = \frac{1}{2 w}  \int_x \sqrt{g} \, C^2,
\end{align}
 with $w$ being a dimensionless coupling. \\
Alternatively, conformal invariance can be achieved in any action by means of the introduction of extra fields such as the dilaton, see, e.g., \cite{Wetterich:2019qzx,Percacci:2011uf,Codello:2012sn,Pagani:2013fca,Shaposhnikov:2018jag,Shaposhnikov:2018xkv,Mooij:2018hew,Shaposhnikov:2018nnm}. 
 For further discussions on the role of scale symmetry as a key ingredient of a fundamental theory for quantum gravity, see, e.g., \cite{tHooft:2009wdx,tHooft:2011aa,Hooft:2014daa}.

It is curious to observe that promoting conformal transformations\footnote{Throughout the paper we refer to the local rescaling of the metric by a conformal factor as a conformal transformation, and caution that this should not be confused with the global action of the conformal group on flat spacetime.} to gauge transformations does not require to introduce a new gauge field in order to write a gauge-invariant action. Nevertheless, one can of course introduce such a gauge field, the so-called Weyl photon, as part of the gravitational connection, see, e.g., \cite{Smolin:1979uz,Cheng:1988zx} for older work and \cite{Ghilencea:2018dqd,Ferreira:2018itt} and references therein for more recent work. We will not focus on this version of Weyl-gravity or conformal gravity here. For clarity we refer to the theory we study as Weyl-squared gravity.\\
Weyl-squared gravity is power-counting renormalizable \cite{Stelle:1976gc}, but most importantly also asymptotically free \cite{Fradkin:1981iu}, such that it could potentially be a candidate for a UV complete description of gravity. Yet, there is a major problem that requires a solution:
 Around flat spacetime, upon expansion of $C^2$ to second order, a ghost mode, i.e., a mode with negative kinetic term, propagates, signalling an inconsistency of the theory about flat spacetime \cite{Lee:1982cp,Riegert:1984hf}. Upon spontaneous breaking of the conformal symmetry, this mode acquires a mass that might be high enough to render this mode phenomenologically irrelevant. We will not focus on this question in this paper. For recent discussions how to potentially evade such unitarity problems in curvature-squared gravity, see, e.g., \cite{Holdom:2016xfn,Donoghue:2017fvm,Becker:2017tcx,Anselmi:2018ibi}.

Vacuum solutions to the Einstein equations, where the Planck mass drops out since the energy-momentum tensor vanishes, actually turn out to be solutions of Weyl-squared gravity, including the Schwarzschild solution. Moreover, it has even been argued that Weyl-squared gravity might reproduce the observed galactic rotation curves \cite{Mannheim:2010xw}, although it is not clear whether further evidence for dark matter, coming, e.g., from the spectrum of CMB fluctuations, can actually be reproduced.

 Invariance under conformal transformations -- as expected -- precludes the existence of dimensionful couplings. At a first glance this could appear to be phenomenologically problematic, as gravity at IR scales clearly contains a mass scale, the Planck mass. 
Introducing a scalar field with a conformal coupling to gravity, i.e., a $\phi^2\, R$-term, conformal symmetry can be broken spontaneously once the scalar acquires a vacuum expectation value, thereby generating an Einstein term $\phi^2\, R \rightarrow M_{\rm Pl}^2 R+...$, see, e.g., \cite{Bars:2013yba,deCesare:2016mml,Oda:2018zth} and references therein.  Clearly, the theory space of Weyl-squared gravity differs significantly from the ones explored for ``standard" and unimodular quantum gravity.

\section{Setup \label{Setup}}

\subsection{Functional Renormalization Group for quantum gravity}

In this paper we study quantum-gravity matter systems within a quantum-field theoretic setting. 
We aim at discovering whether the formal path integral can be defined in a predictive, UV-complete fashion. To that end, we explore the change of the dynamics for matter under coarse-graining steps, and search for scale-invariant points  in the space of the dynamics for matter, providing a UV completion. The existence and location of these points depend on the gravitational couplings which we treat as free parameters in this work. Accordingly, the microscopic gravitational dynamics is constrained by demanding a UV complete matter sector, as discussed, e.g.,  in \cite{Eichhorn:2017eht}. 

The functional renormalization group (FRG) has been extensively employed in the asymptotic-safety program for quantum gravity, based on the seminal work \cite{Reuter:1996cp}, see \cite{Percacci:2017fkn,Eichhorn:2018yfc,Reuter:2019byg,Pereira:2019dbn} for recent reviews. The key point of the FRG is the introduction of an IR-cutoff which allows us to capture the scale dependence of the dynamics. As it should distinguish UV from IR modes in a local coarse-graining scheme, it requires the introduction of a background metric $\bar{g}_{\mu\nu}$. We will mostly focus on  the exponential split introduced in $2+\epsilon$ dimensions in \cite{Kawai:1992np} and first used in the context of the functional RG in \cite{Eichhorn:2013xr, Nink:2014yya,Percacci:2015wwa} which has the form 
\be
g_{\mu\nu} = \bar{g}_{\mu\kappa}\,[{\rm exp}(h_{\,\,\,\cdot}^{\cdot})]^{\kappa}_{\,\,\,\nu},
\label{paramexpdef}
\ee
where $g_{\mu\nu}$ is the full metric and $h_{\mu\nu}$ is a fluctuation field (of arbitrary amplitude). In Sec.~\ref{sec:ConGrav}, where we study the parameterization dependence of the results in Weyl-squared gravity, we also explore more general splits  \cite{Gies:2015tca,Ohta:2016npm,Ohta:2016jvw,deBrito:2018jxt}.

The quantization is performed by integrating over the fluctuation field $h_{\mu\nu}$. This background field approach allows us to define an IR-suppression term that is added to the action appearing in the Euclidean generating functional 
\be
Z_k [J]= \int \mathcal{D}h\mathcal{D}\bar{C}\mathcal{D}C\, e^{-S[g_{\mu\nu}] - S_{\rm gauge-fixing} - S_{\rm ghost} -\Delta S_k - \int_x \sqrt{\bar{g}} h_{\mu\nu}J^{\mu\nu},}
\label{genfunct1}
\ee
where $J_{\mu\nu}$ is  an external source and $(\bar{C},C)$ are the Faddeev-Popov ghosts.
The suppression term takes the form

\begin{align}
\Delta S_k = \frac{1}{2}\int_x \sqrt{\bar{g}} \, h_{\mu\nu}\, [\textbf{R}_k(-\bar{D}^2)]^{\mu\nu\alpha\beta}\,h_{\alpha\beta} \,.
\end{align}
It suppresses quantum fluctuations in $h_{\mu\nu}$ based on the  spectrum of the covariant background Laplacian $-\bar{D}^2$:  Modes with eigenvalue $\lambda_l$ of $-\bar{D}^2$ lower than the momentum scale $k^2$ are suppressed, i.e., $k$ acts as an IR cutoff scale. The function $\textbf{R}_k(-\bar{D}^2)$ has to satisfy several requirements: in order for it to act as an IR suppression term, it has to vanish for $\lambda_l>k^2$ and take a finite value for $\lambda_l<k^2$. Further, it should diverge in the limit $k^2\rightarrow \infty$ such that $\Gamma_k$, which we will define just below, approaches $S$ in that limit, see, e.g., \cite{Reuter:1996eg}. This leaves some freedom in the form of $\textbf{R}_k(-\bar{D}^2)$. A popular choice is the so-called Litim cutoff \cite{Litim:2001up,Litim:2000ci}, but, e.g., \cite{Reuter:2001ag,Lauscher:2002sq,Groh:2010ta,Eichhorn:2010tb} highlight that the use of a different shape function does not qualitatively alter the results for the Reuter fixed point. 
This allows to define the flowing action $\Gamma_k$ as a modified Legendre transform of the scale-dependent Schwinger generating functional $W_k (= \ln Z_k)$,
defined as follows
\be
\Gamma_k[h_{\mu\nu}, \bar{g}_{\mu\nu}] = \underset{J}{\rm sup} \left(\int_x\sqrt{\bar{g}}J^{\mu\nu}h_{\mu\nu} - {\rm ln}Z_k [J] \right) - \Delta S_k.
\ee
Note that we slightly abuse the notation here, as the arguments of the flowing action are the expectation values of the 
fields, which we denote by the same variable as the fields that are integrated over in the path integral. 
The flowing action $\Gamma_k$ interpolates between a microscopic (bare) UV action  $\Gamma_{k\to\Lambda} = S_\Lambda$, with $\Lambda$ being a UV cutoff, and the full quantum effective action $\Gamma_{k \to 0} = \Gamma$. 
As the scale $k$ plays the role of an IR cutoff, $\Gamma_k$ contains the effect of modes with generalized momentum (i.e., $\lambda_l$) higher than $k^2$. 
The key advantage of this setting is that $\Gamma_k$ obeys an exact flow equation of one-loop structure, the Wetterich equation \cite{Wetterich:1992yh,Morris:1993qb,Ellwanger:1993mw}, formally written as
\begin{align}\label{Wetterich}
\pt_t \Gamma_k = \frac{1}{2} \STr \left[ \left( \Gamma_k^{(2)} + \textbf{R}_k\right)^{-1}  \pt_t \textbf{R}_k \right],
\end{align}
where $\pt_t = k \pt_k$, $\Gamma^{(2)}_k=\delta^2 \Gamma_k/\delta\Phi\delta\Phi$ is the Hessian and $\STr$ denotes the supertrace which contains a negative sign for Grassmann-valued fields and a factor of 2 for complex fields. As $\textbf{R}_k (-\bar{D}^2)$ vanishes for modes with $\lambda_l>k^2$, its scale derivative actually acts as an \emph{ultraviolet} cutoff in Eq.~\eqref{Wetterich}. Accordingly, the physical interpretation of the flow equation is the following: Under a change of the momentum scale -- intuitively speaking to be thought of as the ``resolution" scale of the theory -- the effective dynamics changes. The main contribution to the right-hand-side of Eq.~\eqref{Wetterich} comes from modes with momenta close to the scale $k$, i.e., the change of the dynamics at $k$ is driven by quantum fluctuations with momenta close to $k$. This translates the Wilsonian idea of performing the path integral in a momentum-shell-wise fashion into an equation that is structurally one-loop, a fact that greatly simplifies practical calculations.

While Eq.~\eqref{Wetterich} is formally exact, in practice it requires an approximation. 
In other words, a truncation of the dynamics has to be used. In this work, we will work under the assumption that an asymptotically safe fixed point in four-dimensional Euclidean quantum gravity is a near-Gaussian one. This implies that the relevant terms in the dynamics are those with positive, vanishing or only slightly negative canonical scaling dimensions. This motivates truncations that follow canonical power-counting and neglect most higher-order operators.  This assumption is corroborated by several recent results, see \cite{Niedermaier:2009zz,Niedermaier:2010zz,Falls:2013bv,Falls:2014tra,Falls:2017lst,Eichhorn:2018akn,Falls:2018ylp,Eichhorn:2018ydy,Eichhorn:2018nda}. 
For a recent demonstration of the quantitative reliability of the FRG and the fast convergence of truncations for interacting fixed points in the non-gravitational context, see \cite{Balog:2019rrg}.

\begin{figure}[!t]
		\includegraphics[width=0.8\linewidth]{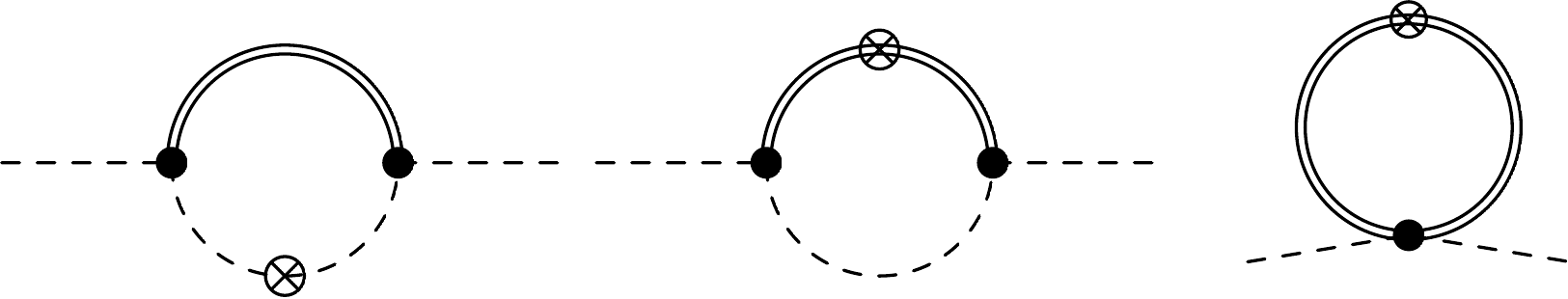}\\
		\includegraphics[width=0.8\linewidth]{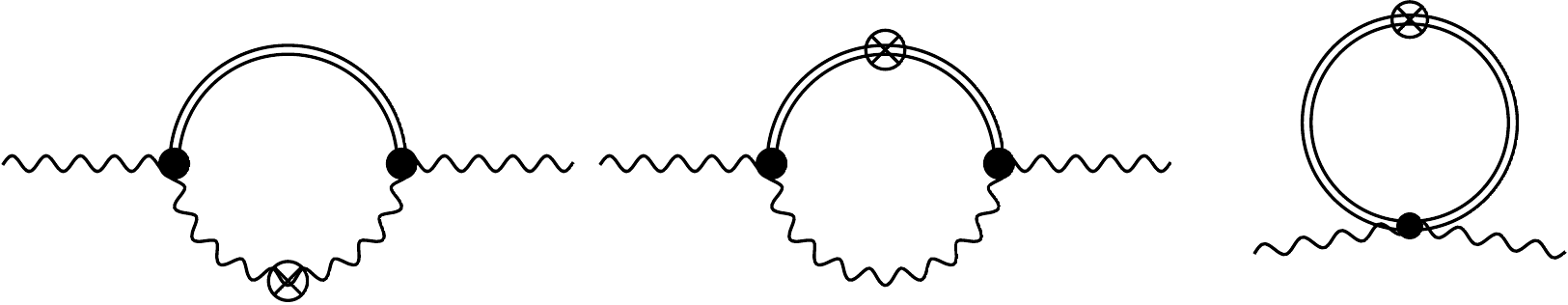} \\
		\includegraphics[width=0.8\linewidth]{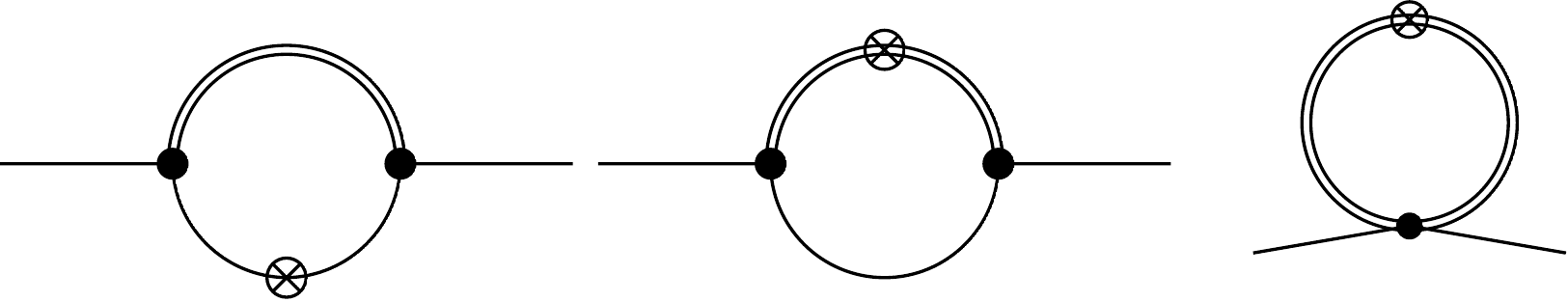}
		\caption{\footnotesize Diagrams encoding the quantum-gravity contributions to the anomalous dimensions of the matter fields. Double lines correspond to the metric propagator, the dashed line represents a scalar field, the wiggly line denotes a gauge field and the single solid line stands for a fermionic field. The crossed circle denotes the regulator insertion in Eq.~\eqref{Wetterich}.}
		\label{diagrams}
\end{figure}

\begin{figure}[!t]
	\begin{center}
		\includegraphics[scale=.7]{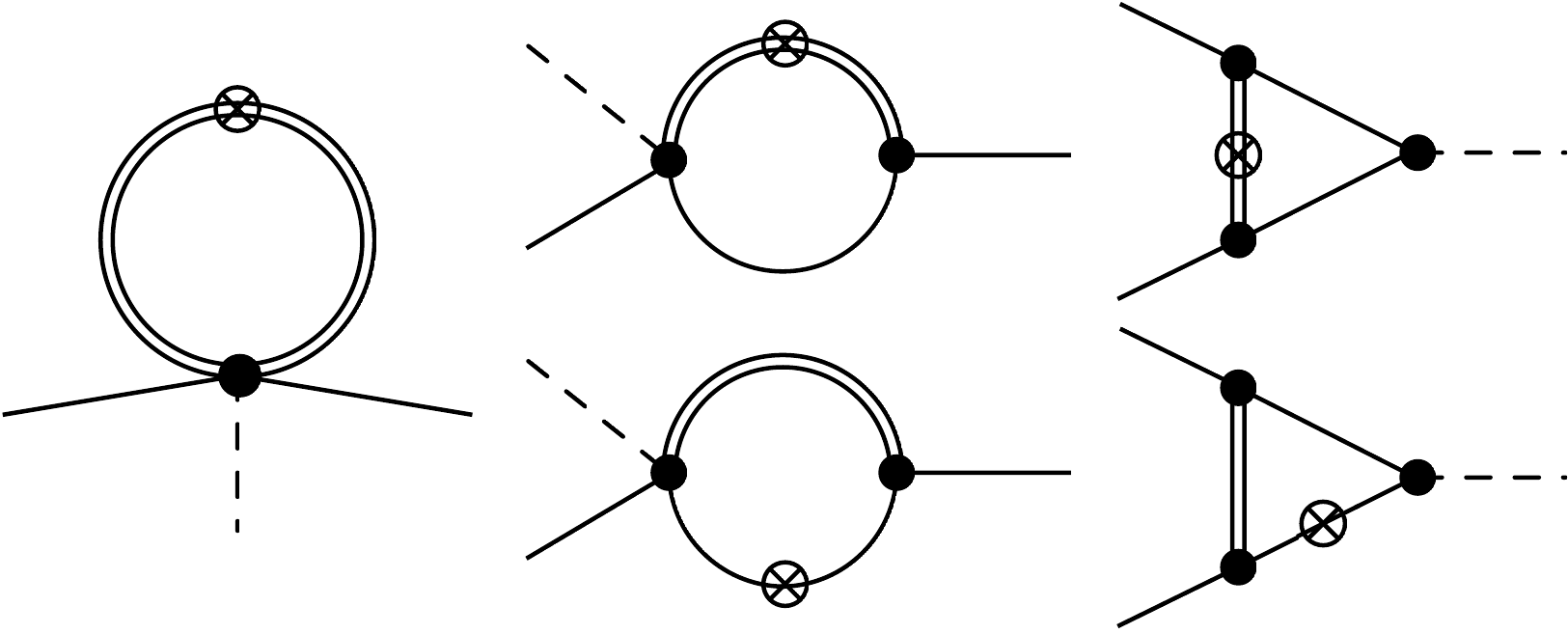}
		\caption{\footnotesize Further diagrams contributing to the running of the Yukawa coupling (denoted as $\mathcal{D}_y \, y$). In the unimodular gravity framework only the triangle diagrams in the last column are non-zero. In the Weyl-squared gravity case the only non-vanishing diagram is the tadpole in the first column. Within the standard asymptotically safe quantum gravity (ASQG) framework all the diagrams depicted above may lead to non-trivial contributions (depending on the gauge choice).}
		\label{diagrams_2}
	\end{center}
\end{figure}

Plugging a given truncation into the flow equation \eqref{Wetterich}, we can extract the anomalous dimensions and the running of dimensionless couplings by taking functional derivatives and applying appropriate projection rules (see App. \ref{Convention}). Due to the one-loop structure of the flow equation, the right-hand side of Eq.~\eqref{Wetterich} can be expressed diagrammatically as in Figs.~\ref{diagrams} and \ref{diagrams_2}, which should not be confused with Feynman diagrams.

Let us add two cautionary remarks: 
Firstly, we stress that our results are obtained in Euclidean gravity, and there is no straightforward way to extract implications in a Lorentzian setting, as the Wick-rotation is in general ill-defined in quantum gravity, see, e.g., \cite{Visser:2017atf,Baldazzi:2018mtl}.\\
 We further stress that we explore the dependence of couplings on the RG cutoff scale $k$, not on physical momenta.  A priori, the cutoff scale $k$ is not a physical scale, although external momenta can of course act similarly, i.e., as an IR cutoff, e.g. in scattering processes. We
 use that in a scaling regime, the dependence of couplings on physical momenta is expected to mimick the $k$-dependence simply because a scaling regime by definition does not feature any characteristic physical scale. Hence, fixed points in $k$ provide indications whether asymptotic safety is realized in a physical sense.

\subsection{Functional Renormalization Group for unimodular gravity}

Unimodular quantum gravity (UQG) is characterized by a restriction on the configuration space such that the metric determinant is non-dynamical, more specifically
\begin{align}
\det g_{\mu\nu} = \omega \,,
\end{align} 
where $\omega$ denotes a fixed scalar density \footnote{Let us note, as already hinted at in \cite{Eichhorn:2013xr}, that this could provide an interesting vantage point from which to develop a more background-independent flow equation: In the implementation of the RG as a local coarse-graining, one only requires a definition of a local ``patch" over which to average fluctuations. The fixed background density provides this, as it is sufficient to define a local volume. Whether this is sufficient to derive a flow equation is an intriguing open question.}.  This is not the same as imposing $\det g_{\mu\nu} = \omega$ as a gauge-condition in a path integral over all components of the metric. The difference lies in the symmetry group, which is trans\title{A link that matters: Towards phenomenological tests of unimodular asymptotic safety}

verse diffeomorphisms, \textit{TDiff}, in the unimodular case instead of the full diffeomorphism group.
Formally, unimodular quantum gravity is based on a definition of the functional measure in such a way that the integration is performed over the space of metrics satisfying the unimodularity condition. In this sense, we are interested in computing functional integrals of the form
\begin{align}
Z_{\rm UQG} = \int (\mathcal{D}g)_{\rm UQG} \,e^{ iS_{\rm UQG}[g]} \,.
\end{align}
The application of functional RG techniques makes it necessary to focus on the Euclidean version of the path-integral, 
\begin{align}
Z_{{\rm UQG}, \, k} = \int (\mathcal{D}g)_{\rm UQG} \,e^{-S_{\rm UQG}[g] - \Delta S_k[\bar{g};h]} \,.\label{eq:ZEu}
\end{align}
We apply the background field method by means of the  exponential parameterization, see Eq.~\eqref{paramexpdef}.
Since the exponential parameterization allows us to express $\det(g_{\mu\nu}) = \det(\bar{g}_{\mu\nu})\, e^{h}$, one can impose the unimodularity condition on the background metric (namely $\det \bar{g}_{\mu\nu} = \omega$). As a consequence, the same condition on the full metric can be achieved by setting the trace mode to zero, i.e., removing $h= \bar{g}^{\mu\nu} h_{\mu\nu}$ from the path integral. Therefore, the functional measure for UQG can be defined in such a way that the integration is performed over the space of traceless fluctuations \cite{Eichhorn:2013xr,Eichhorn:2015bna,Ardon:2017atk,Percacci:2017fsy}.

In this work we employ a truncation containing matter fields (scalar, vector and spinor fields) minimally coupled to gravity in the unimodular setting including all operators compatible with the symmetries with positive or vanishing canonical dimension of the corresponding couplings. Our truncation is given by
\begin{align} \label{EAA_UG}
\Gamma_k^{\textmd{UG}} &= \frac{1}{16\pi \,G_N} \int_x \sqrt{\omega} \left( -R +\bar{a}\, R^2 + \bar{b} \, R_{\mu\nu}R^{\mu\nu} \right)  + \Gamma_k^{\textmd{g.f.}} +  \Gamma_k^{\textmd{gh.}} \nn\\
&+ \frac{1}{2}\int_x \!\sqrt{\omega} \left({ Z_{\phi}} g^{\mu\nu} \pt_\mu \phi \pt_\nu \phi {+2 V(\phi^2)}\right)+  \int_x\!\!\sqrt{\omega}\,( { Z_{\psi}}i \bar{\psi} {\slashed{\nabla}} \psi + i \, y \phi \bar{\psi} \psi)\nn\\
& + \frac{{Z_A}}{4} \int_x\!\! \sqrt{\omega} \,g^{\mu\alpha}g^{\nu\beta} F_{\mu\nu} F_{\alpha\beta}
 \,.
\end{align}
An explicit mass term for fermions is incompatible with the discrete ``chiral" symmetry $\psi \rightarrow e^{i\pi/2 \gamma_5} \psi$, $\bar{\psi}\rightarrow \bar{\psi}e^{i\pi/2 \gamma_5}$, under which the scalar transforms as $\phi \rightarrow -\phi$.  The fermion is coupled to gravity by means of the vielbein and the spin-connection, which, for spaces with vanishing torsion, can be expressed in terms of the fluctuation field (see App.~\ref{Fermions_exp} for further details). \\

In setting up this truncation, we follow a canonical power-counting scheme, as we expect most canonically irrelevant (i.e., higher-order) interactions to also be irrelevant at the interacting fixed point since we expect that anomalous scaling dimensions are roughly $\mathcal{O}(1)$. Thus, the canonically least irrelevant operators might be shifted into relevance at the interacting fixed points, but this ordering principle makes it unlikely that, e.g., operators of mass dimension beyond  6 become relevant.
Within this scheme, a leading-order understanding of gravity-matter systems is based on the direct quantum-gravity contribution to the scale dependence of gauge couplings and Yukawa sectors (as well as the scalar potential).\\
 We therefore assume that at this leading order, induced higher-order matter interactions and non-minimal couplings can be neglected.
 A subset of those, selected by their global symmetries as discussed in \cite{Eichhorn:2017eht}, is generically nonzero at  an interacting fixed point  of the gravity-matter system, as has been studied for standard asymptotically safe gravity \cite{Eichhorn:2011pc,Eichhorn:2012va,Meibohm:2016mkp,Eichhorn:2016esv,Eichhorn:2017eht,Eichhorn:2017sok,Christiansen:2017gtg,Eichhorn:2018nda,Eichhorn:2019yzm}. More specifically, derivative interactions are induced in the presence of gravitational interactions. For example, this includes a $F^4$-term for gauge fields \cite{Christiansen:2017gtg,Eichhorn:2019yzm} and interactions of the schematic form $\bar{\psi}\slashed{\nabla}\psi \partial_{\mu}\phi\partial^{\mu}\phi$ for the Yukawa sector \cite{Eichhorn:2016esv,Eichhorn:2017eht}, as well as non-minimal derivative interactions \cite{Eichhorn:2017sok,Eichhorn:2018nda}.
  We expect this property to persist in the case of unimodular gravity. Given that around a flat background, the TT propagator at vanishing cosmological constant is the same in unimodular as in standard gravity in the linear parameterization, the TT-approximation of, e.g., \cite{Eichhorn:2017eht} carries over to the unimodular case and supports the presence of higher-order interactions at the fixed point. In our context it is important that these higher-order interactions are expected to be sub-leading compared to the direct quantum-gravity contributions that we calculate here, see \cite{Eichhorn:2017eht, Christiansen:2017gtg,Eichhorn:2019yzm} for a study of this in standard gravity.  This holds, as long as gravity is sufficiently weakly coupled such that the induced fixed points in \cite{Eichhorn:2016esv,Eichhorn:2017eht,Eichhorn:2017sok,Christiansen:2017gtg,Eichhorn:2018nda} remains at real values. This motivates our choice of truncation which neglects such higher-order operators.

We work with dimensionless couplings which are given by
\bea
G= G_N\, k^{ 2-d}, \quad a= \bar{a}\,k^{ 2}, \quad b= \bar{b}\,k^{2}.
\eea
Further, we introduce the anomalous dimensions
\be
\eta_{\phi/\psi/A}= - \partial_t \ln Z_{\phi/\psi/A}.
\ee

The gauge-fixing part is given by \cite{Eichhorn:2013xr}
\begin{align}
\Gamma_k^{\textmd{g.f.}} =&\, \frac{1}{32\pi \alpha\, G_N} \int_x \sqrt{\omega}\, \bar{g}^{\mu\nu} F^T_\mu[h] F^T_\nu[h] +\frac{1}{2\zeta}\int_x \sqrt{\omega} \, (\bar{g}^{\mu\nu} \bar{\nabla}_\mu A_\nu)^2 \,,
\end{align}
where $F^T_\mu[h]= P_{\T,\mu\nu} \,\bar{g}^{\nu\alpha}\bar{\nabla}^\beta h_{\alpha\beta}$   defines a transverse gauge fixing condition, with $P_{\T,\mu\nu} = \bar{g}_{\mu\nu} - \bar{\nabla}_\mu (\bar{\nabla}^2)^{-1} \bar{\nabla}_\nu$ being the transverse projector. The parameters $\alpha$ and $\zeta$ are gauge parameters for the gravitational and Abelian sectors, respectively.
 The use of the transverse part of the usual gauge fixing condition $F_{\mu}[h] =  \bar{\nabla}^\beta h_{\mu\beta}- \frac{1+\beta}{d} \bar{\nabla}_\mu h \,$  is  required to account for the fact that the symmetry underlying UQG corresponds to the \textit{TDiff} group instead of the \textit{Diff} group. For a discussion of BRST quantization of unimodular gravity see \cite{Upadhyay:2015fna,Alvarez:2015sba}.

The explicit form of the ghost sector will not be relevant for the analysis performed in this paper. For further details on the gauge fixing procedure for UQG see, e.g., \cite{Ardon:2017atk,Eichhorn:2013xr,Benedetti:2015zsw,Eichhorn:2015bna}. 

In addition, we perform a York decomposition \cite{York:1973ia} of the fluctuation field (note the absence of the trace mode due to the unimodularity condition),
\begin{align}
h_{\mu\nu} = h_{\mu\nu}^{\TT} + \bar{\nabla}_\mu \xi_\nu + \bar{\nabla}_\nu \xi_\mu + \bar{\nabla}_\mu \bar{\nabla}_\nu \sigma - \frac{1}{d} \bar{g}_{\mu\nu} \bar{\nabla}^2 \sigma .
\end{align}
It is convenient to adopt the Landau gauge $\alpha \to 0$ which further removes the $\xi_\mu$ degree of freedom such that it does not contribute to the flow of matter couplings \cite{Dona:2015tnf}. Furthermore, we employ the non-local field redefinition $\sigma \mapsto \Big( (-\bar{\nabla}^2)^2 \!+\! \frac{d}{d-1} \bar{\nabla}^\mu \bar{R}_{\mu\nu} \bar{\nabla}^\nu \Big)^{\!\!\!-1/2} \!\!\sigma\,$ in order to cancel the $\sigma$-part of the Jacobian in the generating functional arising from the York decomposition.

Let us briefly discuss the structure of the (flat) propagator obtained from the unimodular truncation given by Eq.~\eqref{EAA_UG}. The propagators for the TT- and $\sigma$-modes  are, respectively, given by
\begin{subequations}
	\begin{align}
	G_\TT^{\mu\nu\alpha\beta}(p^2) =& \frac{1}{ Z_\TT \left(  p^2+  \bar{b} \,p^4 \right)}P_\TT^{\mu\nu\alpha\beta} \,,\\
	G_{\sigma\sigma}(p^2) =& -\frac{d^2}{(d-2)(d-1)}
	\frac{1}{Z_\sigma \left( p^2  - \frac{4\,\bar{a}\,(d-1)+ \bar{b} \,d}{d-2} p^4\right)}\, ,
	\end{align}
\end{subequations}
 with $Z_\TT$ and $Z_\sigma$ defined by Eq.~\eqref{zttandzsigma}. The first point to be emphasized is the appearance of massive poles in both the TT- and $\sigma$-sectors, namely
\begin{align}
p^2 = - \,\bar{b}^{-1} \qquad \textmd{and} \qquad 
p^2 = \frac{d-2}{4\,\bar{a}\,(d-1)+ \bar{b} \,d}\,.
\end{align}
In the framework of perturbative curvature-squared quantum gravity, the existence of such poles is problematic. In particular, according to the usual perturbative treatment, the pole at $p^2 = - \,\bar{b}^{-1}$ corresponds to either a massive ghost-like particle if $\bar{b} < 0$ or a tachyon if $\bar{b} > 0$, therefore leading to unitarity or causality problems. Recent proposals on how to avoid such ghosts include, e.g., \cite{Holdom:2015kbf,Holdom:2016xfn,Donoghue:2017fvm,Donoghue:2018lmc,Anselmi:2018tmf}. 

From the FRG perspective, on the other hand, the association of such poles with ghosts/tachyons cannot directly be made. The presence of instabilities (non-unitarity) should be analyzed at the level of the full effective action, i.e., $\Gamma_{k \rightarrow 0}$.  On the other hand, the propagators shown above were obtained within a \emph{truncation} of the flowing action within a derivative expansion. Further, we are only focusing on the fixed-point regime at large $k$. Accordingly, it is not clear whether such higher-order terms will be present or not in the full non-perturbative result for the effective action $\Gamma_{k\rightarrow 0}$. Let us stress that the presence of higher-order terms in the effective action does not automatically result in instabilities/non-unitarity. This becomes obvious by considering the effective action in cases like QED or QCD, where higher-order terms are certainly present, but do not signal any inconsistency. Moreover, already the analysis of classical instabilities by Ostrogradsky, see \cite{Woodard:2015zca} for a pedagogical review, is based on the assumption of a finite number of higher-order terms. 
In the case of asymptotic safety one expects infinitely many higher-order terms to be present that enter the full propagator of the physical theory at $k \rightarrow 0$. We stress that even for a function $\Gamma^{(2)}(p^2)$ which only features a simple zero at $p^2=0$, a Taylor expansion to finite order generically features additional zeros. Therefore truncations of $\Gamma_k$ to finite order in the derivative expansion are not suitable to address the question of instability/unitarity.
Therefore no automatic conclusion can be drawn on unitarity from the presence of higher-order terms in a truncation for $\Gamma_k$. Analyzing the full propagator is beyond the scope of this work.

The above propagators results in the presence of the following contributions to the running of matter couplings and anomalous dimensions
\begin{align}
\frac{1}{(b + 1)^{\#_1}} \qquad \textmd{and} \qquad 
\frac{1}{\left(d-2 - 4\,a\,(d-1) - b \,d \,\right)^{\#_2}}\, ,
\end{align}
with $\#_{1,2}\geq 1$.
 The gravitational parameter space in our truncation therefore features two pole lines in $a$ and $b$, namely
\begin{align}
b + 1 = 0 \qquad \textmd{and} \qquad d-2 - 4\,a\,(d-1) - b \,d = 0 \,.
\end{align}
The flow cannot cross these pole lines (or the corresponding generalizations in an enlarged parameter space.)
The current experimental bounds on $a$ and $b$, see, e.g., \cite{Calmet:2017rxl,Chen:2019stu,Kim:2019sqk} and references therein, do not lead to significant restrictions on the parameter space, and do not provide any guidance as to which side of the poles is of phenomenological interest.

\subsection{Functional Renormalization Group for Weyl-squared gravity}

We explore, taking into account Weyl-squared gravity along with Standard-Model-like interactions, the viability of a UV completion for the matter sector in the gravitational parameter space, which here is spanned by the coupling $w$. 

An interesting point about a conformal gravity description of the fixed-point regime is  that if we restrict ourselves to the subspace of local terms, then there is only a finite number of Weyl-invariant operators. Under this assumption, the most general (local) truncation for Weyl-squared gravity (WG) coupled to a fermion, a scalar and a gauge field is given by 
\begin{align} \label{EAA_CG}
\Gamma_k^{\textmd{WG}} &= \int_x \sqrt{g} \,\bigg[ \, \frac{1}{2\,w}  C^2 
+ \frac{{Z_{\phi}}}{2} g^{\mu\nu}\partial_\mu \phi \,\partial_\nu \phi + \frac{\lambda_4}{4!} \phi^4 + \frac{\chi}{2} \phi^2 R  +\nn\\
&\qquad\qquad \, + \frac{{Z_A}}{4}\,g^{\mu\alpha}g^{\nu\beta} F_{\mu\nu} F_{\alpha\beta} + {Z_{\psi}}i \bar{\psi} { \slashed{\nabla}} \psi + iy \,\phi \bar{\psi} \psi \, \bigg] + \Gamma_k^{\textmd{g.f.}} +  \Gamma_k^{\textmd{gh.}} . 
\end{align} 
 Weyl symmetry requires the renormalized non-minimal coupling $\tilde{\chi}= \chi/Z_{\phi} =1/6$. As we will only explore the Yukawa and gauge sector, the non-minimal coupling and scalar potential are actually irrelevant for us.\\
As an alternative to the above, one can introduce a Weyl gauge field that allows to render the kinetic term for the scalar  field Weyl invariant on its own, such that the coupling of the non-minimal term can be arbitrary, see, e.g., \cite{Ferreira:2018itt}. As a phenomenologically important consequence, this allows to absorb the dilaton that arises as the Goldstone boson of spontaneous scale symmetry breaking in the longitudinal mode of the corresponding massive vector, thereby evading fifth-force constraints.

For our explicit calculation we consider the following gauge-fixing sector
\begin{align}
\Gamma_k^{\textmd{g.f.}} =&\, \frac{1}{2\alpha} \int_x \sqrt{\bar{g}}\, F_\mu[h] Y^{\mu\nu} F_\nu[h] +  \frac{\mu_0^4}{2\tilde{\alpha}} \int_x \sqrt{\bar{g}} \, \,h^2 +\frac{{Z_{A}}}{2\zeta}\int_x \sqrt{\bar{g}} \, (\bar{g}^{\mu\nu} \bar{\nabla}_\mu A_\nu)^2 ,
\end{align}
where $Y_{\mu\nu} = \gamma_1 \,\bar{g}_{\mu\nu} \bar{\nabla}^2 + \gamma_2 \,\bar{\nabla}_\mu \bar{\nabla}_\nu$ and $F_{\mu}[h] =  \bar{\nabla}^\nu h_{\mu\nu}- \frac{1+\beta}{4} \bar{\nabla}_\mu h$. The first term is a gauge-fixing for the diffeomorphisms. We have chosen a higher-derivative gauge-fixing in order to avoid the introduction of a mass scale in this sector. The second term is the gauge-fixing term for the Weyl symmetry. In this case, we introduce an arbitrary mass parameter $\mu_0$. In the Landau gauge, $\alpha \to 0$ and $\tilde{\alpha} \to 0$, the graviton propagator becomes independent of the parameters $\gamma_1$, $\gamma_2$, $\beta$ and $\mu_0$. As in the unimodular case, the ghost sector is not relevant for the computations performed in this paper. For further details on the gauge-fixing procedure in the framework of Weyl-squared gravity, see \cite{Ohta:2015zwa,Ohta:2016jvw}.

We perform a York decomposition \cite{York:1973ia} of the fluctuation field
\begin{align}
h_{\mu\nu} = h_{\mu\nu}^{\TT} + \bar{\nabla}_\mu \xi_\nu + \bar{\nabla}_\nu \xi_\mu + \bar{\nabla}_\mu \bar{\nabla}_\nu \sigma - \frac{1}{4} \bar{g}_{\mu\nu} \bar{\nabla}^2 \sigma 
+ \frac{1}{4} \bar{g}_{\mu\nu} h . 
\end{align}
In terms of York decomposed variables, the Landau gauge $\alpha \to 0$ and $\tilde{\alpha} \to 0$ entails an interesting simplification. In this case, the vector and scalar sectors ($\xi_\mu$, $\sigma$ and $h$) do not contribute to the running of the matter couplings. Therefore, all computations can be done by taking into account only the TT-mode.
The Hessian in the gravitational sector is given by
\begin{align}
\left[\Gamma_{k,h^\TT\! h^\TT}^{(2)}\right]^{\mu\nu\alpha\beta} =   Z_\TT \,p^4 \,P_\TT^{\mu\nu\alpha\beta} \,.
\end{align}
The regulator associated with the $\TT$ sector is
\begin{align}
[\textbf{R}_k^\TT(p^2)]^{\mu\nu\alpha\beta} =   Z_\TT \,[ P_k(p^2)^2 - p^4 ]\,P_\TT^{\mu\nu\alpha\beta} \,,
\end{align}
where $P_k(p^2) = p^2 + R_k(p^2)$, for a generic shape function $R_k(p^2)$.
The other relevant objects for computations within the Weyl-squared gravity framework, such as the Hessians associated with the matter sector and the gravity-matter vertices, are basically the same as in the UQG case. These can be obtained from the expressions reported in App.~\ref{Further_UG} (with the replacement $\kappa \to \sqrt{2 \,w}$.)

Although the model exhibits  Weyl invariance, the introduction of the regulator function in the FRG framework breaks this symmetry. Intuitively, the regulator is a momentum-dependent mass term and therefore is not invariant under Weyl rescalings\footnote{This can be dealt with in principle by introducing a dilaton, see  \cite{Percacci:2011uf,Codello:2012sn}.}. Hence, the flow generates couplings which break the Weyl  symmetry and which are controlled by appropriate modified Ward identities. This falls outside the scope of the present paper and we restrict the ansatz for the flowing action to be Weyl  invariant. The situation is analogous to that of other local symmetries, reviewed, e.g., in \cite{Pawlowski:2005xe,Gies:2006wv}.\\
  We also highlight that higher-order matter interactions, as induced by gravity without Weyl symmetry \cite{Eichhorn:2011pc,Eichhorn:2012va,Meibohm:2016mkp,Eichhorn:2016esv,Eichhorn:2017eht,Eichhorn:2017sok,Christiansen:2017gtg,Eichhorn:2018nda,Eichhorn:2019yzm}, are all dimension-5 or higher operators, which introduce explicit mass-terms and are accordingly incompatible with Weyl symmetry. In the flow-equation setup, they are presumably present as a cutoff-artifact, and also subject to modified Ward-identities.

\section{Quantum-gravity induced ultraviolet completion of Standard-Model like theories \label{sec:results}}

The Abelian hypercharge sector and the Higgs-Yukawa sector of the Standard Model exhibit Landau poles in perturbation theory, most likely rendering the Standard Model UV incomplete. This is a consequence of the fact that the free fixed point is IR attractive in these couplings. In this work, we explore the question whether quantum gravity can solve this problem by inducing a (near-) perturbative ultraviolet completion. \\
Results obtained in the last few years indicate that asymptotically safe quantum-gravity effects might induce a UV completion of the Standard Model and might even allow to predict (or retrodict) the values of several couplings which are free parameters in the Standard Model without gravity \cite{Shaposhnikov:2009pv,Harst:2011zx,Eichhorn:2017ylw,Eichhorn:2017lry,Eichhorn:2017als,Eichhorn:2018whv}. Here, we extend this study to unimodular as well as Weyl-squared gravity. For the latter, the dimensionless nature of the gravitational coupling suggests that the leading-order contribution is universal. Our study is the first implementation of this calculation in the FRG framework, providing an explicit test of one-loop universality by comparison with previous perturbative results.\\ 
The quantum-gravity contribution is generically linear in Standard-Model-like matter couplings, in accordance with symmetry considerations, see \cite{Eichhorn:2017eht} and as  follows from the corresponding diagrammatic expressions.
Therefore, the quantum-gravity contribution to the running matter couplings $g_i$ (e.g., gauge and Yukawa), takes the form
\be
\beta_{g_i}{|_{\rm grav}} = -f_{g_i}\, g_i +...,
\ee
where $f_{g_i}$ is a function of the gravitational couplings. As gravity is ``blind" to internal symmetries, this direct quantum-gravity contribution to the scale-dependence of gauge couplings is independent of the gauge group. Similarly, there is no flavor-dependence of the direct quantum-gravity contribution to the scale-dependence of Yukawa couplings, nor a dependence of the direct quantum-gravity contribution on the internal symmetries of a scalar sector. Thus, $f_{g}$ is the same for all gauge couplings and $f_y$ is the same for all Yukawa couplings.
 To induce an ultraviolet completion for gauge and Yukawa couplings, the gravity contribution needs to be antiscreening, i.e., $f_{g}>0$ and $f_y>0$. In fact, such a linear term is also present in $d\neq4$ spacetime dimensions, where gauge and Yukawa interactions are not marginal. The case $f_{g_i}>0$ is analogous to an effective dimensional reduction (though not necessarily to an integer spacetime dimension). \\
For the Higgs self-coupling, the situation is slightly different: A screening quantum-gravity contribution actually results in a prediction of the Higgs mass close to the experimental value \cite{Shaposhnikov:2009pv}. This is a consequence of the fact that a Higgs mass of about 125 GeV is connected to a near-vanishing Higgs quartic coupling \footnote{The exact value features a delicate dependence on the mass of the top quark \cite{Buttazzo:2013uya,Bezrukov:2014ina}.}, which in turn follows from a screening quantum-gravity contribution, as found in \cite{Narain:2009fy,Oda:2015sma,Percacci:2015wwa,Hamada:2017rvn,Eichhorn:2017ylw,Eichhorn:2017als,Pawlowski:2018ixd}. 
On the other hand, an antiscreening contribution would result in the Higgs self-coupling being a free parameter of the theory, such that the model would also be UV complete, albeit less predictive.

In the following, we therefore focus on evaluating the leading-order quantum gravity contribution (i.e., linear in the respective matter couplings) to the beta functions of a (non-)Abelian gauge coupling and a Yukawa coupling.

\subsection{(Non-)Abelian gauge couplings \label{Gauge_Viability}}

\begin{figure}[htb!]
	\begin{center}
		\includegraphics[width=6.0cm]{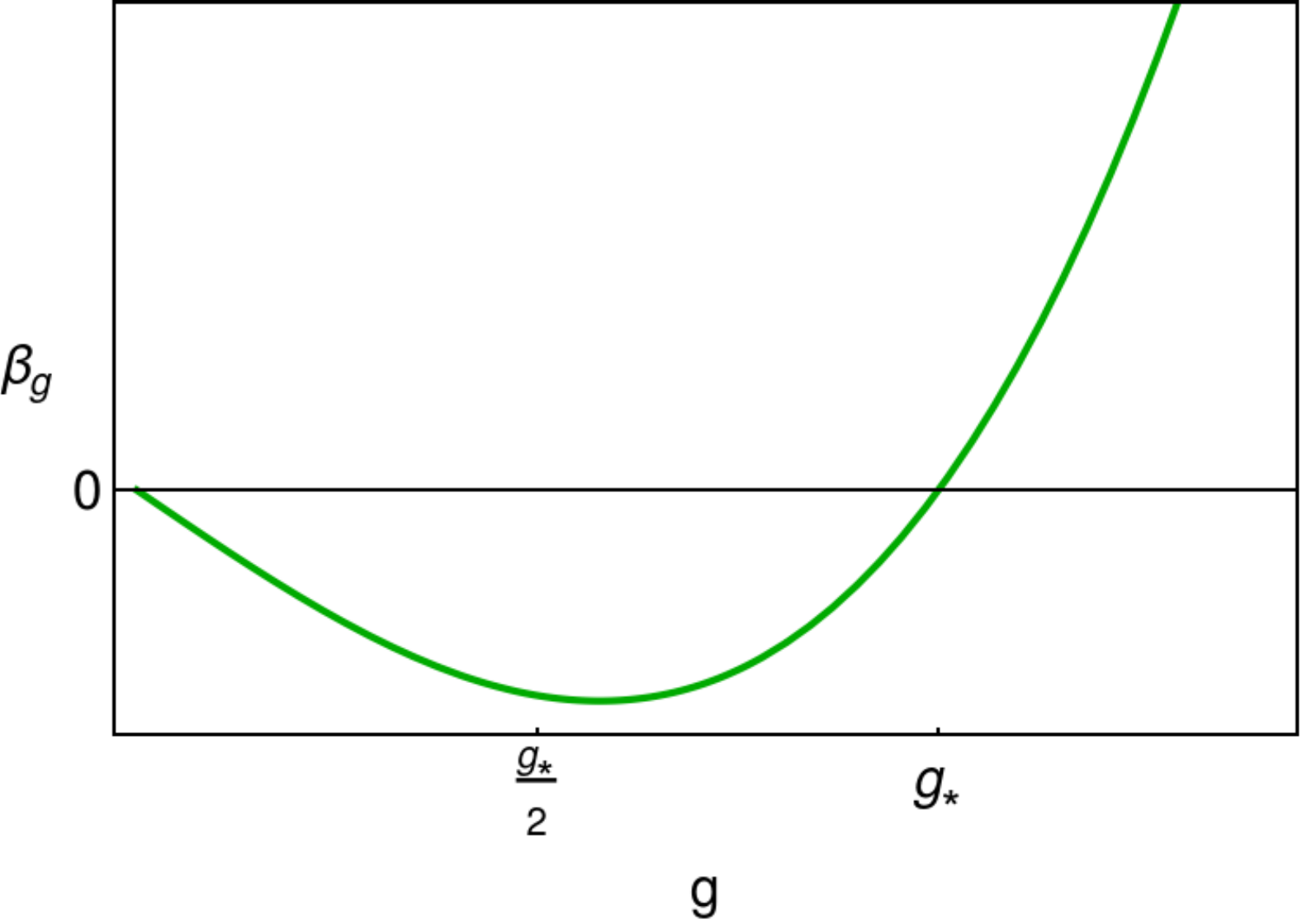}
		\quad\quad
		\includegraphics[width=6.7cm]{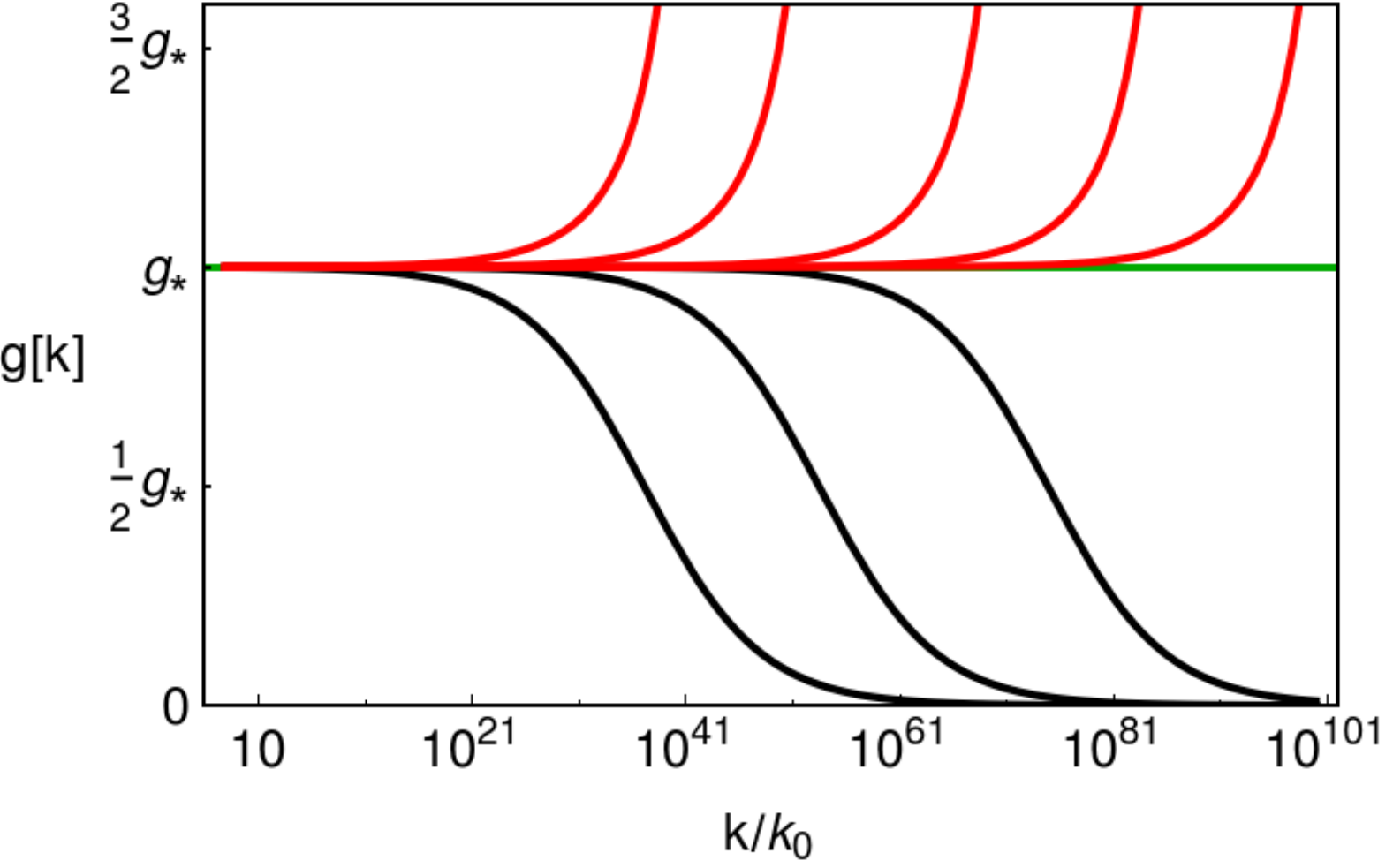}
		\caption{\label{fig:ill_upperbound} We illustrate the beta function (left panel) arising from $f_g<0$ in Eq.~\eqref{eq:etaA_schem} and the corresponding RG trajectories (right panel). The interaction fixed point at $g_{\ast}$ generates an upper bound on low-energy values of $g$: No value above the green line in the right panel can be reached on a UV complete trajectory.}
	\end{center}
\end{figure}

The quantum-gravity contribution to the running of a minimal (non-)Abelian gauge coupling, denoted as $g$, is related to the anomalous dimension $\eta_A$,
\begin{align}
\beta_{g^2} = \eta_A \, g^2.
\label{beta_e2}
\end{align}
It receives contributions from the interaction with other matter fields and also from quantum-gravity fluctuations, namely
\begin{align}
\eta_{A} = \eta_A|_{\textmd{matter}} + \eta_A|_{\textmd{grav}} \,.\label{eq:etaA_schem}
\end{align}
The first contribution, i.e., $\eta_A|_{\textmd{matter}}$ is positive for Abelian gauge fields due to the screening impact of charged matter fields and starts at quadratic order in $g$, indicating a single UV repulsive free fixed point in the absence of gravity. For instance, in the presence of a single charged fermion, $\eta_A|_{\textmd{matter}} = g^2/(6\pi^2)$.
The situation may potentially change if the quantum-gravity part is taken into account as discussed in \cite{Harst:2011zx,Eichhorn:2017lry,Eichhorn:2018whv,Eichhorn:2017muy}. If the latter admits a region in the space of parameters such that $ \eta_A|_{\textmd{grav}} < 0$, corresponding to $f_g>0$, then two possibilities for a UV completion are generated, illustrated in Fig.~\ref{fig:ill_upperbound}: 
\begin{enumerate}
	\item an asymptotically free Abelian gauge coupling;
	\item an asymptotically safe Abelian gauge coupling with uniquely calculable IR value.
\end{enumerate}

This fixed-point structure follows from a competition of the two terms in Eq.~\eqref{eq:etaA_schem}. Due to the quantum-gravity contribution, the fixed point at $g=0$ is IR repulsive. This allows to reach \textit{finite} IR-values of the gauge coupling along UV-complete RG trajectories, thereby solving the triviality problem. At the same time, the competition between the antiscreening quantum-gravity contribution and the screening matter contribution results in an IR attractive fixed point at a finite value $g_{\ast}$. Due to its IR attractive nature it generates an upper bound on IR values of the gauge coupling: The unique IR value of $g$ that is reached along the trajectory emanating from the interacting fixed point is the largest IR-value of $g$ that follows from any UV complete trajectory. Larger IR-values of $g$ cannot be reached starting from the free fixed point, as the critical trajectory emanating from $g_{\ast}$ cannot be crossed.

For non-Abelian theories with an appropriate matter content, as, e.g., in the Standard Model, it holds that $\eta_A|_{\textmd{matter}}<0$, such that a non-Abelian sector is already UV complete without gravity. As long as $\eta_A|_{\textmd{grav}}<0$, this situation does not change once gravity is included. For grand unified theories with a large number of matter fields, such that asymptotic freedom is lost,  $\eta_A|_{\textmd{grav}}<0$ could result in a UV completion, see \cite{Eichhorn:2017muy}.

\subsection{Yukawa terms \label{Yukawa_Viability}}

Quantum-gravity fluctuations could also play an important role for the Yukawa sector.
As we have mentioned before, an antiscreening contribution coming from graviton fluctuations may induce a UV completion for the Yukawa coupling. 
In fact, considering the leading gravitational contribution\footnote{By ``leading contribution'' we mean that we are not considering contributions coming from induced fermion-scalar-fermion interactions, see refs.~\cite{Eichhorn:2016esv,Eichhorn:2017eht} for a discussion of such terms.}, the structure of the Yukawa  beta function is given by
\begin{align}
\beta_y = \left( {\frac{1}{2}}\eta_\phi|_{\textmd{grav}} + \eta_\psi|_{\textmd{grav}}+ \mathcal{D}_{y}\right) y +\beta^{(1)} y^3 = -f_y\, y+ 
\beta^{(1)}
y^3,
\label{betay}
\end{align}
where $\beta^{(1)}$ is the universal (i.e., RG-scheme independent) one-loop contribution from matter, with $\beta^{(1)}>0$. It also includes the matter contributions to the anomalous dimensions.
We distinguish between the gravitational contribution to the anomalous dimensions, and the direct gravitational contribution to the flow of the vertex, encoded in $\mathcal{D}_{y}$, where we adopt the notation from \cite{Eichhorn:2017eht}.  Both contribute at $\mathcal{O}(y)$. Thus,
the antiscreening nature of the gravitational contribution can be characterized by the following inequality
\begin{align} 
- f_y=\frac{1}{2} \eta_\phi|_{\textmd{grav}} + \eta_\psi|_{\textmd{grav}}  + \mathcal{D}_{y} < 0 \,.
\end{align}
 
The resulting fixed-point structure is as in the Abelian gauge sector: For $f_y>0$, the free fixed point is IR repulsive, rendering the Yukawa coupling asymptotically free. Due to a competition between antiscreening quantum-gravity fluctuations and screening matter fluctuations, an IR attractive interacting fixed point exists. It shields the free fixed point from large values of the Yukawa coupling in the IR, i.e., trajectories emanating from the free fixed point can only reach a finite range of IR values. That range is bounded from above by the unique value that is reached along the trajectory emanating from the interacting fixed point.

 Our goal in this paper is the first estimation of $f_y$ and $f_g$ for unimodular gravity and Weyl-squared gravity from functional RG techniques. 
 For unimodular gravity, the gravitational contribution to Yukawa systems in the perturbative setting has been studied in \cite{Gonzalez-Martin:2018dmy}. We will in particular interpret our results in the context of unimodular asymptotic safety, whereas the perturbative literature does not assume the existence of a fixed point for gravity.
 \\
For Weyl-squared gravity, the absence of mass-scales in theory implies that the one-loop results must be universal. It is one of our goals to explicitly test this within the FRG setup. We will therefore explore the regulator-, gauge- and parameterization-dependence of the beta functions in the Weyl-squared case to confirm that the FRG, as expected, reproduces universal results also in a gravitational setting.

\section{Results for Unimodular Gravity}\label{sec:UGres}

\subsection{Higgs-Yukawa and (non-)Abelian gauge sector in unimodular gravity}

In this section, we explore where in the gravitational parameter space a UV completion for Yukawa and gauge sectors in the framework of unimodular gravity is possible. The relevant quantities for this analysis are $\eta_\phi$, $\eta_A$, $\eta_\psi$ and $\mathcal{D}_y$ (see Secs.~\ref{Gauge_Viability} and \ref{Yukawa_Viability}). Within our truncation for unimodular gravity, Eq.~\eqref{EAA_UG}, we find the following results
\bea
\label{Anomalous_phi_UG}
\eta_\phi|_{\textmd{grav}} &=&\frac{G}{40\pi} \bigg[ \frac{25\, ( 2 +3\,b ) }{\left(1+b\right)^2}  + \frac{4(5-33a-11b)}{(1-6a-2b)^2} \bigg]\,,\\
\label{Anomalous_A_UG}
\eta_A|_{\textmd{grav}} &=& - \frac{G}{90 \pi} \bigg[\frac{5\,(10+7b)}{(1+b)^2} 
- \frac{4\,(5-21a-7b)}{(1-6a-2b)^2}\bigg]\, ,\\
 \label{Anomalous_psi_UG}
\eta_\psi|_{\textmd{grav}} &=& \frac{G}{160\pi}\bigg[\frac{25\, (2 + 3 b) }{(1+b)^2} 
-\frac{2\,(31-246 a - 82 b)}{(1-6a - 2b)^2}\bigg] \, ,\\
\label{D_y_UG}
\mathcal{D}_y|_{\textmd{grav}}& = &\frac{G}{20\pi} \frac{5 - 39 a - 13 b}{(1-6a-2b)^2} \,.
\eea
In the following, we assume a gravitational fixed-point in $G, a,b$, and explore for which values of these couplings it results in $f_{y/g}>0$. 
As all quantities are linear in $G$, the sign of $f_y$ and $f_g$ does not depend on $G$, as long as gravity is attractive, i.e., $G>0$.
Accordingly, these results determine regions in the space of parameters $a$ and $b$ where the conditions required for a UV completion of Yukawa and gauge sectors, $f_g>0$ and $f_y>0$ are satisfied within our truncation. The shape and location of the boundary of these regions can additionally depend on higher-order couplings neglected in our truncation. Further, within a truncation, non-universality of gravitational corrections to matter beta functions implies that the shape and location of the boundary depends on the regulator  function. Therefore, it is not meaningful to combine a determination of the boundary within one specific choice of regulator function, with a determination of gravitational fixed-point values within a different choice. Of course, such unphysical dependences must cancel in physical results, at least up to the systematic uncertainties arising within an approximation. 

\begin{figure}[htb!]
	\begin{center}
	\includegraphics[width=7.5 cm]{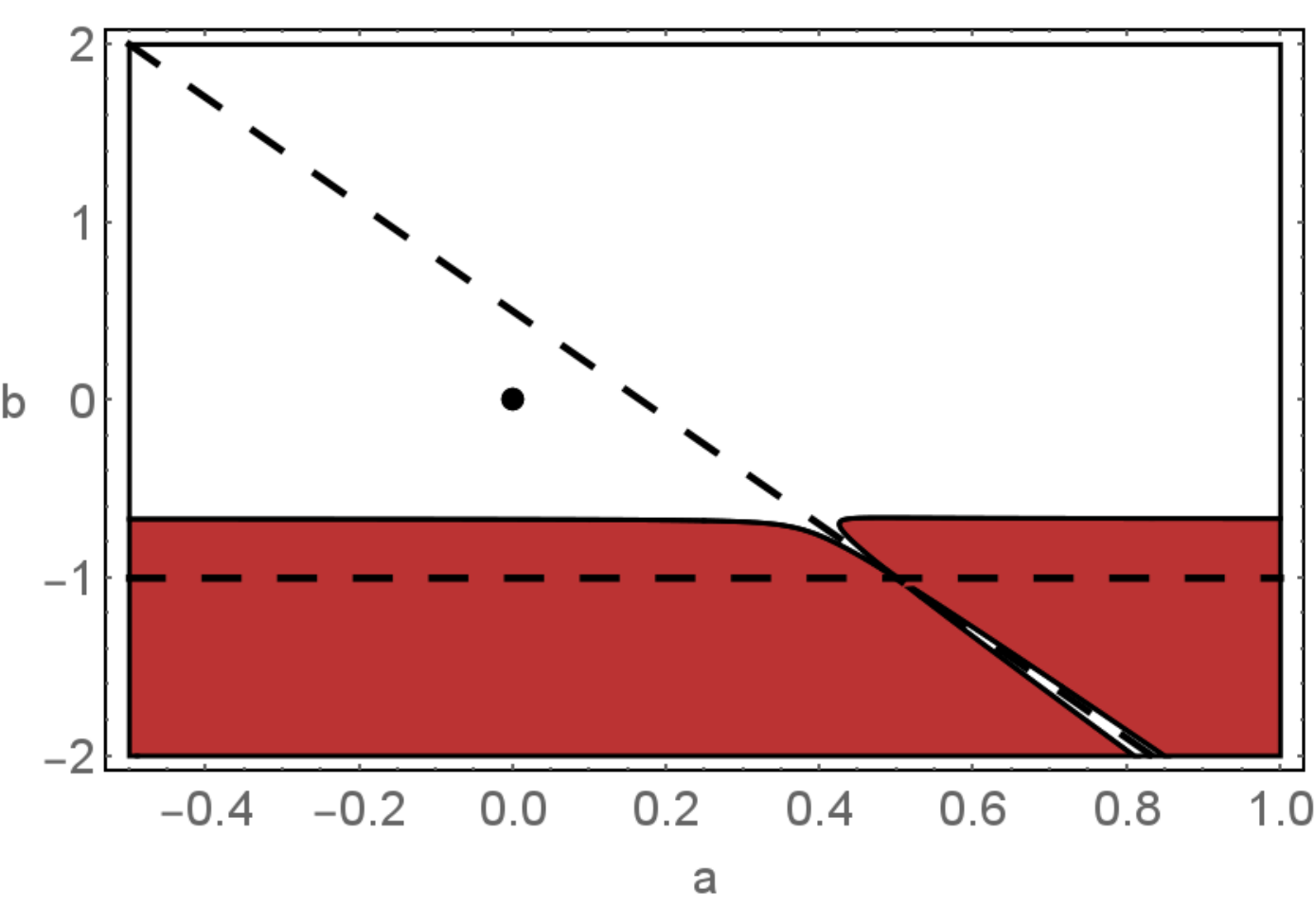}
	\caption{\footnotesize The red region corresponds to the sub-space of higher curvature parameters, $a$ and $b$, where the viability condition for a UV completion of the Yukawa sector, \textit{i.e.}, $-f_y=\left(\eta_\psi + \frac{1}{2} \eta_\phi + \mathcal{D}_y\right)_{\textmd{grav}} \!< 0$, is satisfied. The dashed lines indicate the poles $1+b = 0$ ($\TT$-mode) and $1-6a-2b$ = 0 ($\sigma$-mode).  The black dot marks the Einstein-Hilbert truncation.}
	\label{Yukawa_viability}
	\end{center}
\end{figure}

In the Yukawa sector, the viability condition for a quantum-gravity induced UV completion is given by $-f_y=\left(\eta_\psi + \frac{1}{2} \eta_\phi + \mathcal{D}_y\right)_{\textmd{grav}} \!< 0$. This results in the following constraint on the space of curvature-squared couplings (assuming $G > 0$)
\begin{align}\label{Condition_Yukawa}
\frac{75\,(2+3b)}{(1+b)^2} + \frac{2\,(9-42 a -14 b)}{(1-6a-2b)^2} < 0 .
\end{align}  
In Fig.~\ref{Yukawa_viability} we plot the region where this inequality holds. A gravitational fixed point in that region would generate an antiscreening contribution to the Yukawa beta function. There is a region at negative $b$, with only a sub-leading dependence on $a$, satisfying this condition.
Except for the  vicinity of the pole $3 a + b = 1/2$, the viable region can be approximated by $b \lesssim -0.7$. \\
The point corresponding to the Einstein-Hilbert truncation ($a = b = 0$) does not belong to the viable region.  One might have expected this result from the analogous result in the standard gravity case: There, the presence of the cosmological constant is crucial in the absence of higher-order couplings: At vanishing cosmological constant (and for $a=0=b$), the transverse traceless contribution to $\beta_y$ dominates, and yields $f_y<0$. At sufficiently negative cosmological constant, a reweighing of contributions to $\beta_y$ occurs, such that the transverse traceless contribution is actually subdominant and $f_y>0$ can be realized, see \cite{Eichhorn:2017eht} for a comprehensive discussion. In the unimodular case, the cosmological constant no longer appears in the metric propagators. Accordingly, the results can be expected to be similar to those in the linear parameterization for standard gravity at vanishing cosmological constant (of course, the correspondence is not exact).
The inclusion of higher-order terms opens up a larger parameter space, where the nature of gravitational contributions can change from screening to antiscreening. 

\begin{figure}[htb!]
	\begin{center}
	\includegraphics[width=7.5 cm]{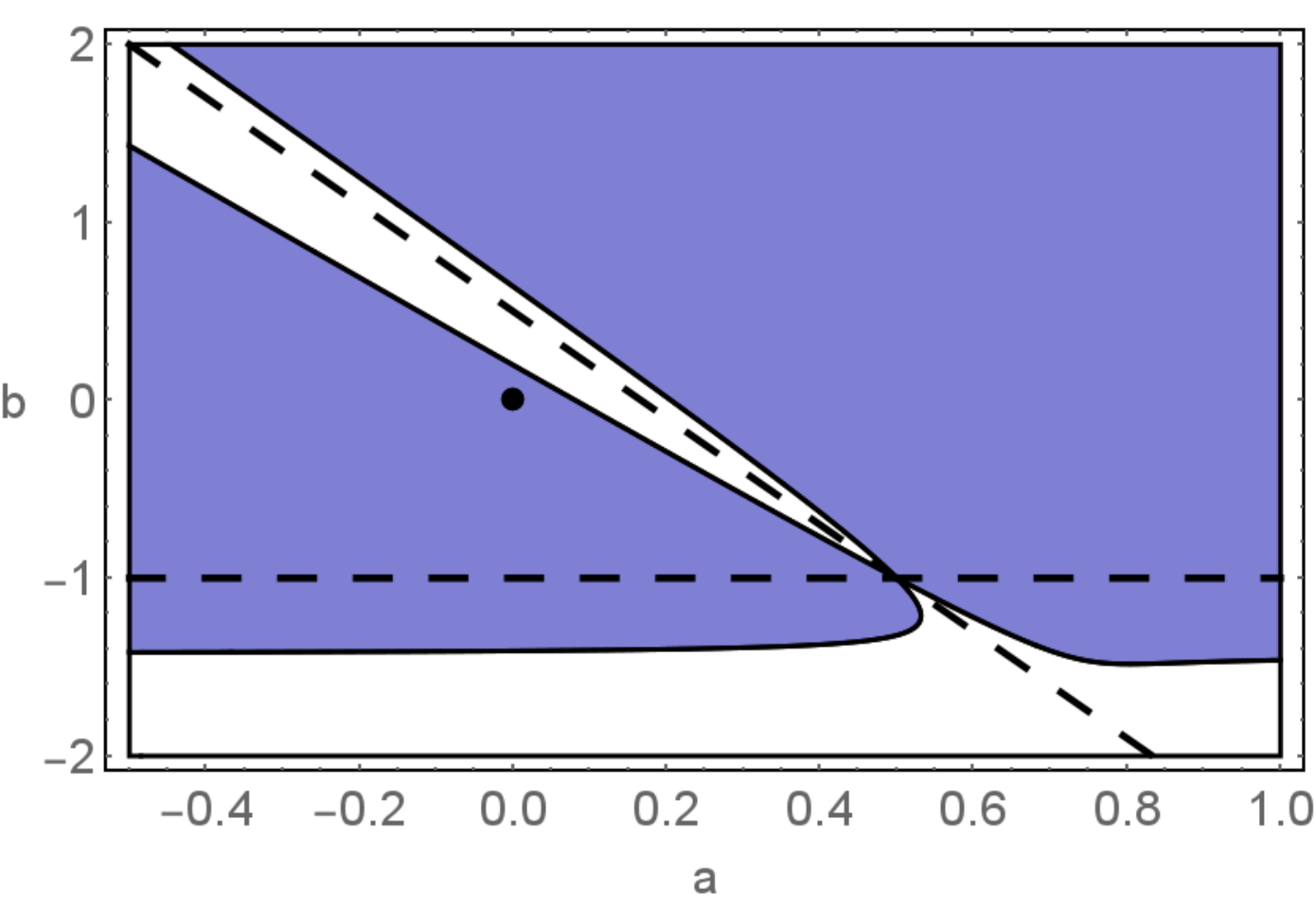}
	\caption{\footnotesize The blue region indicates where the inequality $-f_g = \eta_A|_{\textmd{grav}}\!< 0$, is satisfied. The dashed lines indicate the poles $1+b = 0$ ($\TT$-mode) and $1-6a-2b$ = 0 ($\sigma$-mode).}
	\label{abelian_viability}
	\end{center}
\end{figure}

The viable region for a UV complete Abelian gauge coupling ($-f_g=\eta_A|_{\textmd{grav}} < 0$) is characterized by
\begin{align}
\frac{5\,(10+7b)}{(1+b)^2} - \frac{4\,(5-21a-7b)}{(1-6a-2b)^2} > 0 \, ,\label{inequ:fg}
\end{align}
see Fig.~\ref{abelian_viability}.
Similarly to what happens in the Yukawa sector, the sign of the gravitational contribution to $\eta_A$ is mostly dictated by the coefficient of $R_{\mu\nu}^2$. This result can be made plausible by a counting argument: The coefficient of  $R_{\mu\nu}^2$ enters the propagator of the transverse traceless graviton (which losely speaking counts like 5 scalars), whereas that of $R^2$ only appears in the scalar mode. This argument holds in those regions of parameter space where an enhancement of scalar contributions due to the nontrivial denominators in the inequality \eqref{inequ:fg} is avoided.
In the present case, the viable region can be approximated by $b \gtrsim -1.4$, except in the neighborhood of the pole line $3 a + b = 1/2$. In contrast to the Yukawa sector, the Einstein-Hilbert point ($a=b=0$) belongs to the viable region for a UV completion of the gauge sector. Again, this can be plausibilized by the results in the standard-gravity case, where $f_g>0$ holds at vanishing cosmological constant, see, e.g., \cite{Folkerts:2011jz,Eichhorn:2017lry,Christiansen:2017cxa}.

\begin{figure}[htb!]
	\begin{center}
		\includegraphics[width=7.5 cm]{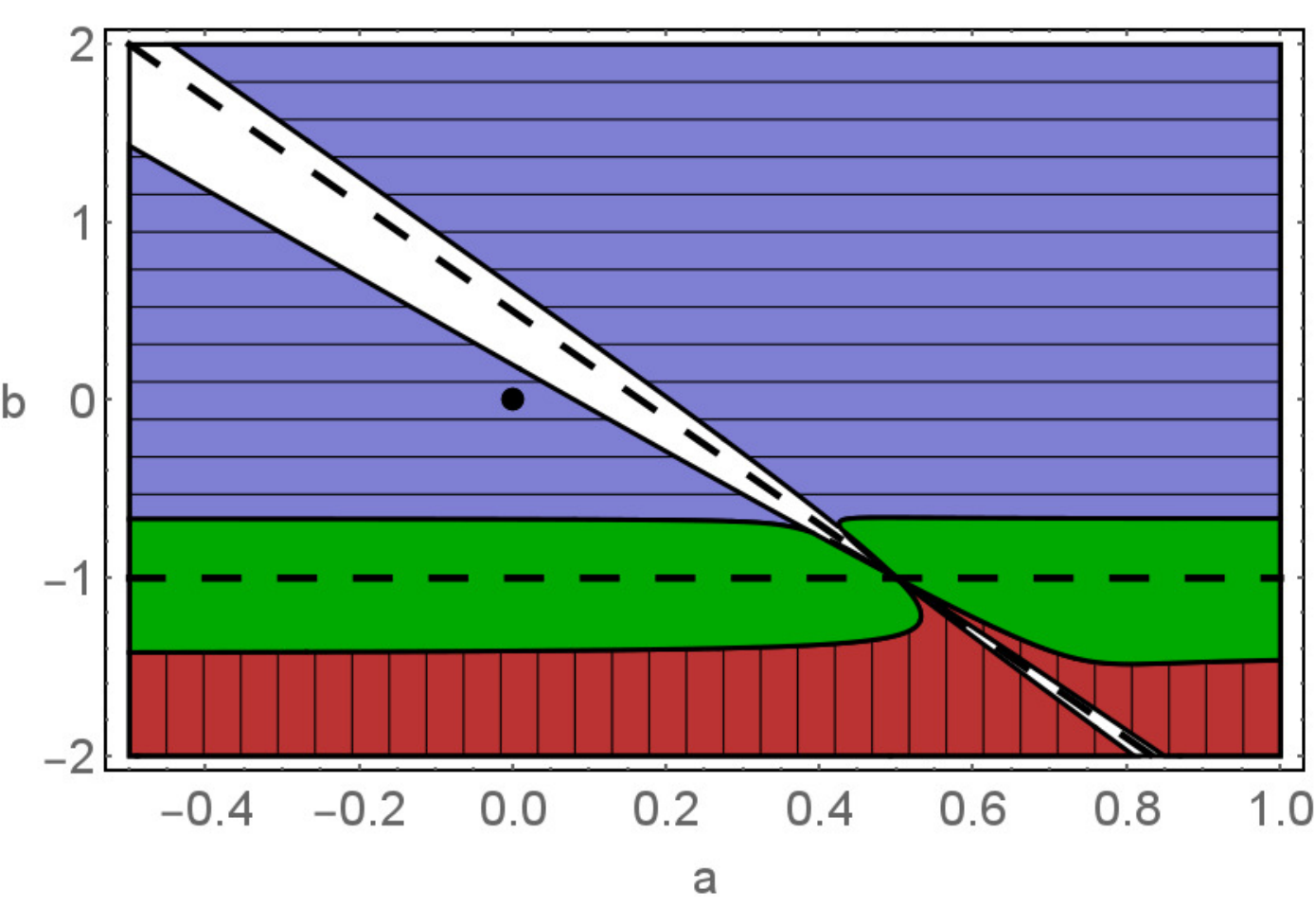}
		\caption{\footnotesize Combined plot showing the regions where the quantum gravitational contribution to gauge and Yukawa interactions is antiscreening. The red region (with vertical lines) corresponds to the sub-space of higher-curvature  couplings where only $f_y>0$ can be achieved. The blue region (with horizontal lines) indicates values of higher-curvature couplings where only $f_g>0$ holds. In the green region, $f_g>0$ and $f_y>0$ both hold. The dashed lines indicate the poles $1+b = 0$ ($\TT$-mode) and $1-6a-2b$ = 0 ($\sigma$-mode).}
		\label{Yuka_Gauge_viability}
	\end{center}
\end{figure}

\begin{figure}[htb!]
	\begin{center}
		\includegraphics[width=7.0 cm]{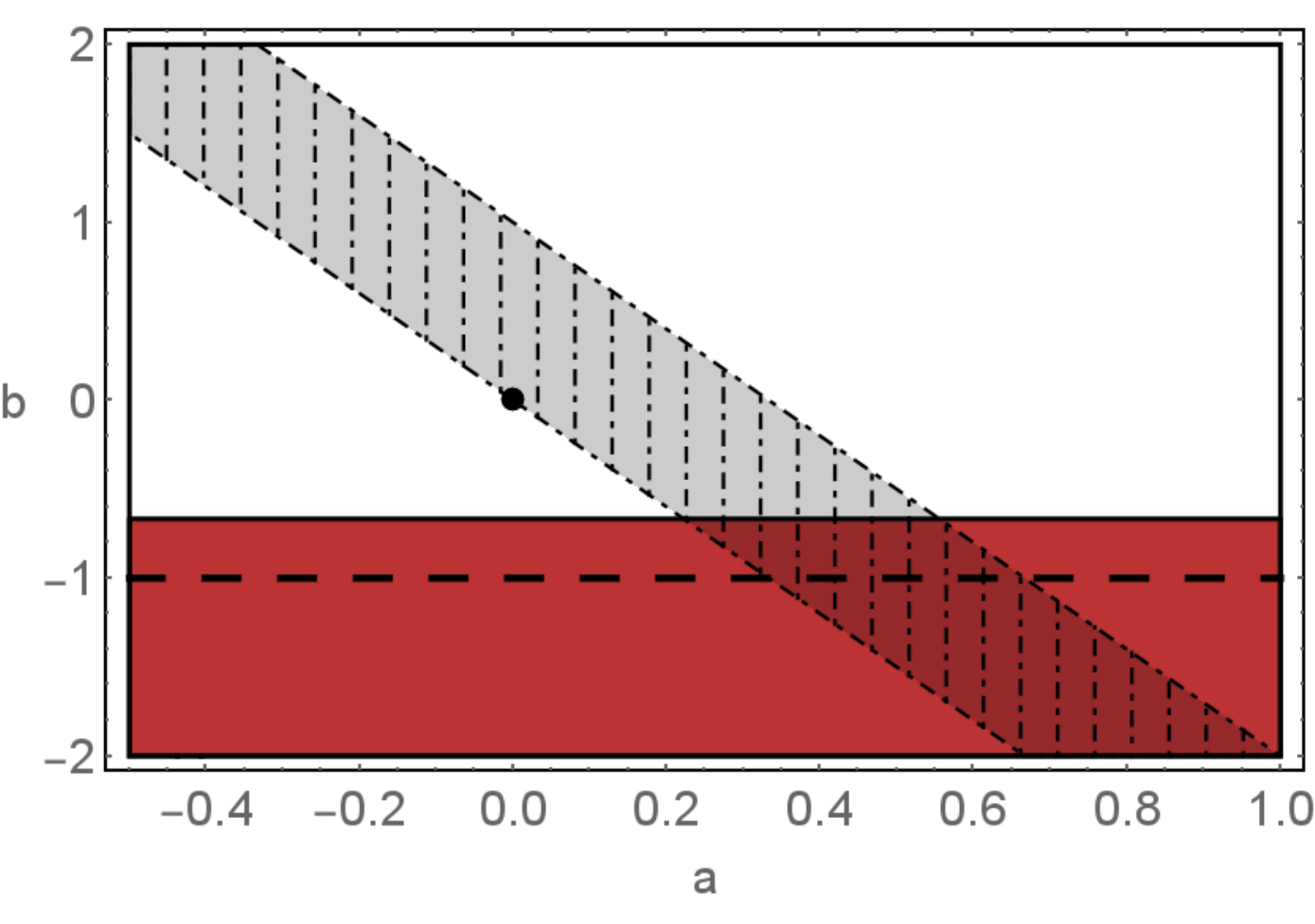}
		\includegraphics[width=7.0 cm]{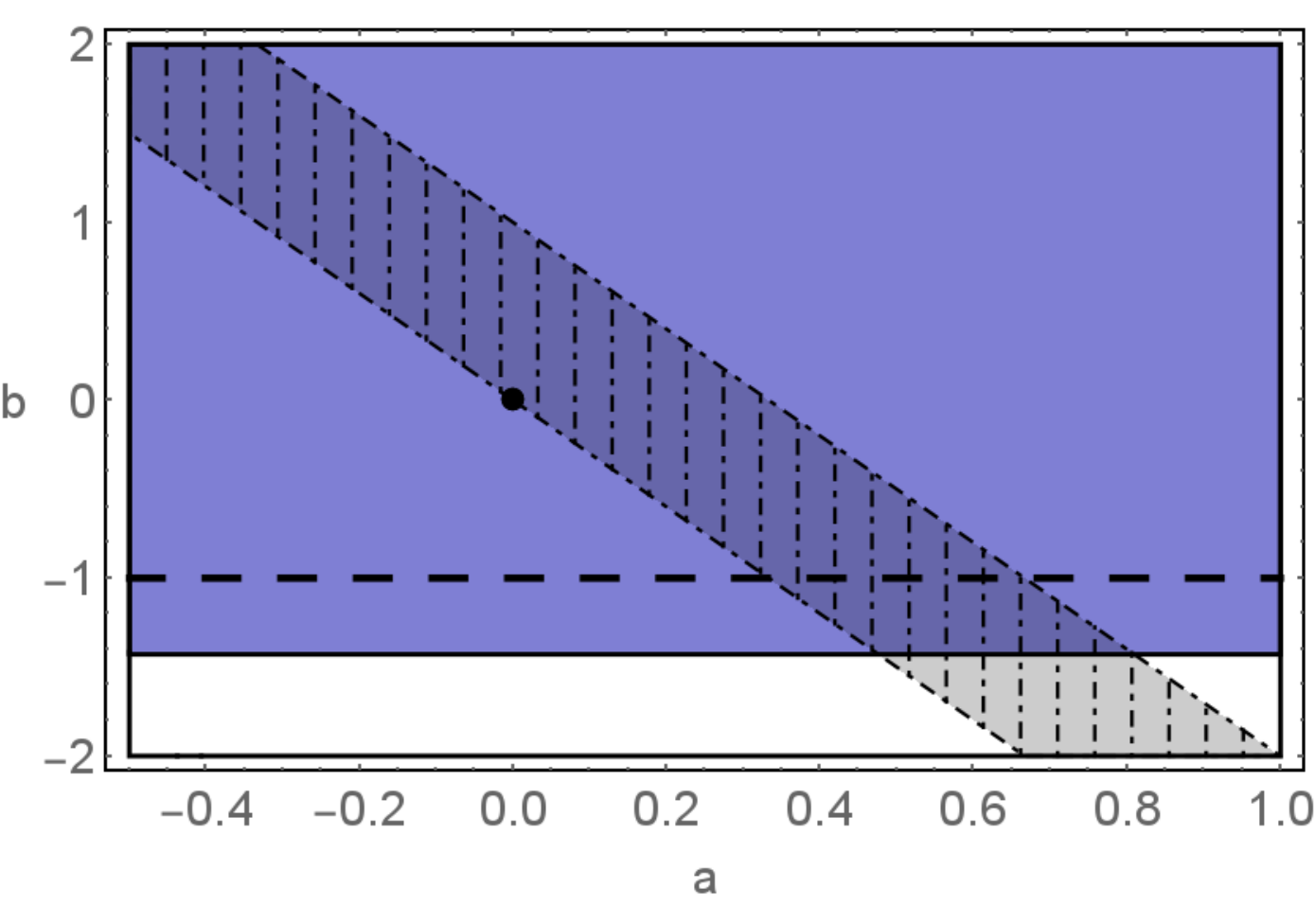}
		\caption{\footnotesize We show the regions where $f_y>0$ (red, left panel) and $f_g>0$ (blue, right panel). In both panels, the gray strip (with vertical dot-dashed lines) show the region were the TT-approximation differs from the full result. The horizontal dashed line indicates the pole $1+b = 0$.}
		\label{Yuka_Gauge_viability_TT}
	\end{center}
\end{figure}

In Fig.~\ref{Yuka_Gauge_viability} we present the combined  constraints on the gravitational parameter space arising from $f_g>0$ and $f_y>0$. Far away from the pole line $3 a + b = 1/2$, the viable region can be approximated by $-1.4 \lesssim b \lesssim -0.7$. This approximated behavior can be understood in terms of a dominance of the transverse traceless mode, 
``TT-dominance'' for short, as discussed above. Fig.~\ref{Yuka_Gauge_viability_TT} shows the viable region in the TT-approximation, which is obtained by neglecting contributions from diagrams containing the $\sigma$-modes, i.e., by neglecting quantum fluctuations of the scalar mode. \newline

So far, we have focused on exploring where in the gravitational parameter space a UV completion of an Abelian gauge sector and a Yukawa sector could be possible. Yet, the thus-defined viable region is not automatically phenomenologically viable, given the measurements of IR values of the corresponding couplings for the Standard Model. If there was only the free fixed point, then any IR value of the couplings could be reached by the RG flow. However, for marginally irrelevant couplings, such as the Abelian gauge and Yukawa coupling, there is automatically  a second fixed point of the beta functions in Eq.~\eqref{beta_e2} and \eqref{betay}. It arises from the competition of the screening one-loop term in the beta functions with the antiscreening gravity term in the case where $f_{y/g}>0$. The larger $f_y$ and $f_g$, the larger the corresponding fixed-point value. This fixed point is necessarily IR attractive. Accordingly, it acts as an upper bound for trajectories emanating from the free fixed point, see \cite{Eichhorn:2017ylw,Eichhorn:2017lry}. Therefore, given a set of values for $f_y$ and $f_g$, only IR-values of $g_Y$ and $y$ up to an upper bound can be reached. To illustrate this, we supplement $\beta_y$ by the one-loop contribution for a single Yukawa coupling from the Standard Model,  see \cite{Cheng:1973nv}. We neglect all terms related to the other Yukawa couplings and the non-Abelian gauge fields, and only keep the Abelian hypercharge contribution, putting in the hypercharge for an up-type flavor (top, charm, up),
\begin{align}
\beta_y = -f_y\, y + \frac{9}{32\pi^2}y^3- \frac{1}{16\pi^2} \frac{17}{12}y \, g_Y^2 .
\end{align}
Similarly, we supplement the one-loop term in the beta function of the Abelian gauge coupling in the Standard Model (i.e., including all charged fermions) with the gravity contribution, obtaining
\begin{align}
\beta_{g_Y} = - f_g\, g_Y + \frac{41}{6}\frac{g_Y^3}{16\pi^2}.
\end{align}
This set of beta functions features the following fixed points (we neglect fixed points at negative values)
\begin{align}
y_{\ast}&=0,\qquad g_{Y\, \ast}=0,\\
y_{\ast}&=0,\qquad g_{Y\, \ast}=\sqrt{\frac{6 \cdot 16\pi^2}{41}f_g},\\
y_{\ast}&= \sqrt{\frac{32\pi^2}{9}f_y}, \qquad g_{Y\, \ast}=0,\\
y_{\ast}&= \frac{4\pi}{3}\sqrt{\frac{17 f_g+82 f_y}{41}}, \quad g_{Y\, \ast}=4 \pi \sqrt{\frac{6 f_g}{41}}.\label{eq:fp_toy_fullint}
\end{align}

\begin{figure}
	\begin{center}
	\includegraphics[width=7.5cm]{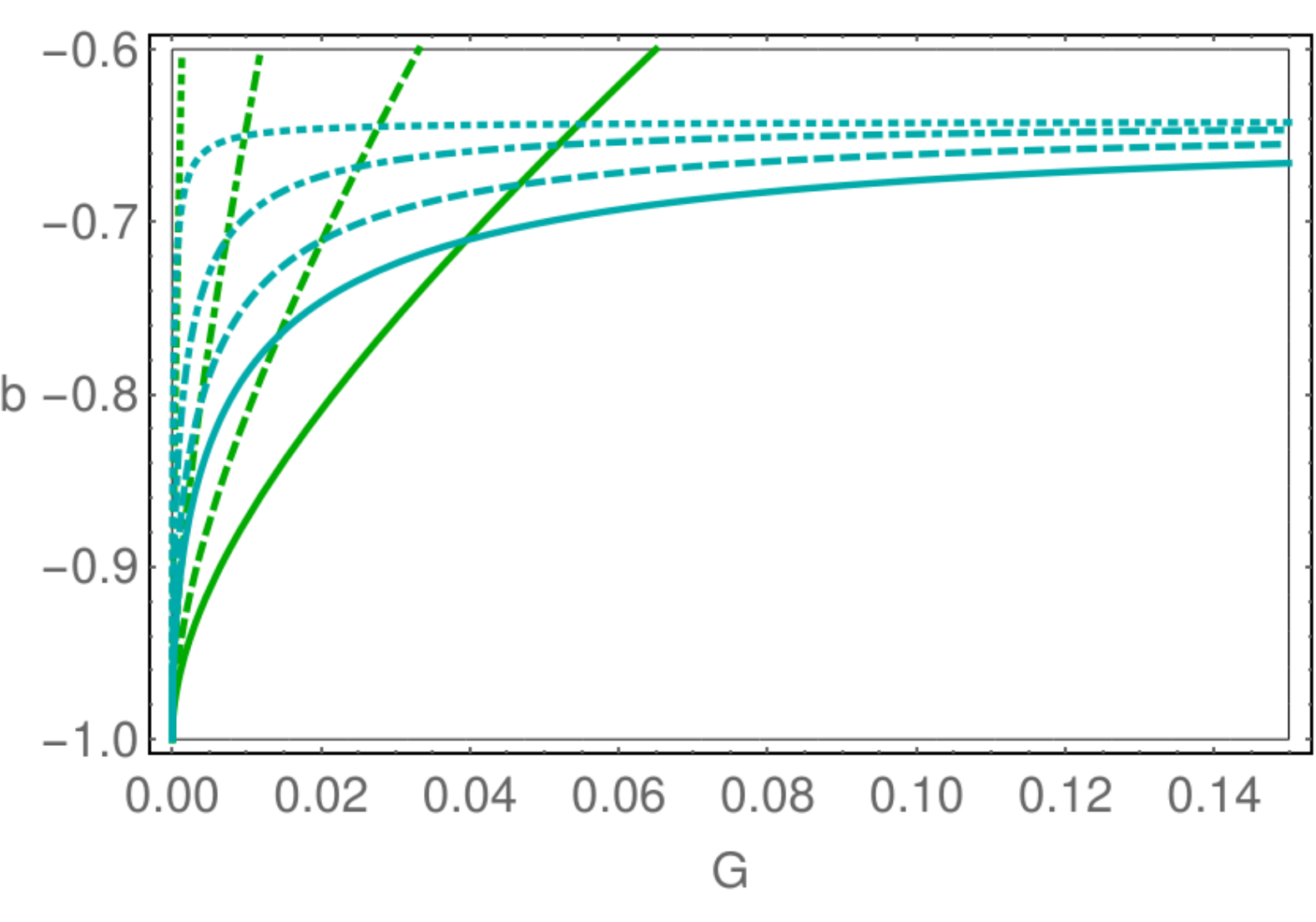}
	\end{center}
	\caption{\label{fig:contours_ygy} We show contour lines in the $b$, $G$ plane for $a=0$, where the fixed-point value of $y$ is given by 0.1 (cyan, dotted), 0.3 (cyan, dot-dashed), 0.5 (cyan, dashed) and 0.7 (cyan, continuous), and contours where the fixed-point value of $g_Y$ is given by 0.1 (green, dotted), 0.3 (green, dot-dashed), 0.5 (green, dashed) and 0.7 (green, continuous).
	Both are evaluated at the fully-interacting fixed point \eqref{eq:fp_toy_fullint}, i.e., the values for the $y$-contours holds for finite values of $g_Y$.}
\end{figure}

The most predictive fixed point is given in \eqref{eq:fp_toy_fullint}, where both $y$ and $g_Y$ correspond to IR attractive directions. For an IR repulsive direction, a deviation from the fixed-point value can set in at any scale (corresponding to the free parameter linked to a relevant direction). This is not the case for an IR attractive direction. In the presence of $f_y>0$, $f_g>0$, $y$ and $g_Y$ must stay at their fixed-point values if Eq.~\eqref{eq:fp_toy_fullint} is chosen as the UV fixed point. Since $f_g$ and $f_y$ depend on gravitational couplings and are proportional to the Newton coupling, a realistic flow will exhibit a sharp decrease of $f_g$ and $f_y$ below the Planck scale. There, $y$ and $g_Y$ start to run as well. Their low-energy values are uniquely fixed by the initial condition at the Planck scale, where they have to assume their fixed-point values. Therefore, a specific value for the prediction of $y$ and $g_Y$ in the IR is tied to a hypersurface in the gravitational parameter space, cf.~Fig.~\ref{fig:contours_ygy}. 

 Gravitational contributions to the beta functions of Yukawa and quartic scalar couplings were computed in the context of unimodular gravity in the framework of effective field theory \cite{Gonzalez-Martin:2017bvw,Gonzalez-Martin:2017fwz,Gonzalez-Martin:2018dmy}. The gravitational action was taken to be the (leading) Einstein-Hilbert term\footnote{ The unimodularity condition is imposed differently from the procedure we adopt in this work. It is unclear if such different prescriptions lead to inequivalent theories at the quantum level.}.  For the particular choice of scheme adopted in \cite{Gonzalez-Martin:2017bvw}, involving a non-multiplicative field redefinition, a vanishing gravitational correction is obtained for the beta functions of the Yukawa and scalar quartic couplings.  We emphasize that, just as in the Standard Model beyond two loops, non-universality is present in gravitational contributions starting at one loop. Accordingly one should not expect our functional RG results to agree with those from \cite{Gonzalez-Martin:2017bvw,Gonzalez-Martin:2017fwz,Gonzalez-Martin:2018dmy} at the level of unphysical quantities, such as, e.g., beta functions. At the level of physical observables, scheme-dependences must cancel (at least up to the accuracy achievable in a given approximations), and comparisons could become meaningful.

\subsection{The Higgs potential in unimodular gravity}
\begin{figure}
	\begin{center}
		\includegraphics[width=7.0 cm]{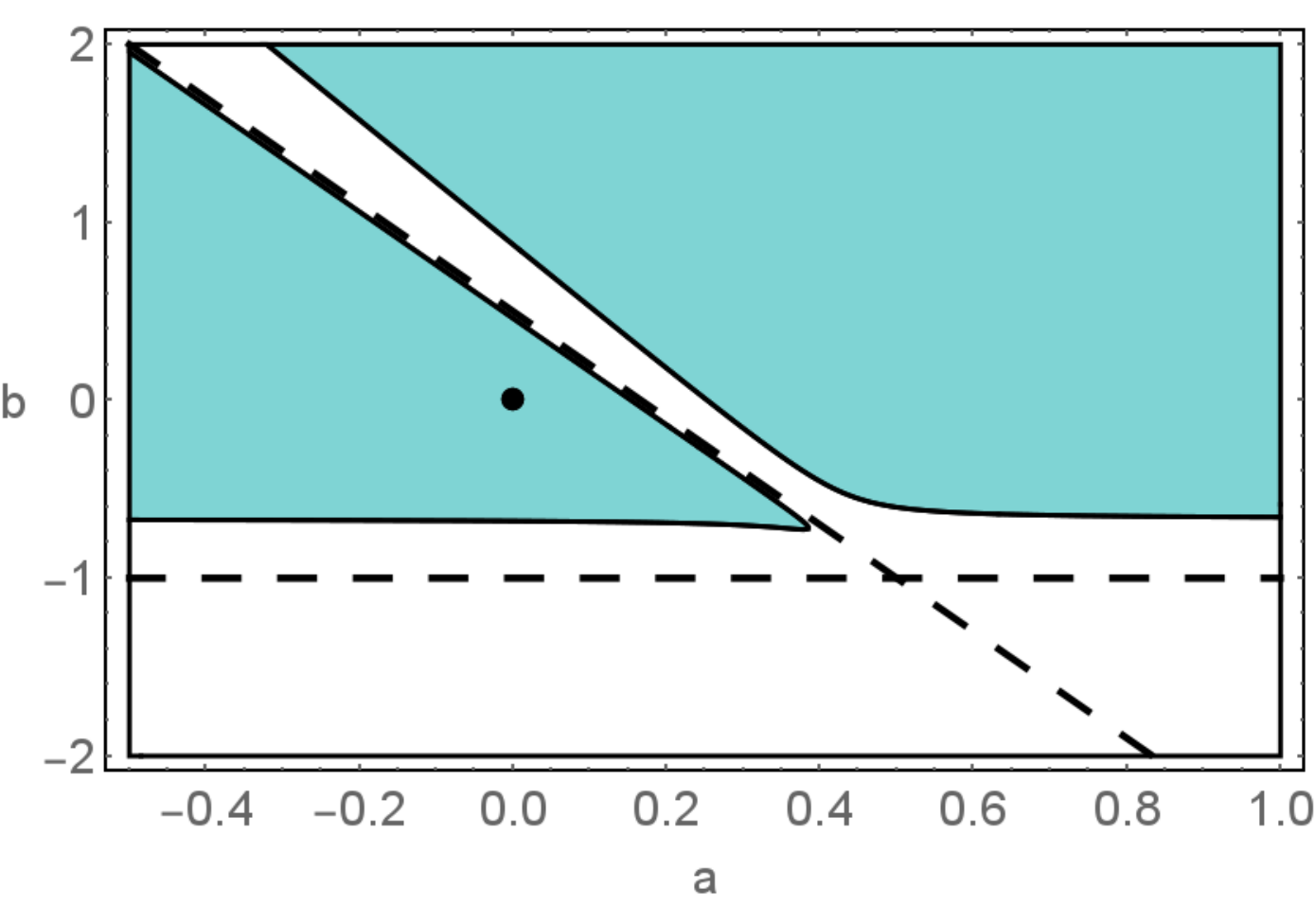}\qquad
		\includegraphics[width=7.0 cm]{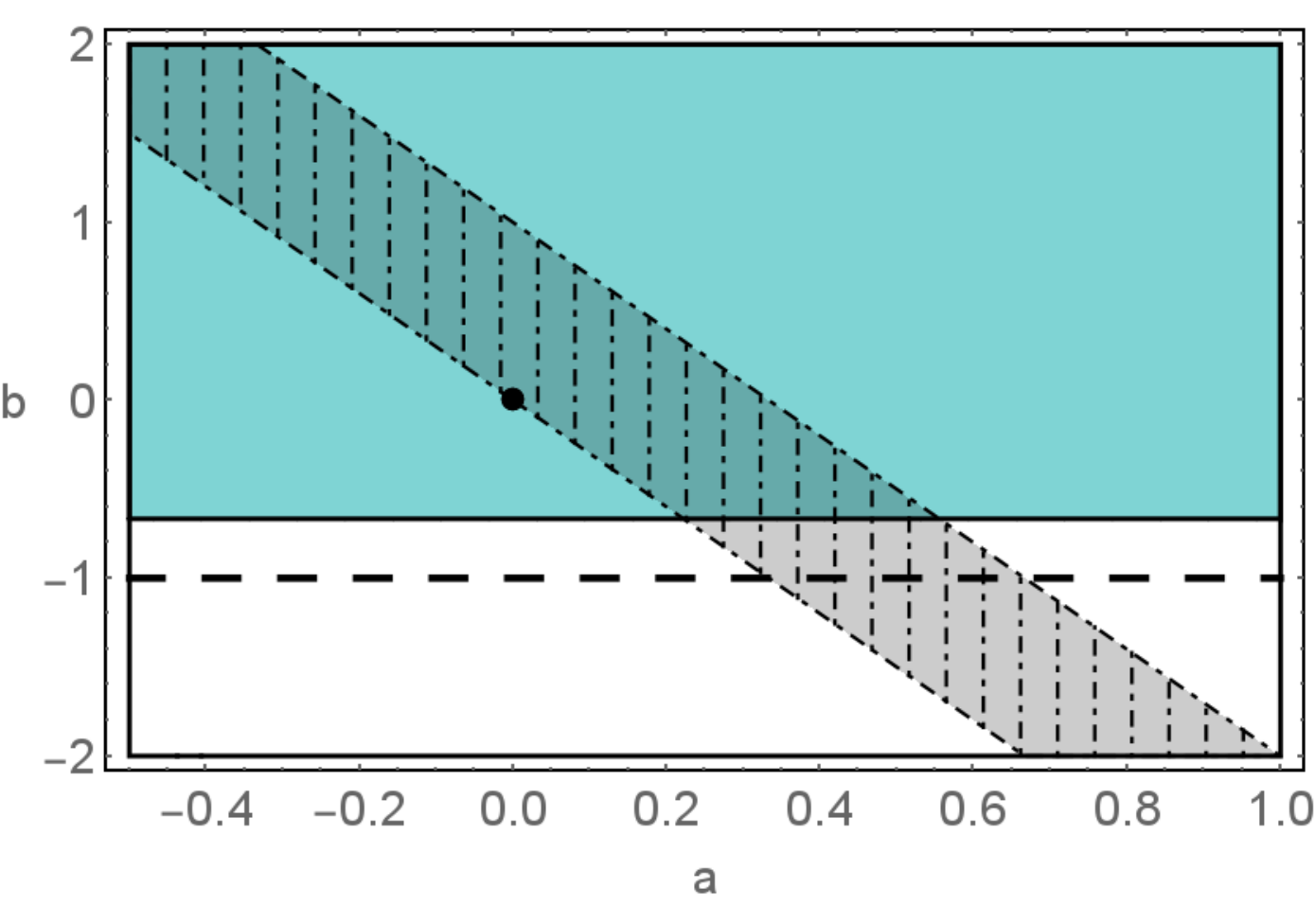}
		\caption{\label{fig:plot_potential} We show the region where $\eta_{\phi}|_{\rm grav}>0$ in cyan as a function of the higher-derivative couplings $a$ and $b$. The left panel shows the full result, whereas the right panel shows the TT-approximation.}
	\end{center}
\end{figure}

In standard gravity, the gravitational contribution to the beta functional of the scalar potential is towards irrelevance at the free fixed point, $V_{\ast}=0$ \cite{Narain:2009fy,Shaposhnikov:2009pv,Oda:2015sma,Hamada:2017rvn,Eichhorn:2017als,Pawlowski:2018ixd}. This implies that (with the possible exception of the mass term, which may remain relevant), all terms in the potential are driven to zero under the impact of quantum-gravity fluctuations \footnote{Note that this can change in the presence of finite fixed-point values for Yukawa and/or gauge couplings, where a non-trivial fixed-point potential is generated by the same terms in the beta function that regenerate scalar self-interactions like the Higgs quartic coupling in the Standard Model, even if it is set to zero at some scale.}, i.e., quantum-gravity fluctuations tend to flatten scalar potentials. In unimodular gravity, the only direct gravitational contribution to the flow of the Higgs potential, or more generally any scalar potential $V$, comes through the gravitational contribution to the anomalous dimension. This is very different from the standard gravity case, where, unless a particular choice of parameterization and gauge is made, the main contribution comes from a direct gravitational tadpole contribution, arising from the term $\sqrt{g}V$ in the action (see \cite{Percacci:2015wwa} for a discussion of this point).
The quantum-gravity effects on scalar potentials in unimodular gravity are therefore potentially rather different from those in standard gravity and it is of interest to compare the two.
Our result is given by Eq.~\eqref{Anomalous_phi_UG}. For a potential with $n$ scalar fields that is Taylor expanded around the origin in terms of the couplings $\lambda_{i_1...i_n}$, i.e., 
\begin{align}
V[\phi_1,...,\phi_n] = \sum_{i_1,...,i_n}\lambda_{i_1...i_n}\phi_{1}^{2i_1}...\phi_n^{2i_n},
\end{align}
there is a gravitational contribution to the beta functions of the form
\begin{align}\label{beta_scalar_couplings}
\beta_{\lambda_{i_1...i_n}}\Big|_{\rm grav}= (i_1+...+i_n)\,\eta_{\phi}\Big|_{\rm grav}  \,\lambda_{i_1...i_n}.
\end{align}
This expression highlights the ``flavor"-independence of gravity: The gravitational contribution to the anomalous dimension is the same for each of the scalar fields, independently of internal symmetries.
In the phase of unbroken symmetry, there is no scalar contribution to $\eta_{\phi}$. 
Therefore the fixed point at $\lambda_{i_1...i_n}=0$, which is guaranteed to exist in the absence of explicit shift-symmetry-breaking contributions, is infrared attractive for $\eta_{\phi}|_{\rm grav}>0$. 
Under the assumption that this fixed-point structure is realized in the corresponding more involved systems despite the presence of additional fields, this could have potential phenomenological consequences, such as a prediction of the Higgs mass in the vicinity of the observed value \cite{Shaposhnikov:2009pv,Bezrukov:2012sa,Pawlowski:2018ixd}, the decoupling of the Higgs portal to  uncharged scalar dark matter \cite{Eichhorn:2017als}, and in general the vanishing of all quartic scalar couplings which could rule out certain breaking chains in grand unified theories \cite{Held:2019}.\\
In the absence of higher-order terms, $\eta_{\phi}|_{\rm grav}>0$ holds as long as $G>0$. Accordingly, as long as gravity remains attractive, it generates a screening contribution for scalar potentials. We highlight that although the origin of the gravitational contribution differs between the ``standard" (in linear parameterization) and the unimodular case, the result is actually the same, i.e., a screening gravitational contribution.
In the presence of higher-order curvature terms, the sign of the gravitational contribution can change, cf.~Fig.~\ref{fig:plot_potential}, and the effect can become 
antiscreening. 

\begin{figure}[htb!]
	\begin{center}
		\includegraphics[width=14cm]{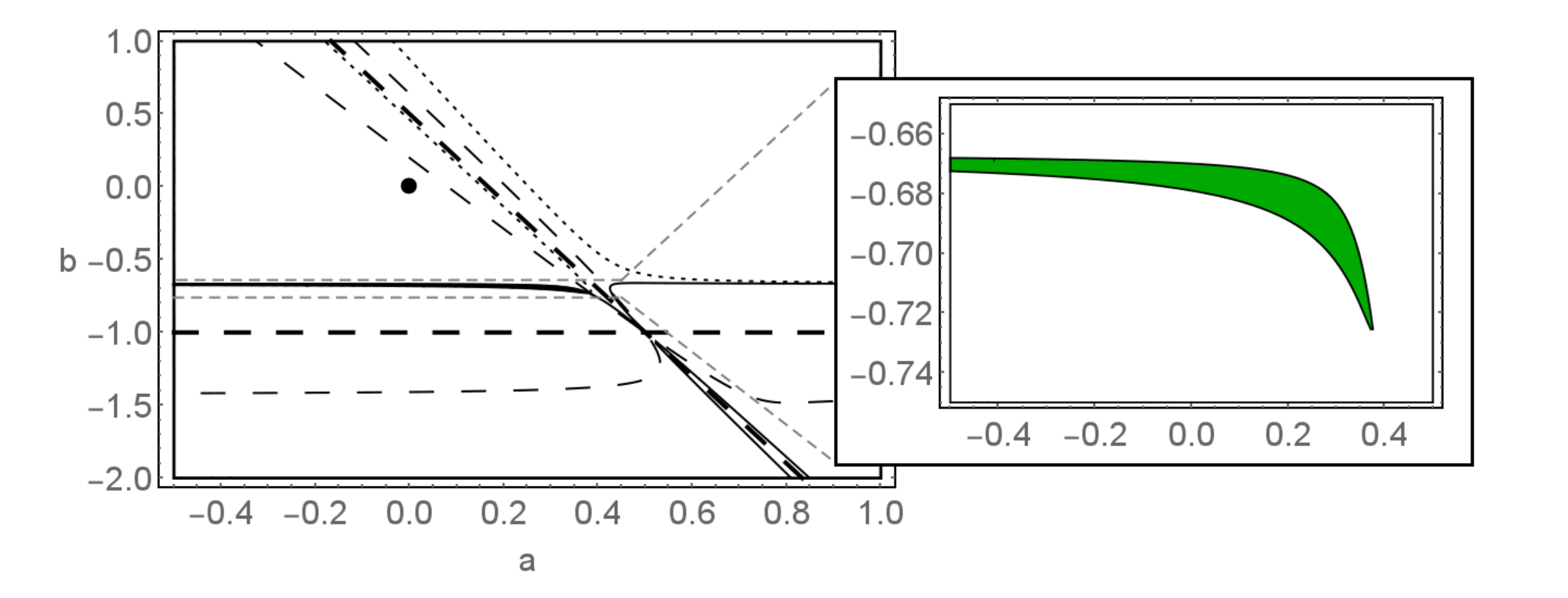}
		\caption{\footnotesize  
 		{We show the boundaries of the regions for a UV completion and predictive Higgs mass: $f_g>0$ above the dashed (thin) line; $f_y>0$ below the continuous line; $\eta_{\phi}|_{\rm grav}>0$ above the dotted line. The thick dashed lines indicates the pole lines $1+b=0$ and $1-6a-2b = 0$. The right-panel zooms in on the overlapping region where the three conditions hold simultaneously.}}
		\label{SM_viability}
	\end{center}
\end{figure}

It is intriguing to understand whether the conditions $f_g>0$ and $f_y>0$, necessary for UV completions of gauge-Yukawa-systems, can be combined with the requirement $\eta_{\phi}>0$, which results in a flat scalar potential at the Planck scale. 
Within our approximation, we find such a region in the gravitational parameter space. The conditions $f_y>0, \, f_g>0, \, \eta_{\phi}>0$ impose very significant restrictions on the space of higher-curvature couplings, see Fig.~\ref{SM_viability}. The origin of this severe restriction can already be seen in the TT-approximation of the beta function for the Yukawa coupling, namely 
\begin{align}
\beta_y|_\TT =  \left(\frac{1}{2}\eta_{\phi}|_\TT+ \eta_{\psi}|_\TT\right)\, y= \frac{15(2+3b)}{32\pi (1+b)^2}\,G\, y = \frac{3}{4}\eta_{\phi}|_\TT\, y ,
\end{align}
where the factor $3/4$ is a consequence of $\eta_{\psi}|_\TT =  \eta_{\phi}|_\TT/4$, since both are only generated by a tadpole diagram.
Hence, the viability condition for a UV completion of the Yukawa coupling becomes $\eta_\phi|_\TT < 0$, in conflict with the requirement for a calculable Higgs mass \footnote{The scalar sector is of course consistent for $\eta_{\phi}|_{\TT}<0$, as the scalar quartic coupling is then asymptotically free. Yet, the scalar mass is then no longer calculable in terms of the scalar vacuum expectation value, but instead becomes a free parameter of the theory.}.  Beyond the TT-approximation, scalar fluctuations generate a region which features $f_y>0$ and $\eta_{\phi}>0$, showing that scalar fluctuations can play an important role in parts of the gravitational parameter space, see Fig.~\ref{SM_viability}.

\section{Comparison with the Reuter fixed point \label{Comparison}}

We contrast the results on unimodular gravity, presented in Sec.~\ref{sec:UGres}, with the corresponding ones obtained in standard asymptotically safe quantum gravity (ASQG), i.e., within the framework where the theory space is defined by full diffeomorphism invariance. 

\begin{figure}[htb!]
	\begin{center}
		\includegraphics[width=4.5cm]{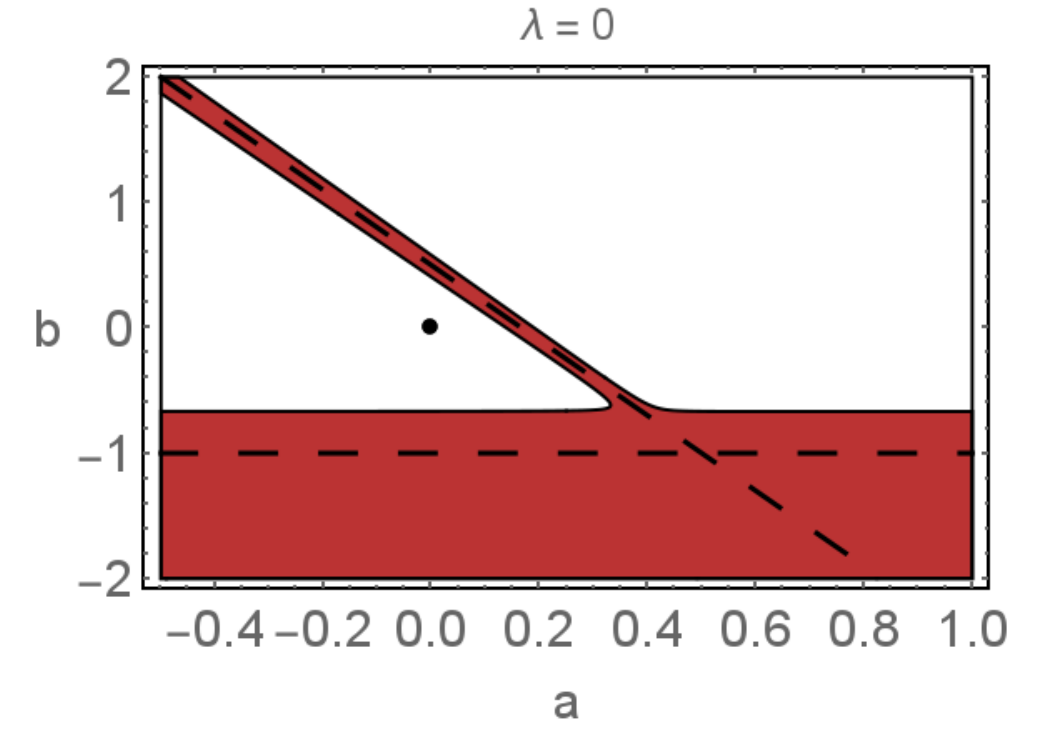} \,\, 
		\includegraphics[width=4.5cm]{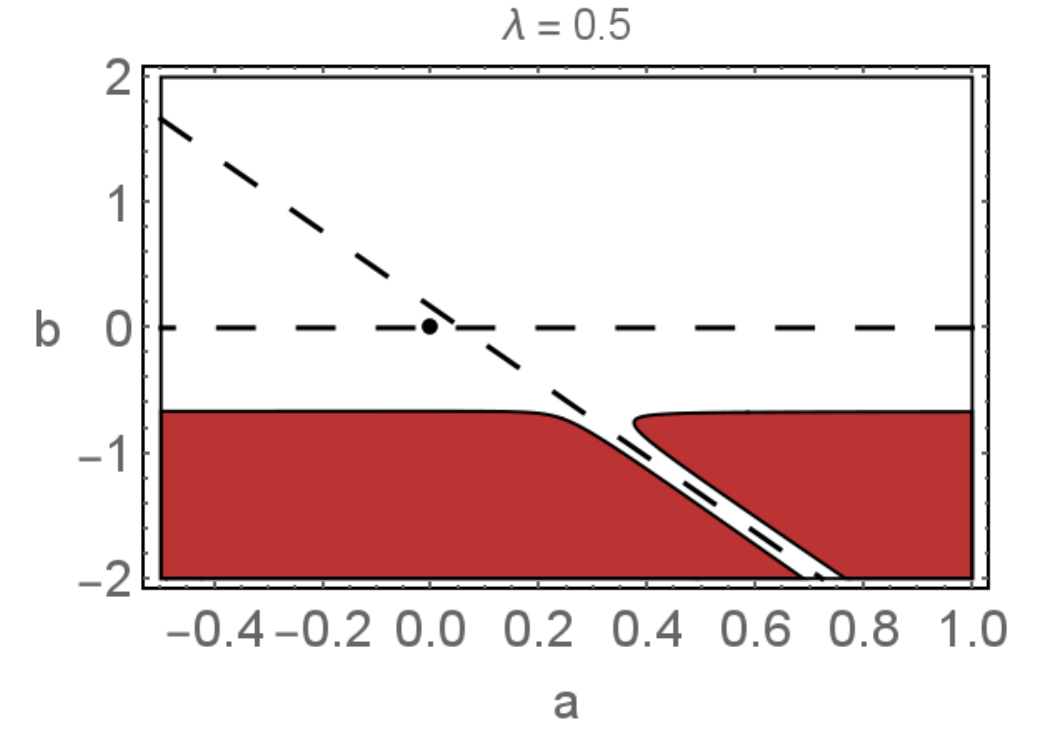} \,\,
		\includegraphics[width=4.5cm]{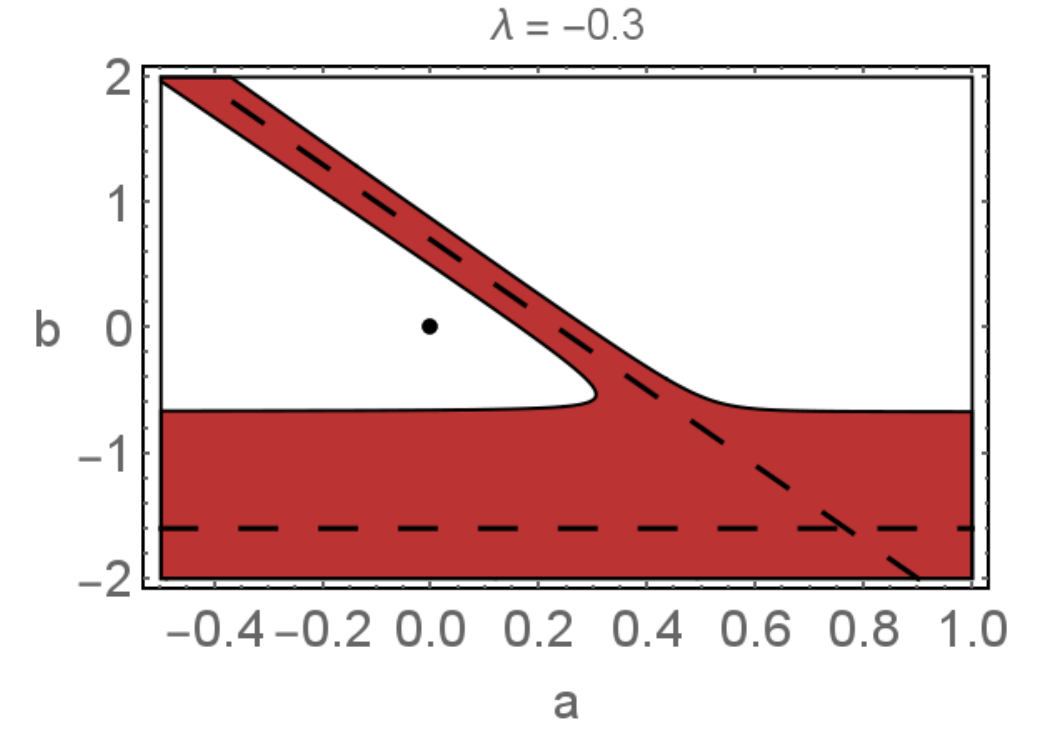} \\ \vspace*{.5cm}
		\includegraphics[width=4.5cm]{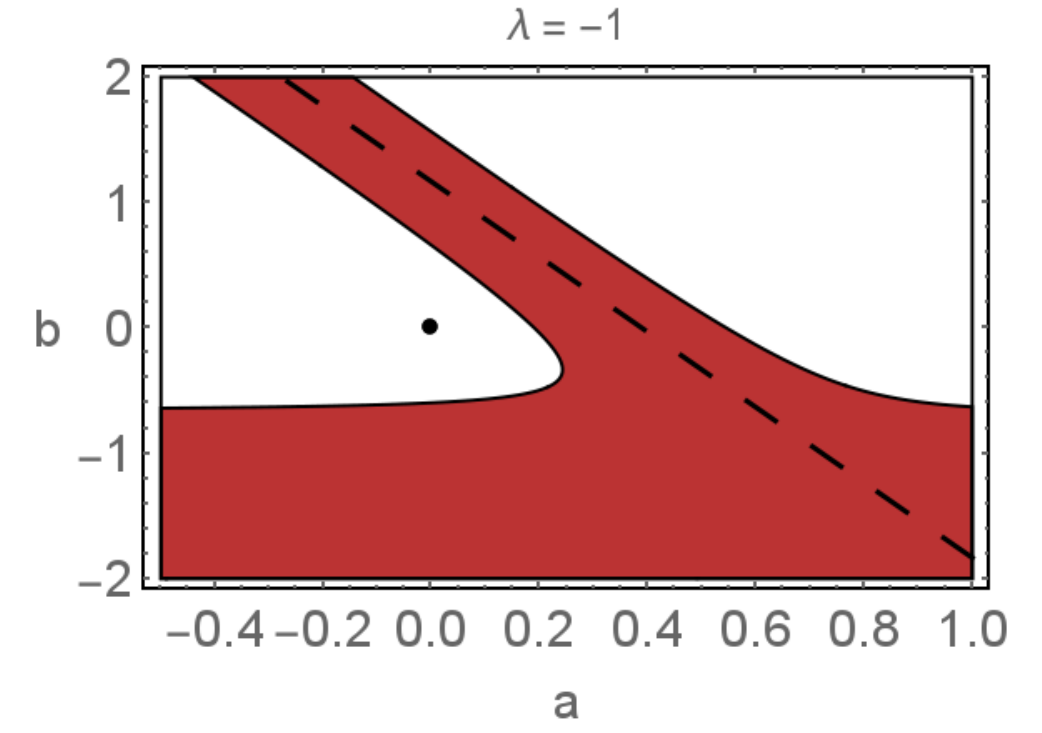} \,\, 
		\includegraphics[width=4.5cm]{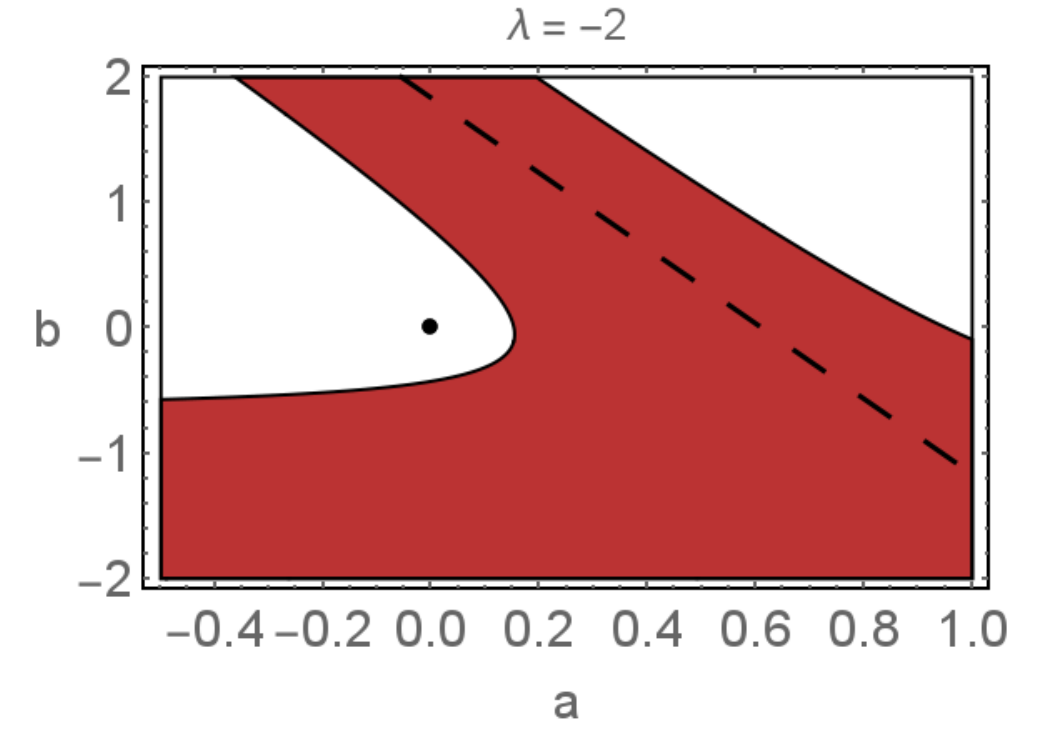} \,\,
		\includegraphics[width=4.5cm]{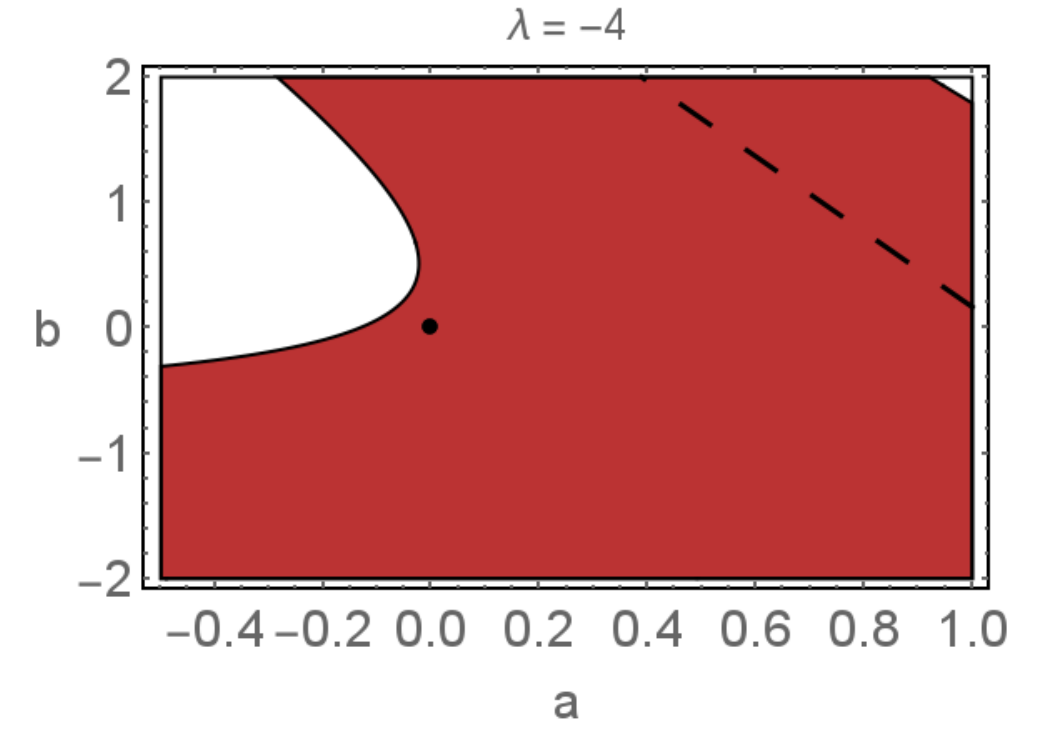}
 		\caption{\footnotesize Viable region (in red) for a quantum-gravity induced fixed point for the Yukawa coupling for several values of the dimensionless cosmological constant within the standard ASQG framework. The dashed lines indicate the poles $1 + b -2\lambda$ (TT-mode) and $1-6a-2b - \frac{4}{3}\,\lambda = 0$ (trace mode).}
		\label{Yukawa_standard_ASQG}
	\end{center}
\end{figure}

Below, we recall the gravitational contribution to the beta function of the Yukawa coupling in standard ASQG (in the Landau gauge) from \cite{Eichhorn:2017eht},
\begin{align}
\beta_y|_{\textmd{grav}} = \bigg[ \frac{15}{32\pi} \frac{2 + 3 b}{(1 + b -2\lambda)^2} \,G - f_{y, \, \rm scalar}(\beta;\lambda,a,b) \bigg]  \,y ,
\end{align}
where $\lambda$ is the dimensionless cosmological constant. We use $f_{y,\, \rm scalar }(\beta;\lambda,a,b)$ to represent the contributions coming from the scalar sector ($\sigma$- and $h$-modes), which depend on the gauge parameter $\beta$ (for explicit expressions see App.~\ref{Standard_ASQG}).
The possibility of a UV completion of the Higgs-Yukawa sector was  studied in \cite{Eichhorn:2016esv,Eichhorn:2017eht, Eichhorn:2017ylw,Eichhorn:2018whv}. In Fig.~\ref{Yukawa_standard_ASQG} we show the viable region for a quantum-gravity induced fixed point for the Yukawa coupling in the $(a, b)$ plane, for several values of the dimensionless cosmological constant (in all cases we set $\beta = 0$). For vanishing dimensionless cosmological constant ($\lambda = 0$), one can observe that there is a  coincidence between the unimodular and the standard setting. This can be plausibilized as a consequence of the $\TT$-dominance in these results. In fact, if we restrict ourselves to the TT-approximation (with $\lambda = 0$) both cases give the same results, namely
\begin{align}
\beta_y|_{\textmd {\rm ASQG,}\TT} = \beta_y|_{\textmd{\rm UQG},\TT} = \frac{15}{32\pi} \frac{2 + 3 b}{(1 + b)^2}  \,G\, y \, .
\end{align}
Note that this result is rather nontrivial, as the various diagrams that contribute to these results differ in the two settings.\\
In the region in gravitational parameter space close to the scalar pole line, on the other hand, the dominant contribution comes from the scalar sector of the fluctuation field $h_{\mu\nu}$. In the unimodular setup, the scalar sector corresponds to the $\sigma$-mode, while in the standard gravity framework the scalar sector is composed of $\sigma$ and $h$\footnote{With the gauge choice $\beta = 0$, only the trace mode contributes to the results.}. Since these different setups receive contributions from different sectors, we observe a quantitative disagreement in the neighborhood of the scalar pole.

For non-vanishing cosmological constant, scalar fluctuations can become more relevant. For positive $\lambda$, there is a screening behavior of metric fluctuations for values of $a$ and $b$ close to the scalar pole. This leads to the absence of viable regions in this regime, cf.~Fig.~\ref{Yukawa_standard_ASQG}. For negative values of $\lambda$, the situation is the opposite. In this regime, scalar fluctuations contribute in an anti-screening manner to the Yukawa beta function, resulting in the enlargement of the viable region. In particular, we note that if $\lambda$ is sufficiently negative, the point corresponding to the Einstein-Hilbert truncation, $a = b = 0$, becomes part of the viable region.

Regarding the (non-)Abelian gauge field sector, we provide the gravitational contribution to (non-)Abelian gauge couplings $g$ in the framework of standard gravity which previously had only been computed with vanishing higher-order couplings.
\begin{align}
\beta_{g^2}|_{\textmd{grav}} = -\bigg[ \frac{1}{18\pi} \frac{10 + 7 b - 40 \lambda}{(1 + b - 2 \lambda)^2} - f_{g}(\beta;\lambda,a,b) \bigg]\,G\,g^2\, .
\end{align}
The first term in the above expression corresponds to the $\TT$ contribution to $\beta_{g^2}|_{\textmd{grav}}$ in standard ASQG, and can be shown to be exactly the same as in unimodular gravity. The second part corresponds to the contribution from the scalar sector ($\sigma$- and $h$-modes) and is characterized by a gauge-dependent function $f_{g}(\beta;\lambda,a,b)$, the definition of which can be extracted from  App.~\ref{Standard_ASQG}. Specific gauge choices which deserve further attention include:
\begin{itemize}
	\item $\beta = 0$: In this case the contributions coming from the scalar sector vanish and, as consequence, the result is completely determined by the $\TT$ approximation. 
	\item $\beta \to \pm \infty$: For this gauge choice and for $\lambda=0$, the contribution from the scalar sector  comes exclusively from the $\sigma$-mode, resulting in the same expression as obtained in unimodular gravity. Therefore, there is a complete agreement between the viable regions in the two settings in the subspace $\lambda=0$ of ASQG.
\end{itemize}

In order to understand possible effects of a non-vanishing cosmological constant, let us consider, for simplicity, the case $\beta = 0$. In this case the gravitational contribution comes from the TT-sector and, as consequence, the possibility for a UV fixed point in the gauge coupling restricts the space of higher-curvature couplings $a$ and $b$ by the inequality $10 + 7 b - 40 \lambda > 0$. Similar to the Yukawa case, negative values of $\lambda$ enlarge the viable region.

\begin{figure}[htb!]
	\begin{center}
		\includegraphics[width=4.5cm]{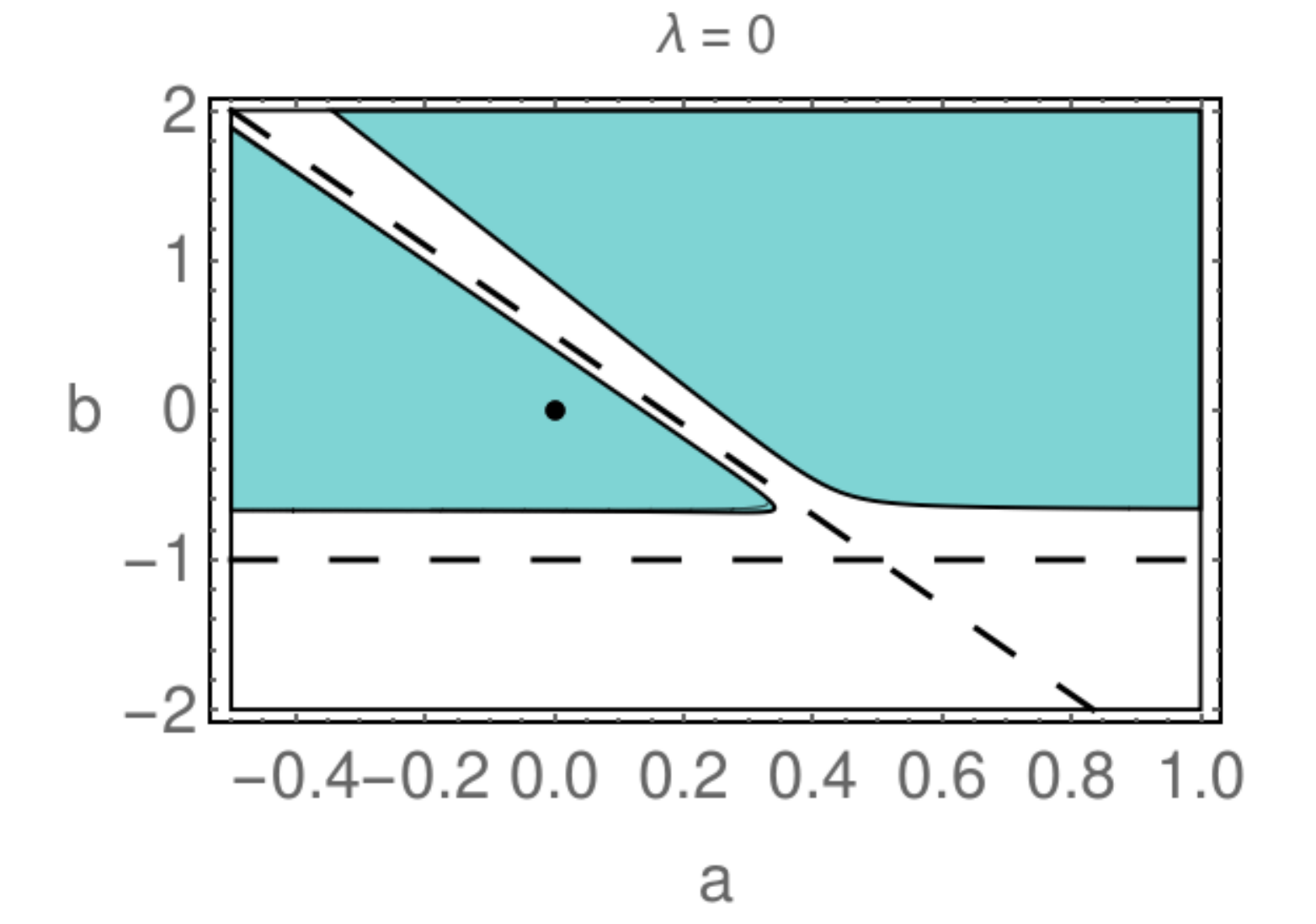} \,\, 
		\includegraphics[width=4.5cm]{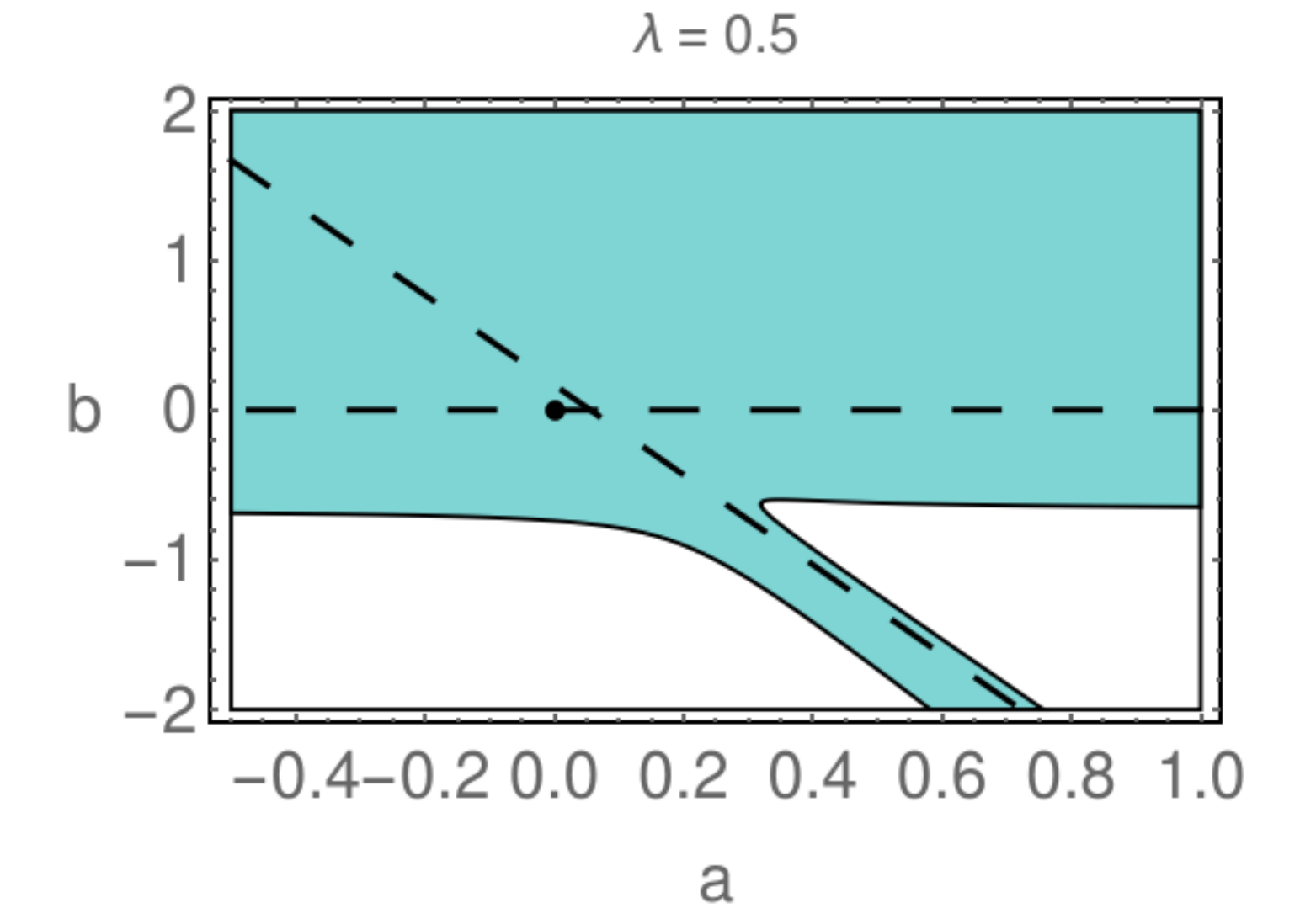} \,\,
		\includegraphics[width=4.5cm]{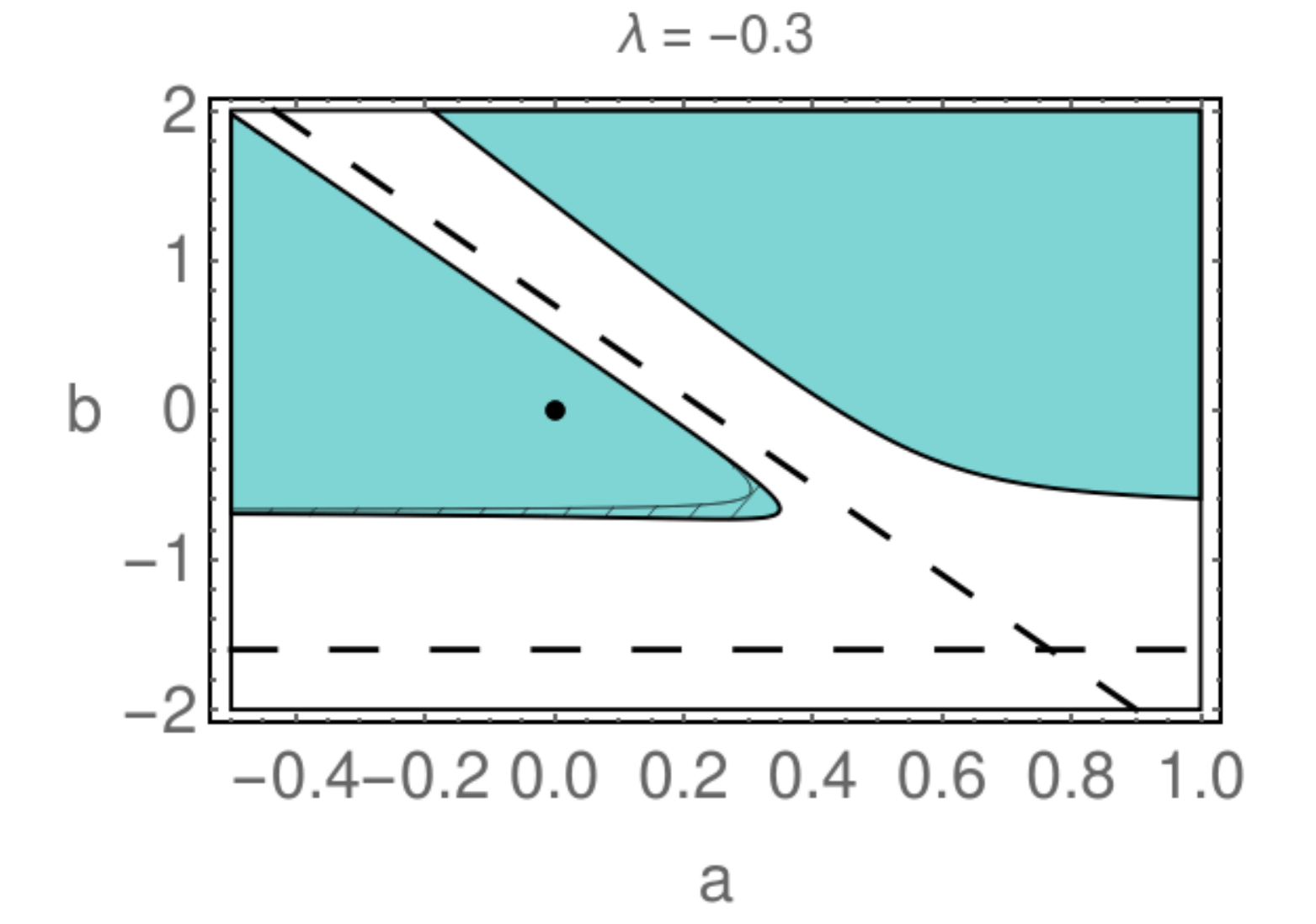} \\ \vspace*{.5cm}
		\includegraphics[width=4.5cm]{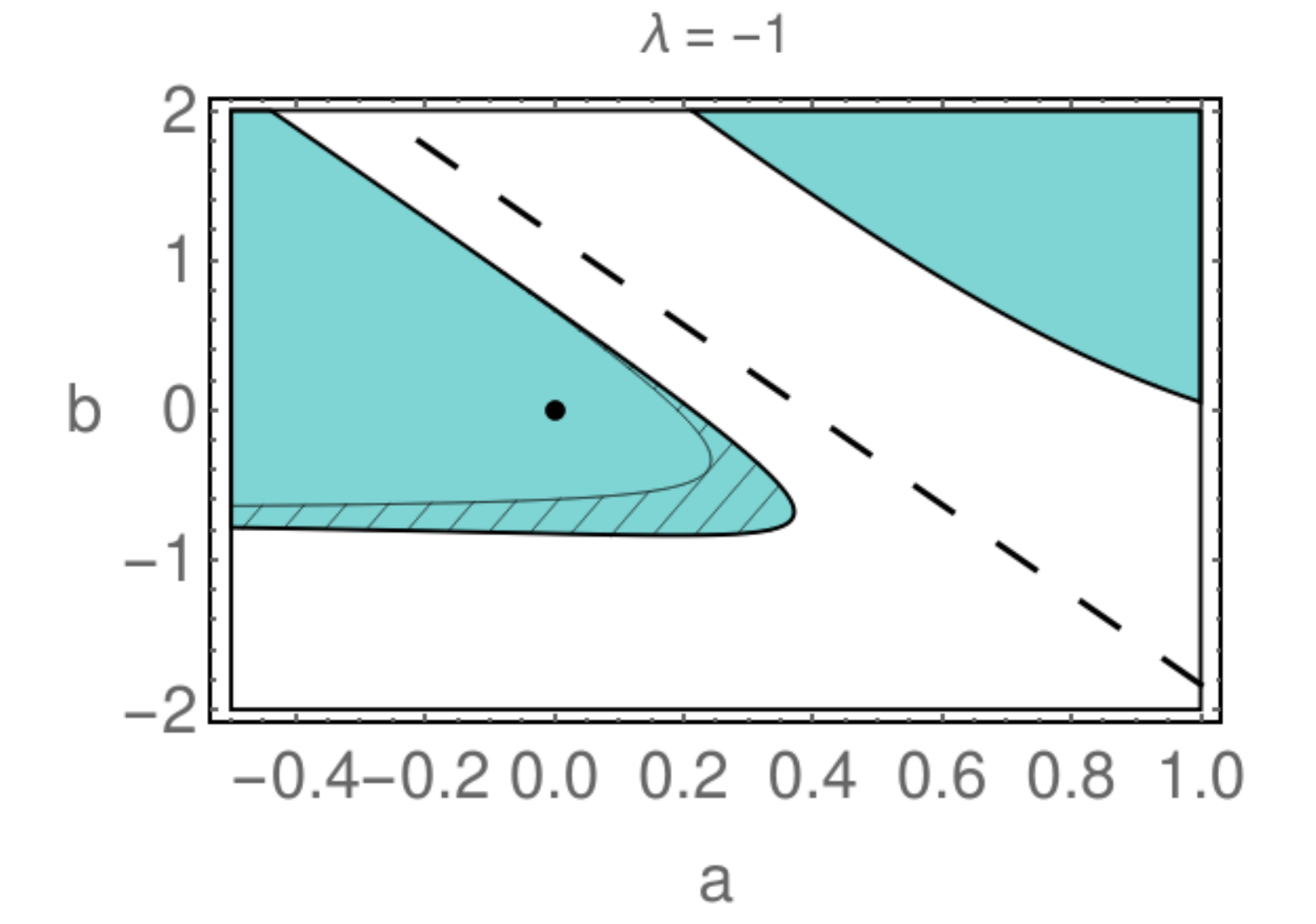} \,\, 
		\includegraphics[width=4.5cm]{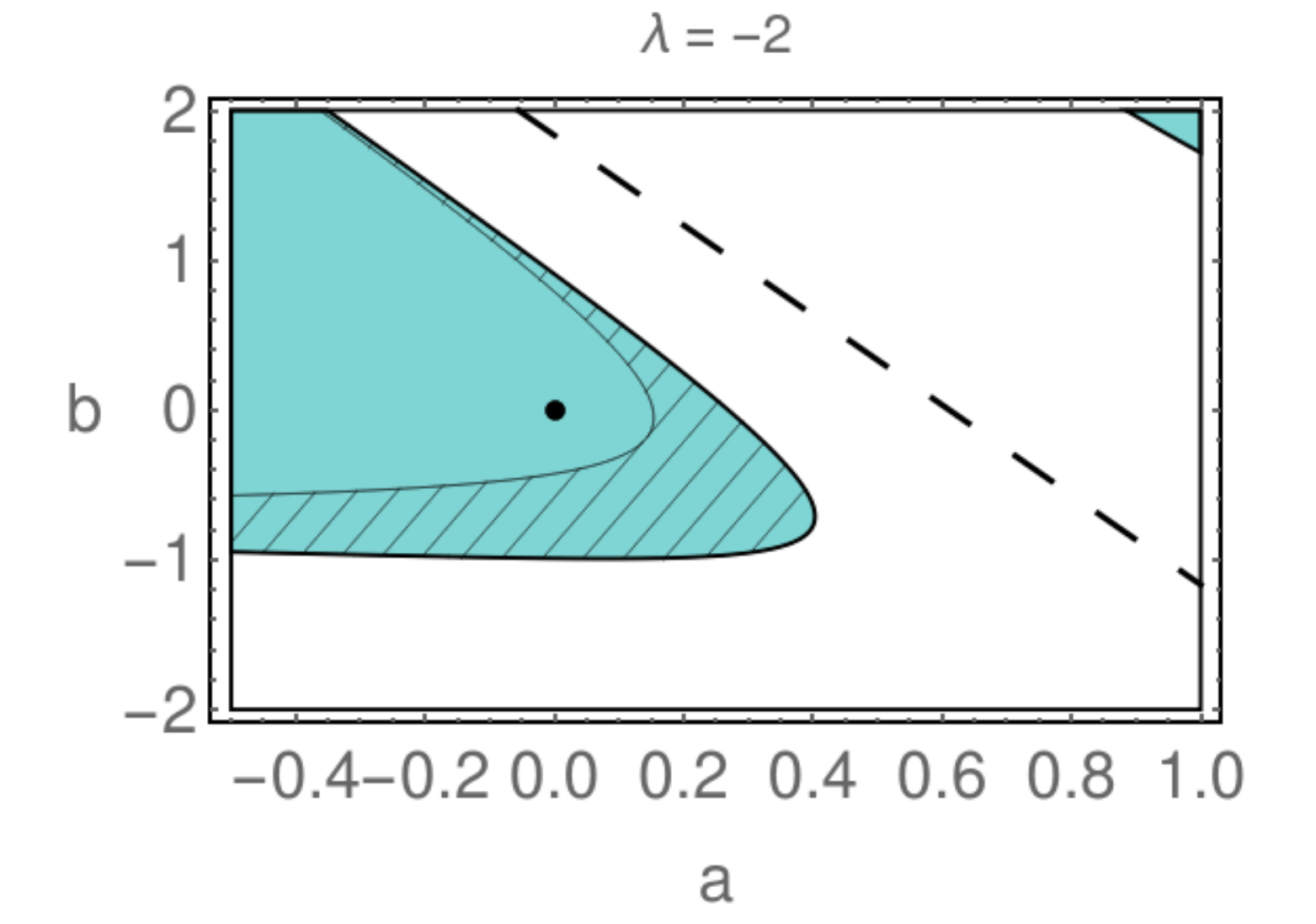} \,\,
		\includegraphics[width=4.5cm]{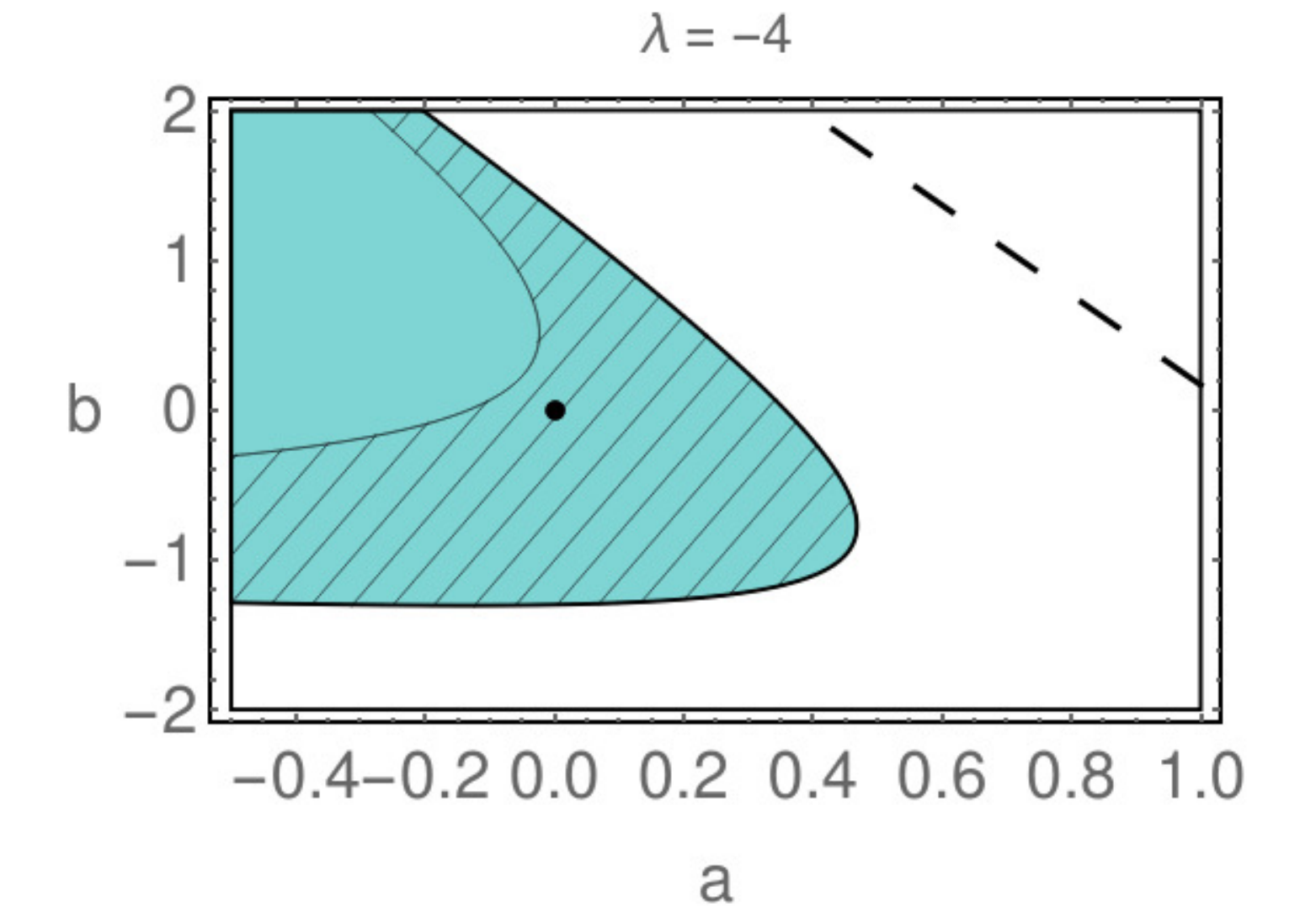}
		\caption{\footnotesize Viable region for predictive Higgs mass for several values of the dimensionless cosmological constant within the standard ASQG framework. The region with diagonal solid lines indicate the set of values where gravity-induced fixed points for the Yukawa and (non-)Abelian gauge coupling can by achieved along with a predictive Higgs mass. The dashed lines indicate the poles $1 + b -2\lambda$ (TT-mode) and $1-6a-2b - \frac{4}{3}\,\lambda = 0$ (trace mode).}
		\label{Scalar_standard_ASQG}
	\end{center}
\end{figure}

Finally, let us discuss quantum-gravity contributions to the scalar potential, more specifically, to the $\lambda_4 \phi^4$ coupling. In the framework of standard ASQG, the running of the scalar quartic coupling receives gravitational contributions coming from the anomalous dimension $\eta_\phi$ and from the tadpole diagram depicted in  Fig.~\ref{tadpole_scalar}. Below, we present the result for the gravitational contribution to the beta function of the scalar quartic coupling in standard ASQG,
\begin{align}
\beta_{\lambda_4}|_{\textmd{grav}} = \bigg[ \frac{5}{4\pi} \frac{2 + 3 b}{(1 + b -2\lambda)^2} +  f_{\lambda_4}(\beta;\lambda,a,b) \bigg]  \,g\lambda_4 \,,
\end{align}
where $f_{\lambda_4}(\beta;\lambda,a,b)$  represents terms coming from the scalar modes of the York decomposition.
 In Fig.~\ref{Scalar_standard_ASQG} we show the region in which the ratio of the Higgs mass to the electroweak scale is predicted for several values of the dimensionless cosmological constant (in all cases we consider $\beta = 0$). In the case with $\lambda = 0$ we observe the same qualitative behavior as in UQG. Once again, this fact can be explained in terms of $\TT$-dominance. For non-vanishing $\lambda$  the results change considerably. In particular, for negative values of the cosmological constant, the region with a calculable Higgs mass is deformed in such a way that if we superpose Figs.~\ref{Yukawa_standard_ASQG} and \ref{Scalar_standard_ASQG}, we observe that the overlap with the region allowing UV fixed points in the Yukawa and (non-)Abelian gauge couplings becomes larger. 

Let us finally highlight that within the truncated theory spaces of UQG and standard ASQG, one can also interpret our results in the unimodular setting as tests of the parameterization-dependence of the results in standard ASQG at vanishing $\lambda$. This is a consequence of the fact that the difference between the two settings lies in the absence of $\lambda$ for UQG and a slight modification of the gauge and Faddeev-Popov ghost sector, to which our truncation is not sensitive, as we do not include induced matter-ghost interactions (see \cite{Eichhorn:2013ug}). Thus it is rather reassuring to observe that the results at $\lambda=0$ in standard ASQG and the results in the unimodular setting are qualitatively as well as quantitatively similar. Such a mild parameterization dependence could be interpreted as a hint for the robustness of the results.

\section{Results for Weyl-squared Gravity}\label{sec:ConGrav}
Let us now turn to the discussion of the gravitational contribution to the beta functions of matter couplings. Our aim is twofold:
\begin{itemize}
\item We explore the one-dimensional gravitational parameter space to discover whether Standard-Model-like theories could be UV complete due to the impact of Weyl-squared gravity. 
\item As the Weyl-squared setting only features dimensionless couplings (in particular also for the gravity sector) the leading-order gravitational contribution to the beta functions should be universal. We show that this holds within the functional RG framework, by showing independence of the choice of regulator shape function, of the parameterization of metric fluctuations and of the additional gauge parameters in Landau gauge for the one-loop contribution. 
\end{itemize}

The gravitational contribution to the beta function of Yukawa couplings is encoded in the gravitational contribution to $\eta_\phi$, $\eta_\psi$ and $\mathcal{D}_y$. Our truncation for Weyl-squared gravity-matter systems gives the following results 
\begin{align} \label{Results_CG}
\eta_\phi = \frac{5 \, w}{32 \pi^2} , 
\qquad\quad \eta_\psi = \frac{5 \, w}{128 \pi^2}, \qquad\quad \textmd{and} \qquad\quad 
\mathcal{D}_y = 0 \,.
\end{align}
Thus, the gravitational contribution to the beta-function for the Yukawa coupling is
\begin{align}
\beta_y|_{\textmd{WG}} = \frac{15}{128 \pi^2} \,w\,y.
\end{align}
Within this approximation the beta function for the Yukawa coupling has an IR repulsive fixed point at $y=0$ if $w < 0$ at the fixed point. In the literature, the only known fixed point for the Weyl-squared theory lies at vanishing value of $w$, and is IR repulsive in $w$ \cite{deBerredoPeixoto:2003pj,Ohta:2013uca,Ohta:2015zwa}.

For the Abelian sector the situation is more subtle. In this case the leading order gravitational contribution vanishes, just as it does in the perturbative calculations \cite{Fradkin:1981iu,Narain:2012te,Narain:2013eea}. Therefore, if we restrict ourselves to this approximation, the screening contribution of charged matter fields is not compensated by an anti-screening effect coming from the gravitational sector. In the non-Abelian  case the same result for the gravitational contribution holds, indicating that asymptotic freedom in that sector is not affected by gravitational fluctuations.  

Beyond the leading-order approximation,  a non-universal contribution arises that depends on the anomalous dimension of gravitational fluctuations. We find the following result for the gravitational contribution to the anomalous dimension of the gauge field by using a regulator that depends on the wave-function renormalization of the metric and is constructed using a Litim-type shape function
\begin{align}
\eta_A|_{\textmd{WG}} = -\frac{7 w \,\eta_\TT }{576\pi^2  - 5 w} \,,
\end{align}
where  $\eta_{\TT}= -\partial_t \ln Z_\TT$ is the anomalous dimension associated with the TT-fluctuation field. The possibility of an anti-screening gravitational contribution depends on the sign of $\eta_\TT$. In the above approximation, a necessary condition for a UV complete Abelian gauge sector requires $\eta_\TT < 0$ at the fixed point if $w<0$, as required in the Yukawa sector. At the fixed point that is known in the literature, namely $w=0$, $\eta_\TT=0$ holds.

Let us now discuss a more technical point, namely the universality of the above results.
The non-universality (i.e., RG-scheme dependence) of gravitational contributions to beta functions in the matter sector has been noted in the literature, see, e.g., \cite{Pietrykowski:2006xy,Toms:2007sk,Ebert:2007gf,Felipe:2011rs,Ellis:2010rw,Anber:2010uj,Toms:2011zza,Nielsen:2012fm,Pietrykowski:2012nc}. Indeed beta functions are never universal, but the onset of non-universality depends on the canonical dimension of the couplings of theory. Due to the dimensionful nature of the Newton coupling, gravitational contributions to beta functions in standard gravity are non-universal already at leading order. This is different for dimensionless couplings, for which non-universality only sets in at 3 loops. Accordingly, in Weyl-squared gravity, where the gravitational coupling is dimensionless, the gravitational contribution to Standard-Model-like matter couplings, which are also dimensionless, should be universal at leading order. 

For clarity of the discussion, let us repeat the corresponding argument for a theory with a single coupling, see, e.g.,  \cite{Weinberg:1996kr}: Two different RG schemes can be understood as two ways of defining the coupling, which we will call $g$ and $\tilde{g}$. We explore the perturbative regime, where the relation between $g$ and $\tilde{g}$ is expressible in terms of a Taylor series,
\be
g(\tilde{g})= \tilde{g}+c_2 \tilde{g}^2+\mathcal{O}(\tilde{g}^3).
\ee
To leading order, the two couplings are the same since there are no quantum effects to that order. We now translate the beta function expressed in terms of $g$, 
\be
\beta_{g}= \beta_1\, g^2+\beta_2\, g^3+ \mathcal{O}(g^4)
\ee
into the corresponding expression in terms of $\tilde{g}$:
\bea
\tilde{\beta}_{\tilde{g}} &=& k \partial_k\, \tilde{g} = \frac{\partial \tilde{g}}{\partial g} \beta_g =\left(1-2 c_2 \tilde{g}\right)\left( \beta_1\, \left(\tilde{g}+ c_2 \tilde{g}^2 \right)^2+ \beta_2 \, \left(\tilde{g}+ c_2 \tilde{g}^2 \right)^3 \right)+ \mathcal{O}(\tilde{g}^4)\nonumber\\
&=&\beta_1 \, \tilde{g}^2 + \left(-2c_2 \beta_1 +  \beta_1\, 2c_2+ \beta_2\right) \tilde{g}^3+ \mathcal{O}(\tilde{g}^4)\nonumber\\
&=& \beta_1\,\tilde{g}^2 + \beta_2\, \tilde{g}^3+\mathcal{O}(\tilde(g)^4).
\eea
Clearly, the two leading-order terms agree, while it is also obvious that higher-order terms do not. In the presence of dimensionful couplings, the relation between $g$ and $\tilde{g}$ can contain the scale explicitly, such that two-loop universality no longer holds.

The above argument is for different RG schemes. Losely speaking, the choice of regulator function can be viewed in a similar way. Further, field reparameterizations can be understood as a different choice of coupling. Accordingly, we expect that the leading-order gravitational contribution to matter beta functions in the Weyl-squared gravity case is
\begin{enumerate}
\item independent of the choice of shape function
\item independent of the choice of field parameterization
\item independent of the choice of gauge parameter (note that we work off-shell).
\end{enumerate}
In fact, a similar result should hold for the gravitational beta functions themselves. Indeed, \cite{Ohta:2015zwa} shows explicitly that to leading order the beta functions are the same for the linear and the exponential parameterization. We will now explore the points 1)-3) and also compare to results obtained within perturbation theory. It is known from pure matter systems that the leading order, i.e., the universal one-loop result, can be obtained from FRG results from a truncation that includes the perturbatively renormalizable couplings as well as a wave-function renormalization $Z$. Typically, the regulator is then  chosen to be $\sim Z$, and the scale-derivative of the regulator on the rhs of the Wetterich equation generates terms $\sim \eta$. Once $\eta$ is expressed in terms of the couplings, it becomes obvious that these terms are  higher-order in the couplings. To recover the universal one-loop result, these terms therefore have to be neglected, see also App.~C of \cite{Eichhorn:2015kea}.

In Weyl-squared gravity in the exponential parameterization, i.e., $g_{\mu\nu} = \bar{g}_{\mu\alpha}  [ e^{h^{\cdot}_{\,\,\, \cdot}} ]^\alpha_{\,\,\,\,\nu}$, the gravitational contributions to the running of Yukawa and (non-)Abelian gauge couplings come exclusively from the anomalous dimension. Our results are given by
\begin{subequations}
	\begin{align}\label{eta_phi_CG}
	\eta_\phi|_{\textmd{grav}} = \frac{5 \,w}{16\pi^2} \Phi_2^3 -
	\frac{5 \,w\,\eta_\TT}{32\pi^2} 
	\left( \tilde{\Phi}_2^3 + 2 \tilde{\Phi}_3^4 \right) \, ,
	\end{align}
	\begin{align}\label{eta_A_CG}
	\eta_A|_{\textmd{grav}} = \frac{5 \,w}{12\pi^2} \left(\frac{}{}\! \Phi_2^3 - 3 \Phi_3^4 \right)
	- \frac{5 w \, \eta_\TT}{24 \pi^2} \left( \tilde{\Phi}_2^3 - 6 \tilde{\Phi}_4^5 \right) 
	+ \frac{5 w \,\eta_A}{12\pi^2} \, \tilde{\Phi}_3^4 \, ,
	\end{align}
	\begin{align}\label{eta_psi_CG}
	\eta_\psi|_{\textmd{grav}} = \frac{5 \,w}{64\pi^2} \Phi_2^3  -
	\frac{5 \,w\,\eta_\TT}{128\pi^2} \left( \tilde{\Phi}_2^3 + 2 \tilde{\Phi}_3^4 \right)  ,
	\end{align}
\end{subequations}
where the threshold integral is defined as 
\begin{subequations}
	\begin{align}\label{threshold_1}
	\Phi_n^p = \frac{1}{\Gamma(n)}\int_0^\infty dy \, y^{n-1} \frac{r(y) - y\,r^\prime(y)}{(y+r(y))^p}\,,
	\end{align} 
	\begin{align}
	\tilde{\Phi}_n^p = \frac{1}{\Gamma(n)}\int_0^\infty dy \, y^{n-1} \frac{r(y)}{(y+r(y))^p}\,,
	\end{align} 
\end{subequations}
with dimensionless shape function $r(y) = k^{-2} R_k(k^2 y)$. It not difficult to verify that the special case $\Phi_n^{n+1}$ is independent of the choice of the shape function, namely $\Phi_n^{n+1} = 1/{\Gamma(n+1)}$. Taking this property into account we find
\begin{subequations}
	\begin{align}\label{eta_phi_CG_2}
	\eta_\phi|_{\textmd{grav}} = \frac{5 \,w}{32\pi^2} \left[\, 1 - 
	\eta_\TT\, \left( \tilde{\Phi}_2^3 + 2 \tilde{\Phi}_3^4 \right) \,\right]\, ,
	\end{align}
	\begin{align}\label{eta_A_CG_2}
	\eta_A|_{\textmd{grav}} = - \frac{5 w \, \eta_\TT}{24 \pi^2} \left( \tilde{\Phi}_2^3 - 6 \tilde{\Phi}_4^5 \right) 
	+ \frac{5 w \,\eta_A}{12\pi^2} \, \tilde{\Phi}_3^4 \, ,
	\end{align}
	\begin{align}\label{eta_psi_CG_2}
	\eta_\psi|_{\textmd{grav}} = \frac{5 \,w}{128\pi^2} \left[\, 1 - 
	\eta_\TT\, \left( \tilde{\Phi}_2^3 + 2 \tilde{\Phi}_3^4 \right) \,\right] \,.
	\end{align}
\end{subequations}

As expected, the leading order terms in $\eta_\phi|_{\textmd{grav}}$ and $\eta_\psi|_{\textmd{grav}}$ are universal with respect  to the shape function. The contributions coming from cutoff insertions, however, are, as expected, non-universal and explicit result depend on the choice of the shape function.

To test the universality with respect to the choice of field parameterization, we also employ the linear metric parameterization, i.e., $g_{\mu\nu} = \bar{g}_{\mu\nu} + h_{\mu\nu}$. In this case, there are additional terms in $\beta_y$ coming from the tadpole diagram represented in Fig.~\ref{diagrams_2} (contributing to $\mathcal{D}_y$). Our results in linear parameterization are given by
\begin{subequations}
	\begin{align}\label{eta_phi_CG_linear}
	\eta_\phi|_{\textmd{grav}}^{\textmd{linear}} = 0 \,,
	\end{align}
	\begin{align}\label{eta_A_CG_linear}
	\eta_A|_{\textmd{grav}}^{\textmd{linear}} = 
	\frac{5 \,w}{12\pi^2} \left(\frac{}{}\! \Phi_2^3 - 3 \Phi_3^4 \right)
	- \frac{5 w \, \eta_\TT}{24 \pi^2} \left( \tilde{\Phi}_2^3 - 6 \tilde{\Phi}_4^5 \right) 
	+ \frac{5 w \,\eta_A}{12\pi^2} \, \tilde{\Phi}_3^4 \, ,
	\end{align}
	\begin{align}\label{eta_psi_CG_linear}
	\eta_\psi|_{\textmd{grav}}^{\textmd{linear}} = -\frac{25 \,w}{64\pi^2}\Phi_2^3  
	+\frac{25 \,w\,\eta_\TT}{128\pi^2} \left( \tilde{\Phi}_2^3 + 2 \tilde{\Phi}_3^4 \right)  \,,
	\end{align}
	\begin{align}\label{tadpole_CG_linear}
	\beta_y|_{\textmd{tadpole}}^{\textmd{linear}} \equiv \mathcal{D}_{y} \, y = 
	\frac{5 \,w \,y}{8\pi^2}\Phi_2^3 -
	\frac{5 \,w \,y\, \eta_\TT}{16\pi^2} \left( \tilde{\Phi}_2^3 + 2 \tilde{\Phi}_3^4 \right)  \, .
	\end{align}
\end{subequations}

We first observe that $\eta_A|_{\textmd{grav}}^{\textmd{linear}}$ gives the same expression as the one obtained in the exponential parameterization, as it should in order to give a universal beta function for the gauge coupling. Furthermore, we note that the anomalous dimensions of the scalars and fermions are different from the corresponding expressions obtained in the exponential parameterization. Despite these differences the gravitational contribution to the beta function for the Yukawa coupling gives the same result in both parameterizations, namely
\begin{align}
\beta_y|_{\textmd{WG}} =  \frac{15 \,w \, y}{64\pi^2} \Phi_2^3 -
\frac{15 \,w \, y\,\eta_\TT}{128\pi^2}  \left( \tilde{\Phi}_2^3 + 2 \tilde{\Phi}_3^4 \right) .
\end{align}
Using the universality of $\Phi_2^3$ we find
\begin{align}
\beta_y|_{\textmd{WG}} =  \frac{15 \,w \, y}{128\pi^2}\left[ 1  -
\eta_\TT \left( \tilde{\Phi}_2^3 + 2 \tilde{\Phi}_3^4 \right) \right] .
\end{align}

 It is worth  mentioning that we also have checked the universality with respect to the choice on the field parameterization by means of an interpolating parameterization  \cite{Gies:2015tca,Ohta:2016npm,Ohta:2016jvw,deBrito:2018jxt}, namely
\begin{align}
g_{\mu\nu} = \delta_{\mu\nu} + h_{\mu\nu} + \omega \,h_{\mu\alpha} h^\alpha_{\,\,\,\nu} + \mathcal{O}(h^3),
\end{align}
where our results only depend on the terms up to quadratic order in $h$.
Within this parameterization, we can interpolate between the linear ($\omega=0$) and the exponential ($\omega=1/2$) parameterization, up to order $\mathcal{O}(h^2)$, by varying the parameter $\omega$. Our computations reveal that each one of the diagrams discussed above is $\omega$-dependent, however, this dependence cancels out when we combine them in order to compute the beta functions for the Yukawa and (non-)Abelian gauge couplings.  

We briefly highlight the question of gauge dependence:
 Within our truncation, the gauge fixing sector has 5 parameters $\alpha$, $\tilde{\alpha}$ $\gamma_1$ $\gamma_2$ and $\beta$. By choosing $\alpha, \tilde{\alpha} \to 0$ (Landau gauge), the results only depend on the gauge-independent TT sector and thus turn out to be independent of the remaining parameters. 

We also compare to the one-loop results from perturbative techniques, and find agreement with our results. The gravity-contribution to the gauge coupling is known to vanish at one-loop even in the more general case of curvature-squared gravity  \cite{Fradkin:1981iu,Narain:2012te,Narain:2013eea}. The gravitational contribution to the running of Yukawa couplings  at one loop can be extracted from the results in \cite{Salvio:2017qkx}.

\section{Concluding Remarks \label{conclusions}}

 In this paper, we compare three different quantum field theories for the metric with respect to their potential observational viability in terms of their impact on matter, namely ``standard'' asymptotically safe gravity, unimodular asymptotic safety and Weyl-squared gravity. We work within a toy model for the Standard Model, which consists of an Abelian gauge field and a simple Yukawa system of one Dirac fermion and a real scalar. Quantum-gravity fluctuations are ``blind'' to internal symmetries, allowing us to deduce implications of our results for more general matter models. In particular, the leading  quantum-gravity contribution to the scale-dependence of all gauge couplings is the same, therefore our results also hold for non-Abelian gauge groups. We explore, where in the parameter space spanned by the microscopic gravitational couplings \footnote{These are not free parameters in the full dynamical matter-gravity theory but are set by demanding a consistent (asymptotically free or safe) microscopic dynamics. Here we treat them as free parameters to explore whether there are regions in this space into which a phenomenologically viable gravitational fixed point could fall.} the gravitational contribution to the beta function for the Yukawa and the gauge coupling is antiscreening.\\
 Within our truncation that includes all local gravitational couplings up to four orders in derivatives \footnote{We stress that our study is subject to systematic errors due to our choice of truncation for the dynamics.}, the following results hold: \\
 i) In the unimodular theory space, there is a restricted range of values for the $R_{\mu\nu}R^{\mu\nu}$ coupling $b$, in which quantum-gravity contributions to the running of the gauge coupling and the Yukawa coupling are antiscreening. This is the phenomenologically viable range, where quantum-gravity fluctuations could solve the Landau pole/ triviality problem in the Abelian hypercharge and Yukawa-sector.\\
 These results are very similar to the results for standard gravity based on the Reuter fixed point at vanishing cosmological constant. We stress that the close agreement is rather nontrivial, as the diagrams underlying the results in the two different cases are quite different. For the Reuter fixed point, the cosmological constant can also be nonzero, opening up a significantly larger viable parameter space region for antiscreening gravity contributions. The absence of the cosmological constant in unimodular asymptotic safety therefore leads to rather severe constraints on the higher-order couplings, which need to fall into a rather narrow range in order to achieve antiscreening gravity contributions.\\
 
 ii) For the Weyl-squared case, we recover the known universal one-loop gravitational contribution with the functional Renormalization Group, explicitly showing the independence from the regulator shape function, as well as the choice of gauge parameters and parameterization of metric fluctuations. Depending on the sign of the Weyl-squared coupling, the Yukawa coupling can become asymptotically free. The universal one-loop contribution to the scale dependence of the gauge coupling vanishes.\\

 This motivates several avenues for the future. Firstly, extensions of the truncation are of course indicated to explore whether the viable regions we find here open up further as more directions are added to the space of couplings. Secondly, the calculation of gravitational fixed-point values under the impact of matter fluctuations is of interest, to show whether the fixed point falls into the viable region or not. Ultimately, the resulting comparison of unimodular asymptotic safety with ``standard" asymptotically safe gravity could allow us to disfavor one of the two quantum-gravity models on phenomenological grounds.

\section*{Acknowledgments}
We acknowledge helpful discussions with A.~Held,  M.~Schiffer, M.~Pauly and S.~Lippoldt, and thank all members of the ITP/CP3-Origins quantum-gravity group for discussions during the Heraeus-Klausurtagung.
G.~P.~B.~is grateful for the support by Capes under the grant no.~88881.188349/2018-01 and CNPq no.~142049/2016-6 and thanks the ITP at Heidelberg University for hospitality.
A.~E.~and A.~D.~P.~acknowledge support by the DFG under grant no.~Ei-1037/1. A.~E.~is also partially supported by a visiting fellowship at the Perimeter Institute for Theoretical Physics and by the Danish National Research Foundation under grant DNRF:90. 
 
\appendix

\section{Conventions \label{Convention}} 

In this appendix we list the conventions and notations used in this paper. Our convention for the Fourier transform is
\begin{align}
\Phi(x) =\int_p \Phi(p) e^{ip\cdot x}.
\end{align}
Conjugated fields, such as $\bar{\psi}$, accordingly carry a minus sign in the exponential factor, namely $e^{-ip\cdot x}$. 

We use a shorthand for $d$-dimensional integrals in position and Fourier space, respectively
\begin{align}
\int_x\,\,\equiv \int d^dx \, \quad \textmd{and} \quad \int_p \,\,\equiv \int \frac{d^dp}{(2\pi)^d} \, .
\end{align}

The transverse and longitudinal projectors (on vector fields) are  defined, around the flat background, in the standard way
\begin{align}
P_\T^{\mu\nu} = \delta^{\mu\nu} - P_\textmd{L}^{\mu\nu} \qquad \textmd{and} \qquad P_\textmd{L}^{\mu\nu} = \frac{p^\mu p^\nu}{p^2} .
\end{align}
For symmetric rank-2 tensors we use the transverse and traceless projector 
\begin{align}
P_\TT^{\mu\nu\alpha\beta}= \frac{1}{2}(P_\T^{\mu\alpha} P_\T^{\nu\beta} + P_\T^{\mu\beta} P_\T^{\nu\alpha}) - \frac{1}{d-1} P_\T^{\mu\nu} P_\T^{\alpha\beta}.
\end{align}
In addition, it is useful to define a momentum dependent tensor given by
\begin{align}
\varrho_{\mu\nu}(p) = \frac{p_\mu p_\nu}{p^2} - \frac{1}{d} \delta_{\mu\nu} .
\end{align}

In order to compute the anomalous dimensions of the matter fields we employ the following projection rules  to the (functional derivatives of the) flow equation 
\begin{subequations}
	\begin{align}
	\eta_\phi=&  -\frac{1}{Z_\phi} \Bigg[ \frac{\pt}{\pt p^2} \bigg( \sum \textmd{Diagrams} \bigg)\Bigg]_ {p^2=0} ,\\
	\eta_A =& -\frac{1}{Z_A} \Bigg[ \!\frac{\pt}{\pt p^2} \bigg(\! \frac{1}{d\!-\!1}  P^{\mu\nu}_\T \sum (\textmd{Diagrams})_{\mu\nu} \!\bigg)\!\Bigg]_ {p^2=0} ,\\
	\eta_\psi =&\,\frac{1}{Z_\psi} \Bigg[ \!\frac{\pt}{\pt p^2} \bigg( \frac{1}{2^{[d/2]}} \slashed{p} \sum \textmd{Diagrams} \bigg)\!\Bigg]_ {p^2=0} .
	\end{align}
\end{subequations}
For the additional contributions for the beta function of the Yukawa coupling coming from the diagrams depicted in Fig.~\ref{diagrams_2}, we use the projection rule
\begin{align}
\mathcal{D}_y \,y = \frac{k^{d/2-2}}{2^{[d/2]}} \frac{1}{i\, Z_\phi^{1/2}\, Z_\psi} 
\sum \textmd{Diagrams} \Big|_ {p=0} .
\end{align}

\section{Fermions and the exponential parameterization \label{Fermions_exp}}

The coupling of fermions to gravity in a setting with vanishing torsion is through the vielbein and the spin-connection. Since our formulation is based on functional quantization of the fluctuation field $h_{\mu\nu}$, we have to express both the vielbein and spin-connection in terms of $h_{\mu\nu}$ in accordance with the exponential parameterization.

We start with the vielbein, denoted as $e_\mu^a$. For our purposes it will be sufficient to expand the vielbein up to second order around a flat background, namely
\begin{align}
e_\mu^a = \delta_\mu^a + \delta e_\mu^a + \frac{1}{2} \delta^2 e_\mu^a + \mathcal{O}(\delta^3 e) ,
\end{align}
where $ \delta_\mu^a$ is the (trivial) flat space vielbein. In order to gauge fix the local $O(d)$ symmetry associated with the definition of the vielbein, we adopt the Lorentz symmetric gauge-fixing given by \cite{Woodard:1984sj,vanNieuwenhuizen:1981uf}
\begin{align}
e_{\mu a} \delta_{b}^\mu - e_{\mu b} \delta_{a}^\mu = 0 \,.
\end{align}  
This condition allows us to obtain the following expressions
\begin{subequations}
	\begin{align}
	\delta e_\mu^a =& \frac{1}{2} \delta^{\nu a} \delta g_{\mu\nu},\\
	\delta^2 e_\mu^a =& \frac{1}{2} \delta^{\nu a} \delta^2 g_{\mu\nu} - \frac{1}{4} \delta^{\nu a} \delta^{\alpha\beta} \delta g_{\mu\alpha} \delta g_{\nu\beta}.
	\end{align}
\end{subequations}
For the exponential parameterization, we have $\delta g_{\mu\nu} = h_{\mu\nu}$ and $\delta^2 g_{\mu\nu} = h_{\mu\alpha} h_\nu^{\,\,\,\alpha}$, resulting in the following expansion for the vielbein
\begin{align}
e_\mu^a = \delta_\mu^a +\frac{1}{2} \delta^{\nu a} h_{\mu\nu} + \frac{1}{8} \delta^{\nu a} h_{\mu\alpha} h_\nu^{\,\,\,\alpha} + \mathcal{O}(h^3) .
\end{align}

For the spin-connection, which is not an independent field in our setting, we use the expression
\begin{align}
\omega_\mu = [\gamma^a,\gamma^b] \left( \delta_{ac} e_\nu^c \pt_\mu e_b^\nu + \delta_{ac} \Gamma_{\mu\alpha}^\lambda e_\lambda^c e_b^\alpha  \right),
\end{align}
in order to express the spin-connection in terms of the fluctuation field $h_{\mu\nu}$. After some manipulations we arrive at the following result
\begin{align}
\omega_\mu &= [\gamma^\alpha,\gamma^\beta] \pt_\beta h_{\mu\alpha} + \frac{1}{2} [\gamma^\alpha,\gamma^\beta] \bigg( \!\!-\! \frac{1}{2} h_\alpha^{\,\,\,\lambda} \pt_\mu h_{\beta\lambda} \,+ \nn\\
&- h_\beta^{\,\,\,\lambda} \pt_\lambda h_{\mu\alpha} - h_\alpha^{\,\,\,\lambda} \pt_\beta h_{\mu\lambda} 
+ \pt_\beta h_{\mu\rho}\, h_\alpha^{\,\,\,\rho} + h_{\mu\rho} \pt_\beta h_\alpha^{\,\,\,\rho}  \bigg) + \mathcal{O}(h^3) .
\end{align}
With these results we can compute all the fermion-gravity vertices used in this paper. An alternative to the use of vielbein in the description of fermion-systems is the spin-base formalism \cite{Gies:2013noa,Gies:2015cka,Lippoldt:2015cea}. At the level of our computations both formalisms render the same results.

\section{Unimodular Gravity-Matter systems in general dimensions \label{Further_UG}}
In this section we report on some additional details and results for unimodular gravity-matter systems. 
All the results presented in the main text were restrict to $d=4$.  Here, we report results for UQG in $d$ dimensions. 

The starting point is the truncation given by Eq.~\eqref{EAA_UG}. In order to compute Hessians and vertices, which are necessary to compute beta functions and anomalous dimensions associated with matter fields, we adopt the following procedure:
\begin{itemize}
	\item[i)] Using the exponential parameterization, we expand the flowing action defined in Eq.~\eqref{EAA_UG} up to second order in the fluctuation field. For simplicity, (and since in our truncation it yields the same results as technically more complicated choices), we use a flat background metric $\bar{g}_{\mu\nu} = \delta_{\mu\nu}$. For the fermionic sector we use the vielbein formalism adapted to the case of exponential parameterization (see App.~\ref{Fermions_exp}). 
	\item[ii)] Redefine the fluctuation field as $h_{\mu\nu} \to \sqrt{32\pi G_N}\, h_{\mu\nu}$. 
	\item[iii)] After performing a York decomposition (with $\alpha \to 0$), we introduce the appropriate wave function renormalization for each one of the fields presented in our setup, namely:
	\begin{align}
	&\qquad h_{\mu\nu}^\TT \mapsto Z_\TT^{1/2} h_{\mu\nu}^\TT \,, \qquad \sigma \mapsto Z_\sigma^{1/2} \sigma \,.
	\label{zttandzsigma}
	\end{align}
	\item[iv)] As a last step, we observe that the introduction of gauge-fixing and cut-off terms in the FRG formulation breaks the original (volume preserving) diffeomorphism invariance and, as consequence, the RG-flow generates terms which are not manifestly invariant under the original symmetry. At the level of the  flowing action, the aforementioned symmetry is encoded in modified Slavnov-Taylor identities (mSTIs) \cite{Ellwanger:1995qf,Reuter:1996cp,Pawlowski:2005xe,
  Pawlowski:2003sk,Manrique:2009uh,Donkin:2012ud}. As an example, the presence of an infrared cutoff terms can generate  mass-like terms for the $\TT$ and $\sigma$ modes (for $k\neq0$). In order to accommodate such a possibility, we add the following  explicitly symmetry-breaking terms to our truncation
	\begin{align}
	\Gamma_{k\, \rm SB}=\frac{Z_\TT m_\TT^2}{2}\int_x  h_{\mu\nu}^{\TT} P_\TT^{\mu\nu\alpha\beta} h^\TT_{\alpha\beta}\,
	-\frac{(d-2)(d-1)}{d^2}\frac{Z_\sigma m_\sigma^2}{2}\int_x  \sigma^2 \,.
	\end{align}
	The numerical factors in the second term were chosen in such a way that, in the absence of higher curvature terms, $p^2 = -m_\sigma^2$ becomes a pole in the $\sigma$-mode propagator. 
\end{itemize} 
Let us comment that in the case of standard gravity such mass-like terms for the fluctuation field arise from the cosmological-constant term. In the unimodular setting, one must not confuse the generation of such mass-like terms with the generation of a cosmological constant. Instead, these are to be understood as a purely symmetry-breaking effect, subjected to mSTIs, with no cosmological-constant counterpart in the symmetry-preserving subspace of theory space.

After these steps, we can extract the necessary Hessians and vertices. Below we present the list of Hessians employed in our computations
\bea
	\left[\Gamma_{k,h^\TT\! h^\TT}^{(2)}\right]^{\mu\nu\alpha\beta} &=&  Z_\TT \,  \left[ \,\bar{b} \,p^4 + p^2 + m_\TT^2 \,\right] P_\TT^{\mu\nu\alpha\beta} \,,\\
	\Gamma_{k,\sigma\sigma} &=&  - \frac{(d-2)(d-1)}{d^2}\,Z_\sigma \,\bigg[\!- \bigg( \frac{4\,\bar{a}\,(d-1)+ \bar{b} \,d}{d-2}\bigg) p^4 + p^2 + m_\sigma^2 \,\bigg] \, ,\\
	\Gamma_{k,\phi\phi} &=& Z_\phi\,p^2 \,,\\
	\Gamma_{k,AA}^{\mu\nu} &=& Z_A \,  p^2  \left( P_\T^{\mu\nu} + \frac{1}{\zeta} P_\textmd{L}^{\mu\nu} \right)\,,\\
	 \Gamma_{k,\psi\bar{\psi}} &=& - Z_\psi \, \gamma^\mu p_\mu \,.
\eea
In the gauge field sector we employ the Landau gauge fixing $\zeta \to 0$.

Gravity-matter vertices can be computed by taking functional derivatives of the following terms in the expansion of the flowing action in power of $h_{\mu\nu}$.  The relevant vertices for this work are listed below,
\bea
	\Gamma^{\phi\phi h^\TT}_k &=& \frac{1}{2} Z_\phi Z_\TT^{1/2} \,\kappa \, \int_{p,q} p^\mu q^\nu \, \phi(p) \phi(q) \,h_{\mu\nu}^\TT(-p-q)  \,,\\
	\Gamma^{\phi\phi \sigma}_k &=&- \frac{1}{2} Z_\phi Z_\sigma^{1/2} \,\kappa \, \int_{p,q} p_\mu q_\nu \,\varrho^{\mu\nu}(-p-q)\, \phi(p) \phi(q) \, \sigma(-p-q) \, ,\\
	\Gamma^{\phi\phi h^\TT h^\TT}_k& =& -\frac{1}{4} Z_\phi Z_\TT \, \kappa^2 \,\int_{p,q,l} \delta^{\nu\alpha} \,p^\mu q^\beta \,\phi(p) \phi(q) h_{\mu\nu}^\TT(l) h_{\alpha\beta}^\TT(-p-q-l)  \,,\\
\Gamma^{\phi\phi \sigma\sigma}_k& =& -\frac{1}{4} Z_\phi Z_\sigma \, \kappa^2 \int_{p,q,l} \delta_{\nu\alpha} \,p_\mu q_\beta \,\varrho^{\mu\nu}(l) \varrho^{\alpha\beta}(-p-q-l)  \,\phi(p) \phi(q) \sigma(l) \sigma(-p-q-l) \nn\,,\\
&{}&\\
	\Gamma^{AA h^\TT}_k &=& \frac{1}{2} Z_A Z_\TT^{1/2} \, \kappa \, \int_{p,q} \Bigl( p^\mu q^\nu \, \delta^{\lambda\rho} 
	- p^\rho q^\nu \, \delta^{\mu\lambda}  \nn \\ 
	&{}& -\, p^\mu q^\lambda \, \delta^{\nu\rho} + p\cdot q \, \delta^{\mu\lambda} \delta^{\nu\rho} \Bigr) 
	A_\lambda(p) A_\rho(q) \,h_{\mu\nu}^\TT(-p-q)  \,,\\
	\Gamma^{AA \sigma}_k &=& -\frac{1}{2} Z_A Z_\sigma^{1/2} \, \kappa \int_{p,q} \Bigl( p^\mu q^\nu \, \delta^{\lambda\rho} 
	- p^\rho q^\nu \, \delta^{\mu\lambda}  -\, p^\mu q^\lambda \, \delta^{\nu\rho}  \nn \\ 
	&{}&+ p\cdot q \, \delta^{\mu\lambda} \delta^{\nu\rho} \Bigr)\varrho_{\mu\nu}(-p-q)\, A_\lambda(p) A_\rho(q) \,\sigma(-p-q)  \,,\\
	\Gamma^{AA h^\TT h^\TT}_k &=& -\frac{1}{4} Z_A Z_\TT \, \kappa^2 \,\int_{p,q,l} \Bigl( 
	p^\mu q^\beta \,\delta^{\nu\alpha} \delta^{\lambda\rho} 
	- p^\rho q^\beta \,\delta^{\nu\alpha} \delta^{\mu\lambda}  \nn \\
	&{}&- p^\mu q^\lambda \,\delta^{\nu\alpha} \delta^{\beta\rho} +\, p\cdot q\, \delta^{\mu\lambda} \delta^{\nu\alpha} \delta^{\beta\rho} + 
	\,p^\mu q^\nu \,\delta^{\alpha\lambda} \delta^{\beta\rho} 
	- p^\alpha q^\nu \,\delta^{\mu\lambda} \delta^{\beta\rho} \\
	&{}&- p^\mu q^\beta \,\delta^{\alpha\lambda} \delta^{\nu\rho} +
	p^\alpha q^\beta \,\delta^{\mu\lambda} \delta^{\nu\rho} \Bigr) 
	A_\lambda(p) A_\rho(q) \,h_{\mu\nu}^\TT(l)  \,h_{\alpha\beta}^\TT(-p-q-l),\\
	\Gamma^{AA \sigma\sigma}_k &=& -\frac{1}{4} Z_A Z_\sigma \, \kappa^2 \,\int_{p,q,l} \Bigl( 
	p^\mu q^\beta \,\delta^{\nu\alpha} \delta^{\lambda\rho} 
	- p^\rho q^\beta \,\delta^{\nu\alpha} \delta^{\mu\lambda}   \\
	&{}& - p^\mu q^\lambda \,\delta^{\nu\alpha} \delta^{\beta\rho} +\, p\cdot q\, \delta^{\mu\lambda} \delta^{\nu\alpha} \delta^{\beta\rho} + 
	\,p^\mu q^\nu \,\delta^{\alpha\lambda} \delta^{\beta\rho} 
	- p^\alpha q^\nu \,\delta^{\mu\lambda} \delta^{\beta\rho} \nn \\
	&{}&- p^\mu q^\beta \,\delta^{\alpha\lambda} \delta^{\nu\rho} +
	p^\alpha q^\beta \,\delta^{\mu\lambda} \delta^{\nu\rho} \Bigr) \varrho^{\mu\nu}(l) \varrho^{\alpha\beta}(k) \,A^\lambda(p) A^\rho(q) \,\sigma(l)  \,\sigma(-p-q-l)  \,,\nn	
\eea
\bea
	\Gamma^{\bar{\psi} \psi h^\TT}_k &=& -\frac{1}{4} Z_\psi Z_\TT^{1/2} \, \kappa \, 
	\int_{p,q} \bar{\psi}(-p) \, [ (p - q)^\nu \gamma^\mu ] \, \psi(q)   \,h_{\mu\nu}^\TT(-p-q) \,,\\
	\Gamma^{\bar{\psi} \psi \sigma}_k &=& \frac{1}{4} Z_\psi Z_\sigma^{1/2} \, \kappa \, 
	\int_{p,q} \bar{\psi}(-p)\,[ \varrho_{\mu\nu}(-p-q) (p - q)^\nu \gamma^\mu ]\,\psi(q)  \, \sigma(-p-q) \,,\\
	\Gamma^{\bar{\psi} \psi h^\TT h^\TT}_k &=& \frac{1}{8} Z_\psi Z_\TT \, \kappa^2 \int_{p,q,l} \,\bar{\psi}(-p)\, \bigg[  \,q^\mu \,\delta^{\nu\beta} \gamma^\alpha + \frac{1}{4}  \Big( \delta^{\mu\beta} \delta_{\rho}^\nu \delta_{\lambda}^\alpha k_\theta + 2\,\delta_{\lambda}^\nu \delta_{\theta}^\alpha \delta_{\rho}^\beta (-p-q-l)^\mu  \nn \\
	&{}&+ 2\,\delta^{\mu\beta} \delta_{\rho}^\nu \delta_{\theta}^\alpha (-p-q-l)_\lambda -
	2\,\delta^{\mu\beta} \delta_{\rho}^\nu \delta_{\theta}^\alpha k_\lambda + 2\,\delta^{\nu\alpha} \delta_{\theta}^\mu\delta_{\rho}^\beta (-p-q-l)_\lambda \Big) \, \gamma^\theta [\gamma^\rho,\gamma^\lambda] \,\bigg]\cdot\nn\\
	&{}&\,\, \cdot \psi(q) \,h_{\mu\nu}^\TT(l)  h_{\alpha\beta}^\TT(-p-q-l)  \,,\\	
	\Gamma^{\bar{\psi} \psi \sigma\sigma}_k &=& \frac{1}{8} Z_\psi Z_\sigma \, \kappa^2 \int_{p,q,l} \,\bar{\psi}(-p) \bigg[  \,q^\mu \,\delta^{\nu\beta} \gamma^\alpha \!+\! \frac{1}{4}  \Big( \delta^{\mu\beta} \delta_{\rho}^\nu \delta_{\lambda}^\alpha k_\theta \!+\! 2\,\delta_{\lambda}^\nu \delta_{\theta}^\alpha \delta_{\rho}^\beta (-p-q-l)^\mu  \nn \\
	&{}&+ 2\,\delta^{\mu\beta} \delta_{\rho}^\nu \delta_{\theta}^\alpha (-p-q-l)_\lambda -2\,\delta^{\mu\beta} \delta_{\rho}^\nu \delta_{\theta}^\alpha (-p-q-l)_\lambda   \\
	&{}&+ 2\,\delta^{\nu\alpha} \delta_{\theta}^\mu\delta_{\rho}^\beta (-p-q-l)_\lambda \Big) \, \gamma^\theta [\gamma^\rho,\gamma^\lambda] \,\bigg] \psi(q) \,\varrho_{\mu\nu}(l)  \varrho_{\alpha\beta}(-p-q-l) \, \sigma(l) \sigma(-p-q-l) . 	\nn
\eea
where we have defined $\kappa = \sqrt{32\pi G_N}$.

We implement the IR cutoff in terms of the following regulator functions
\bea
	[\textbf{R}_k^\TT(p^2)]^{\mu\nu\alpha\beta} &=& Z_\TT \left[ P_k(p^2)- p^2 +\bar{b} \,(P_k(p^2)^2 - p^4) \right] P_\TT^{\mu\nu\alpha\beta}	,\\
	\textbf{R}_k^{\sigma\sigma}(p^2) &=& -\frac{(d-2)(d-1)}{d^2} Z_\sigma \bigg[ P_k(p^2)- p^2 \nn \\
	&{}&- \bigg( \frac{4\bar{a}(d-1)+ \bar{b} \,d}{d-2}\bigg) \,(P_k(p^2)^2 - p^4) \bigg] \,,\nn\\
	\textbf{R}_k^{\phi\phi}(p^2) &=& Z_\phi \left[P_k(p^2)- p^2 \right] \,,\\
	\textbf{R}_k^{AA}(p^2)^{\mu\nu} &=& Z_A \left[P_k(p^2)- p^2 \right] P_\T^{\mu\nu} ,\\
	\textbf{R}_k^{\psi\psi}(p) &=& - Z_\psi \left( \sqrt{P_k(p^2)/p^2} - 1 \right) \slashed{p} \,,
\eea
where $P_k(p^2) = p^2 + (k^2-p^2)\theta(k^2-p^2)$ for the  Litim-type shape function \cite{Litim:2001up}. We note that the regulator associated with the gauge field is proportional to the transverse projector, which is consistent with the gauge choice $\zeta \to 0$.

In the following we report our findings for the gravitational (non-vanishing) contributions to the anomalous dimension of matter fields for arbitrary $d$
\bea
\eta_\phi|_{\sigma-\textmd{sunset}} &=& - \frac{d-1}{(d+6)(d-2)^2}\frac{32\,\pi \, G}{(4\pi)^{d/2} \Gamma\left( \frac{d}{2}+2 \right)} \nn \\ 
& \times& \frac{(12-4d-d^2)(2+\tilde{m}_\sigma^2) - (56 - 44 d -12d^2)a + (14d +3d^2)b}{\left(1+\tilde{m}_\sigma^2 - \frac{4a\,(d-1)+b\,d}{d-2} \right)^2} ,\\
\eta_\phi|_{\TT-\textmd{tadpole}} &=& \frac{(d+1)(d^2-4)}{4\,d\,(d+4)}
\frac{32\,\pi \, G}{(4\pi)^{d/2} \Gamma\left( \frac{d}{2}+2 \right)} 
\frac{d+4 + 2(d+2)b}{(1 + \tilde{m}_\TT^2 + b)^2} \,, \\
\eta_\phi|_{\sigma-\textmd{tadpole}} &=& \frac{d+2}{2(d+4)(d-2)^2}\frac{32\,\pi \, G}{(4\pi)^{d/2} \Gamma\left( \frac{d}{2}+2 \right)} \nn \\  
&\times& \frac{8-2d-d^2 - (16-8d-8d^2)a + (4d+2d^2) b}{\left(1+\tilde{m}_\sigma^2 - \frac{4a\,(d-1)+b\,d}{d-2} \right)^2} \,,
\eea

\bea
\eta_A|_{\TT-\textmd{sunset}} &=& \frac{(2\!-\!d)(d\!+\!1)}{(d\!-\!1)(d\!+\!6)} 
\frac{32\,\pi \, G}{(4\pi)^{d/2} \Gamma\left( \frac{d}{2}+2 \right)}  
\frac{(d\!+\!6)(2+\tilde{m}_\TT^2) + (14\!+\!3d) b}{(1 + \tilde{m}_\TT^2 + b)^2} \,,\\
\eta_A|_{\sigma-\textmd{sunset}} &=& - \frac{2}{(d+6)(d-1)}\frac{32\,\pi \, G}{(4\pi)^{d/2} \Gamma\left( \frac{d}{2}+2 \right)} \nn \\ 
&\times& \frac{(12-4d-d^2)(2+\tilde{m}_\sigma^2) \!-\! (56 - 44 d -12d^2)a \!+\! (14d +3d^2)b}{\left(1+\tilde{m}_\sigma^2 - \frac{4a\,(d-1)+b\,d}{d-2} \right)^2} \,,\\
\eta_A|_{\TT-\textmd{tadpole}} &=& \frac{(d+1)(d+2)(d-2)^2}{2d(d+4)(d-1)} 
\frac{32\,\pi \, G}{(4\pi)^{d/2} \Gamma\left( \frac{d}{2}+2 \right)}  
\frac{d+4 + 2(d+2)b}{(1 + \tilde{m}_\TT^2 + b)^2} \,,\\
\eta_A|_{\sigma-\textmd{tadpole}} &=& \frac{d+2}{(d-1)(d-2)(d+4)}\frac{32\,\pi \, G}{(4\pi)^{d/2} \Gamma\left( \frac{d}{2}+2 \right)} \nn \\ 
&\times& \frac{8-2d-d^2 - (16-8d-8d^2)a + (4d+2d^2) b}{\left(1+\tilde{m}_\sigma^2 - \frac{4a\,(d-1)+b\,d}{d-2} \right)^2} \,,
\eea

\bea
\eta_\psi|_{\sigma-\textmd{sunset}} &=&  \frac{(d-1)(d+2)}{16(d+1)(d+5)(d-2)^2}\frac{32\,\pi \, G}{(4\pi)^{d/2} \Gamma\left( \frac{d}{2}+2 \right)} \nn \\ 
&\times& \Bigg\{ \frac{-10+23d-5d^2-2d^3 + (10d-3d^2-d^3) \tilde{m}_\sigma^2}{\left(1+\tilde{m}_\sigma^2 - \frac{4a\,(d-1)+b\,d}{d-2} \right)^2} + \nn \\
&+& \frac{12(2 - 5 d +2 d^2 + d^3)a -3(2d -3d^2 - d^3)b}{\left(1+\tilde{m}_\sigma^2 - \frac{4a\,(d-1)+b\,d}{d-2} \right)^2}\Bigg\}\,,\nn \\
\eta_\psi|_{\TT-\textmd{tadpole}} &=&  \frac{(d-2)(d+1)(d+2)}{16d(d+4)} 
\frac{32\,\pi \, G}{(4\pi)^{d/2} \Gamma\left( \frac{d}{2}+2 \right)}  
\frac{d+4 + 2(d+2)b}{(1 + \tilde{m}_\TT^2 + b)^2} \,,\nn \\
\eta_\psi|_{\sigma-\textmd{tadpole}} &=& \frac{d+2}{8(d-2)^2(d+4)}\frac{32\,\pi \, G}{(4\pi)^{d/2} \Gamma\left( \frac{d}{2}+2 \right)} \nn \\ 
&\times& \frac{8-2d-d^2 - (16-8d-8d^2)a + (4d+2d^2) b}{\left(1+\tilde{m}_\sigma^2 - \frac{4a\,(d-1)+b\,d}{d-2} \right)^2} \,,
\eea

In addition, since we are interested in the running of Standard Model-like couplings, the beta function for the Yukawa coupling receives a gravitational contribution coming from the triangle diagram represented in Fig.~\ref{diagrams_2},
\bea
\beta_y|_{\textmd{triangle}} &\equiv&  \mathcal{D}_{y} \, y \nn \\
&=& -\frac{d(d-1)}{16(d+6)(d-2)^2}\frac{32\,\pi \, G\, y}{(4\pi)^{d/2} \Gamma\left( \frac{d}{2}+2 \right)} \nn \\ 
&\times& \frac{(12-4d-d^2)(2+\tilde{m}_\sigma^2) - (56 - 44 d -12d^2)a + (14d +3d^2)b}{\left(1+\tilde{m}_\sigma^2 - \frac{4a\,(d-1)+b\,d}{d-2} \right)^2} \,.
\eea
We highlight that in the above set of expressions we have neglected the anomalous dimension contribution coming from cutoff insertions. This approach is sometimes referred as a perturbative approximation of the FRG.

In the main text we have restricted our analysis to the symmetric case with $\tilde{m}_\TT^2 = \tilde{m}_\sigma^2 = 0$. The introduction of symmetry-breaking masses in the UQG setting allows us to mimic the behavior of the results obtained in the standard gravity framework (as long as we do not take into account the symmetry-identities which differ in the two settings). More precisely, in $d=4$, identifying the symmetry-breaking masses with the dimensionless cosmological constant, namely $\tilde{m}_\TT^2 = -2\lambda$ and $\tilde{m}_\sigma^2 = 4\lambda$, the UQG result coincides with the expressions obtained within standard gravity with linear metric parameterization and gauge choice $\beta \to - \infty$. This agreement between exponential and linear parameterization, despite differences at the level of individual diagrams, can be interpreted as a hint for the robustness of the results.
We highlight that this agreement between standard ASQG and UQG is only at the level of the present truncation, where no beta functions for the gravitational couplings are calculated, and where the symmetry-identities are neglected.

\begin{figure}[htb!]
	\begin{center}
		\includegraphics[scale=.4]{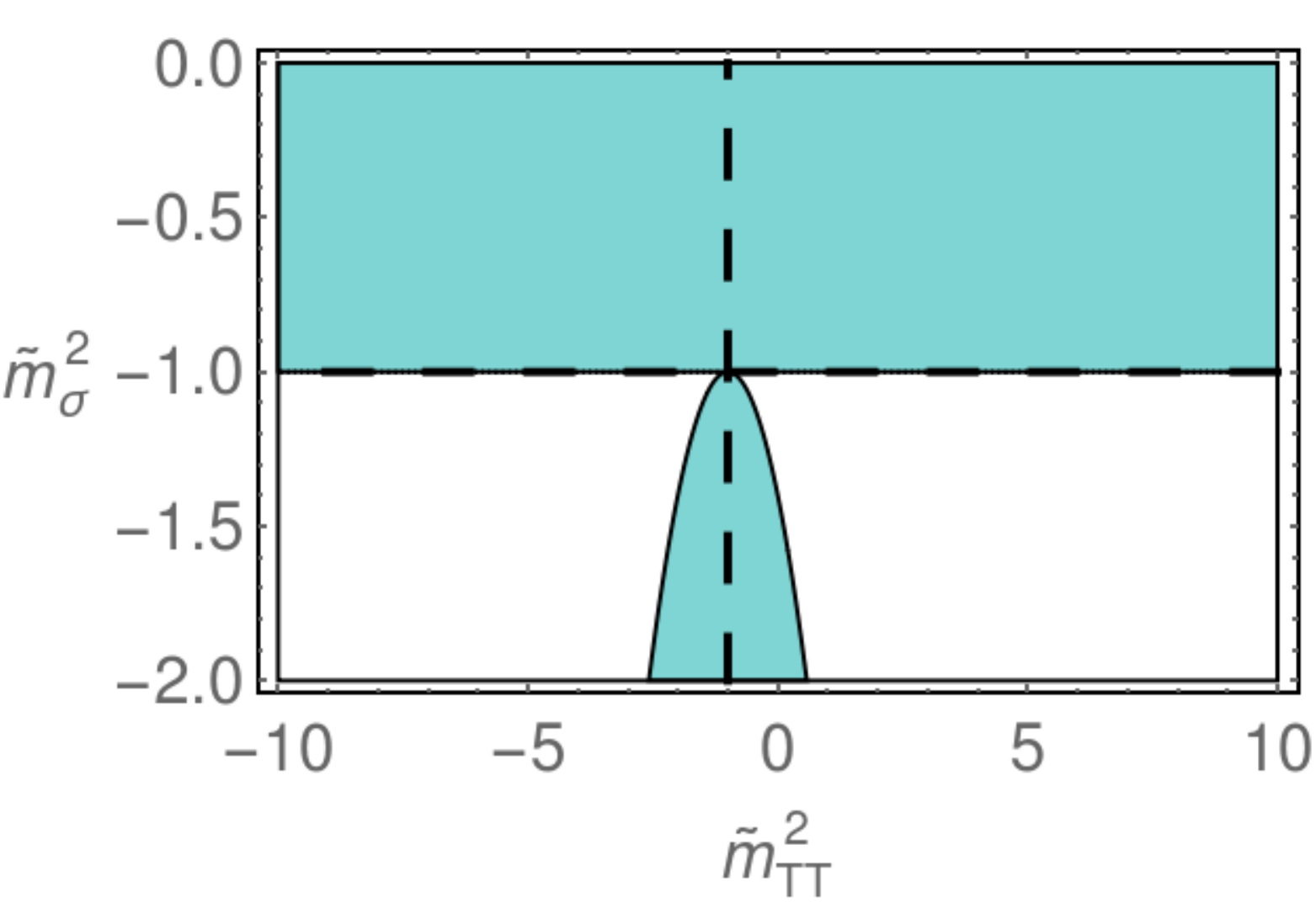} \qquad 
		\includegraphics[scale=.4]{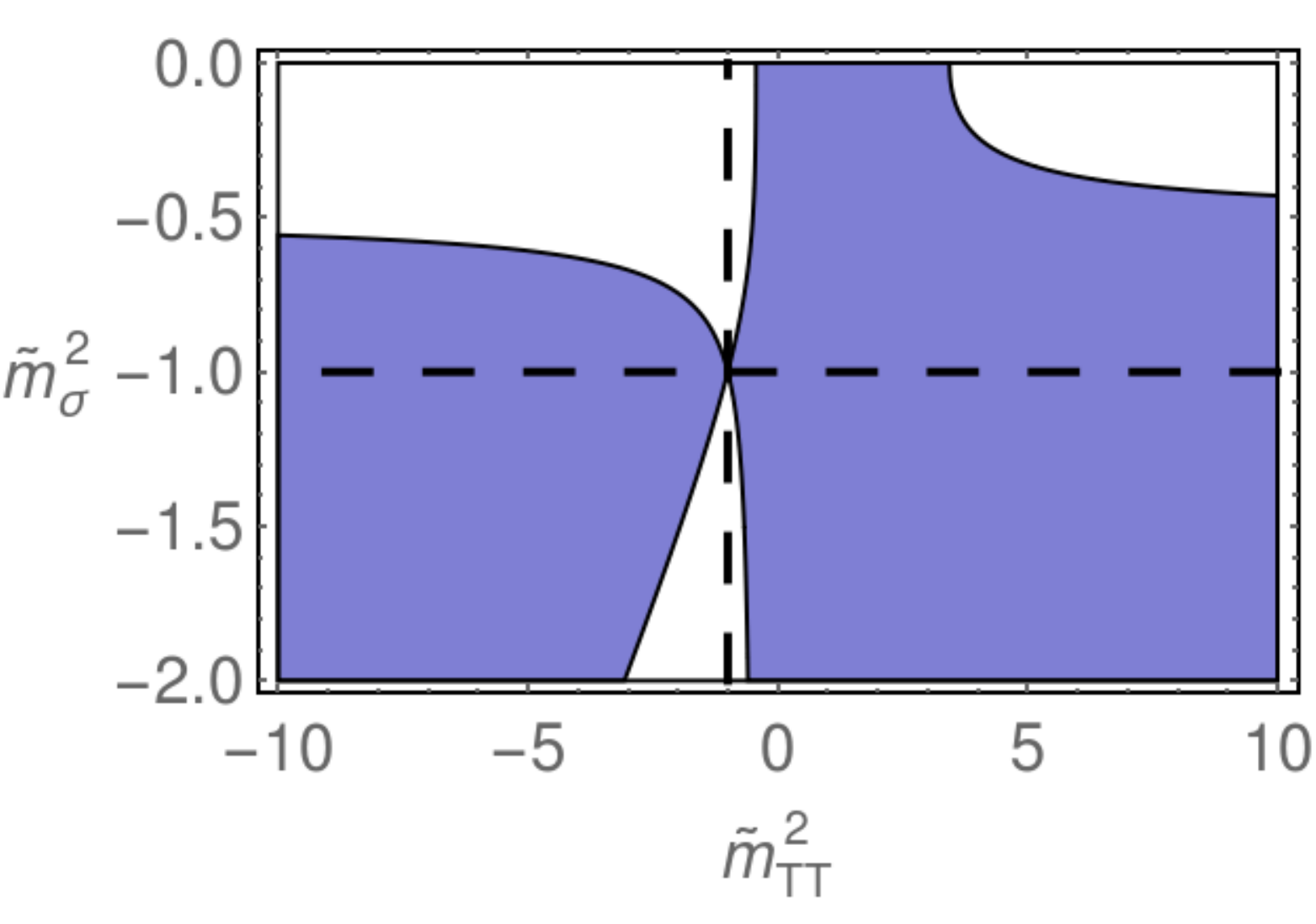} \\
		\includegraphics[scale=.4]{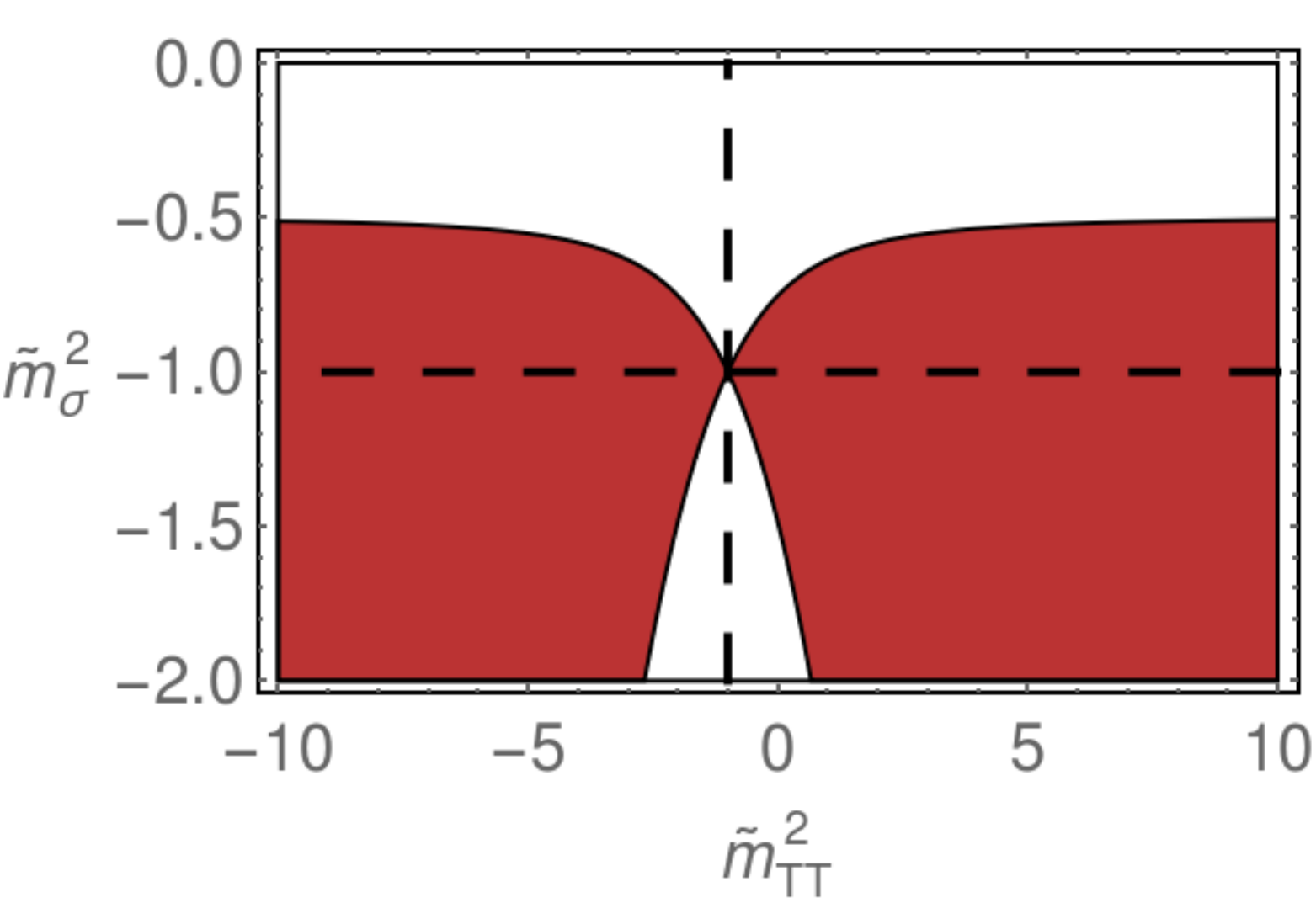} \qquad 
		\includegraphics[scale=.4]{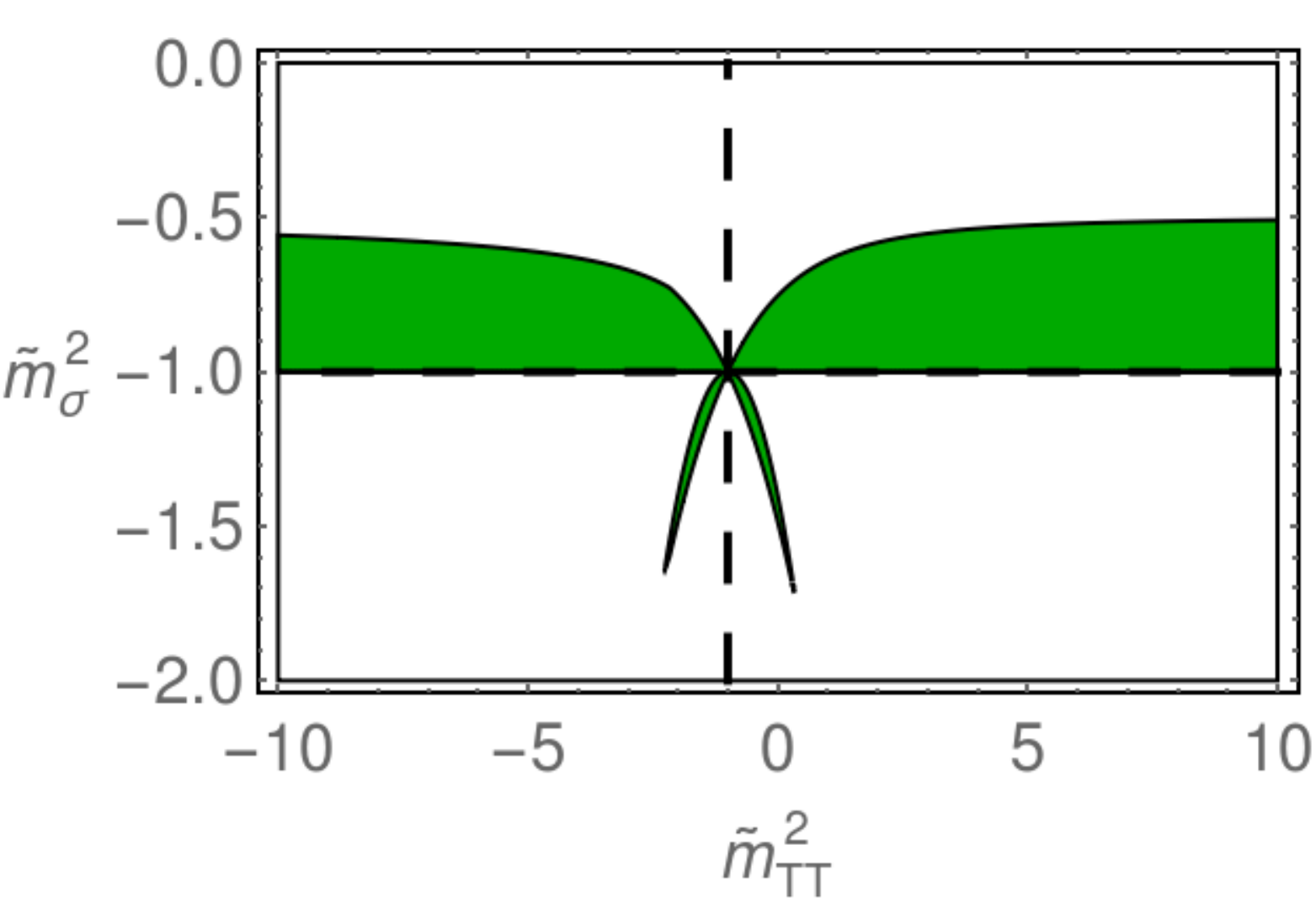}
		\caption{\footnotesize { Region plots for the sign of the gravitation contribution to SM-like couplings in terms of the symmetry breaking masses $\tilde{m}_\TT^2$ and $\tilde{m}_\sigma^2$ (with vanishing curvature squared couplings). From left to right and top to down: 
		i) $\eta_\phi|_{\textmd{grav}}>0$ (cyan region); 
		ii) $-f_g = \eta_A|_{\textmd{grav}}<0$ (blue region);
		iii) $-f_y=\left(\eta_\psi + \frac{1}{2} \eta_\phi + \mathcal{D}_y\right)_{\textmd{grav}} \!< 0$ (dark red region);
		iv) combined plot (dark green) with $\eta_\phi|_{\textmd{grav}}>0$, $-f_g< 0$  and $-f_y<0$.
		}}
		\label{SM_viability_Sym_Break_1}
	\end{center}
\end{figure}

Given the number of free parameters the analysis of results including higher curvature coefficients and symmetric breaking masses can be rather cumbersome. In order to make it simpler, here we switch off the higher curvature terms and focus on the $\tilde{m}_\TT^2 \times \tilde{m}_\sigma^2$ plane. In Fig.~\ref{SM_viability_Sym_Break_1} we plot the viable region for an asymptotically free UV completion of the Yukawa and (non-)Abelian gauge couplings, a predicted ratio of the electroweak scale to the Higgs mass and the intersection of these three conditions. As one can see, even in the absence of curvature squared terms, the symmetry-breaking mass terms induce regions where all three conditions can be satisfied, just as in the case of the linear parameterization.

\section{Explicit Results for the Reuter fixed point  \label{Standard_ASQG}} 

For the sake of completeness in this appendix we report some results obtained in the standard ASQG framework. In order to fix our notation, below we present the truncation used for these computations
\begin{align} \label{EAA_Standard}
&\Gamma_k^{\textmd{Stand.}} = \frac{1}{16\pi \,G_N} \int_x \sqrt{g} \left( 2\Lambda -R +\bar{a}\, R^2 + \bar{b} \, R_{\mu\nu}R^{\mu\nu} \right)  + \Gamma_k^{\textmd{g.f.}} +  \Gamma_k^{\textmd{gh.}} + \\
&+ \int_x \!\sqrt{g} \left( \frac{1}{2} g^{\mu\nu} \pt_\mu \phi \pt_\nu \phi + \frac{\lambda_4}{4!} \phi^4 \right)  + \frac{1}{4} \int_x\!\! \sqrt{g} \,g^{\mu\alpha}g^{\nu\beta} F_{\mu\nu} F_{\alpha\beta} \,+ \int_x\!\!\sqrt{g}\,(i \bar{\psi} \slashed{D} \psi + i \, y \phi \bar{\psi} \psi) ,\nn
\end{align}
with the gauge-fixing part  given by
\begin{align}
\Gamma_k^{\textmd{g.f.}} =&\, \frac{1}{32\pi \alpha\, G_N} \int_x \sqrt{g}\, \bar{g}^{\mu\nu} F_\mu[h] F_\nu[h] +\frac{Z_A}{2\zeta}\int_x \sqrt{g} \, (\bar{g}^{\mu\nu} \bar{\nabla}_\mu A_\nu)^2 ,
\end{align}
where $F_{\mu}[h] =  \bar{\nabla}^\beta h_{\mu\beta}- \frac{1+\beta}{4} \bar{\nabla}_\mu h \, (= 0)$. For the sake of simplicity, we restrict ourselves to the four-dimensional case. In addition, we adopt the linear parameterization for the metric.

\begin{figure}[htb!]
	\begin{center}
		\includegraphics[scale=.6]{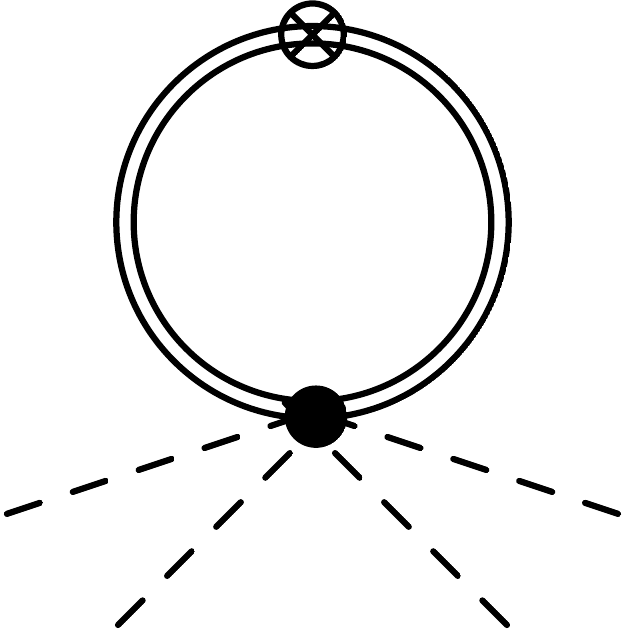}
		\caption{\footnotesize Tadpole diagram contributing to the running of the quartic scalar coupling within the standard ASQG framework.}
		\label{tadpole_scalar}
	\end{center}
\end{figure}

Below we report the results for the gravitational contribution to the   anomalous dimensions of scalars and fermions
\begin{subequations}
	\begin{align}
	\eta_\phi|_{\textmd{grav}} &= \frac{G}{5\pi} \frac{(3-4\beta + \beta^2)^2\, (5-39a-13b) + 10(\beta-1)^2(\beta^2-3) \,\lambda}{\left(  (3-\beta)^2\,(1-6a-2b) - 4(3-\beta^2)\,\lambda \right)^2} \,,\\
	\eta_\psi|_{\textmd{grav}} &= - \frac{25\,G}{32\pi} \frac{2+3b}{(1+b-2\lambda)^2} + 
	\frac{G}{80\pi} \Bigg\{\frac{(3-\beta)^2 \left( 369 - (90-29\beta)\beta \right)}{\left(  (3-\beta)^2\,(1-6a-2b) - 4(3-\beta^2)\,\lambda \right)^2} + \nn \\
	&\,\,- \frac{(3-\beta)^2 \left[ (519-7\beta(20-7\beta))(6a+2b)\right]
	+48 (27-12\beta^2 + \beta^4)\, \lambda}{\left(  (3-\beta)^2\,(1-6a-2b) - 4(3-\beta^2)\,\lambda \right)^2}  \Bigg\}\,.
	\end{align}
\end{subequations}
The running of the Yukawa coupling receives contributions coming from the diagrams represented in Fig.~\ref{diagrams_2}, resulting in the expression
\begin{align}
\beta_y|_{\textmd{Fig.$\,$\ref{diagrams_2}}} &=
\frac{5\,G\,y}{4\pi} \frac{2+3b}{(1+b-2\lambda)^2}-
\frac{G\,y}{20\pi} \Bigg\{ \frac{ (3-\beta)^2 (111 - 885 a - 295 b ) }{\left(  (3-\beta)^2\,(1-6a-2b) - 4(3-\beta^2)\,\lambda \right)^2} \nn \\
&\,+ \frac{ (3-\beta)^2 \,\beta^2 \,(15-141a-47b) + 2 (\beta^2-3)(99-5\beta^2) \,\lambda }{\left(  (3-\beta)^2\,(1-6a-2b) - 4(3-\beta^2)\,\lambda \right)^2} \nn \\
&- \frac{ 2(63-537a-179b)(\beta-3)^2 \beta + 54 \beta\,(\beta^2-3) \lambda }{\left(  (3-\beta)^2\,(1-6a-2b) - 4(3-\beta^2)\,\lambda \right)^2} \bigg\} \,.
\end{align}
Finally, for the scalar quartic coupling, the tadpole diagram represented in Fig.~\ref{tadpole_scalar} gives the following result
\begin{align}
\beta_{\lambda_4}|_{\textmd{tadpole}} = \frac{G\,\lambda_4}{4\pi}
\Bigg[ \frac{5(2+3b)}{(1+b-2\lambda)^2} - \frac{4(\beta-3)^2(\beta^2-3)(1-9a-3b)}{\left(  (3-\beta)^2\,(1-6a-2b) - 4(3-\beta^2)\,\lambda \right)^2} \Bigg] \,.
\end{align}

 The complete result for the gravitational contribution  to the anomalous dimension of (non-)Abelian gauge fields, in an approximation containing higher-curvature terms, is
\begin{align}
\eta_A|_{\textmd{grav}} = - \frac{G}{18\pi} \frac{10+7b-40\lambda}{(1+b-2\lambda)^2}
+ \frac{2\,G}{45\pi} \frac{\beta^2 \left( (3-\beta)^2(5-21a-7b) - 40(3-\beta^2)\,\lambda \right)}{\left(  (3-\beta)^2\,(1-6a-2b) - 4(3-\beta^2)\,\lambda \right)^2}\,.
\end{align}
Although this expression is non-universal, as one could expect due to the dimensionful nature of the gravitational couplings, we observe that it is possible to find universal contributions (with respect to the cutoff shape function) appearing as the coefficient of dimensionless combinations such as $\Lambda \, G_N \equiv \lambda \,G$. In fact, by expanding $\eta_A$ around $\Lambda = 0$ (and setting $a=b=0$), it is possible to verify that the contribution associated with the dimensionless combination $\Lambda \, G_N$ is given by (for a generic shape function)
\begin{align}
\eta_A|_{\textmd{grav}}^{(\Lambda G_N)} = A_\beta \, \left( \Phi_2^3 - 3 \Phi_3^4 \right) \Lambda \,G_N \,,
\end{align}
where $A_\beta$ corresponds to some $\beta$-dependent coefficient and $\Phi_n^p$ is  the threshold integral defined in \eqref{threshold_1}. Given that $\Phi_n^{n+1} = 1/\Gamma(n+1)$  irrespective of the choice for the shape function, the above expression turns out to be universally zero. It is worth emphasizing that such a result depends on the cancellation of universal contributions coming  from different diagrams contributing to $\eta_A$.


\newpage
\bibliography{refs}

\begin{thebibliography}{207}
\expandafter\ifx\csname natexlab\endcsname\relax\def\natexlab#1{#1}\fi
\expandafter\ifx\csname bibnamefont\endcsname\relax
  \def\bibnamefont#1{#1}\fi
\expandafter\ifx\csname bibfnamefont\endcsname\relax
  \def\bibfnamefont#1{#1}\fi
\expandafter\ifx\csname citenamefont\endcsname\relax
  \def\citenamefont#1{#1}\fi
\expandafter\ifx\csname url\endcsname\relax
  \def\url#1{\texttt{#1}}\fi
\expandafter\ifx\csname urlprefix\endcsname\relax\def\urlprefix{URL }\fi
\providecommand{\bibinfo}[2]{#2}
\providecommand{\eprint}[2][]{\url{#2}}

\bibitem[{\citenamefont{Dowker}(2013)}]{Dowker:aza}
\bibinfo{author}{\bibfnamefont{F.}~\bibnamefont{Dowker}},
  \bibinfo{journal}{Gen. Rel. Grav.} \textbf{\bibinfo{volume}{45}},
  \bibinfo{pages}{1651} (\bibinfo{year}{2013}).

\bibitem[{\citenamefont{Maiani et~al.}(1978)\citenamefont{Maiani, Parisi, and
  Petronzio}}]{Maiani:1977cg}
\bibinfo{author}{\bibfnamefont{L.}~\bibnamefont{Maiani}},
  \bibinfo{author}{\bibfnamefont{G.}~\bibnamefont{Parisi}}, \bibnamefont{and}
  \bibinfo{author}{\bibfnamefont{R.}~\bibnamefont{Petronzio}},
  \bibinfo{journal}{Nucl. Phys.} \textbf{\bibinfo{volume}{B136}},
  \bibinfo{pages}{115} (\bibinfo{year}{1978}).

\bibitem[{\citenamefont{Cabibbo et~al.}(1979)\citenamefont{Cabibbo, Maiani,
  Parisi, and Petronzio}}]{Cabibbo:1979ay}
\bibinfo{author}{\bibfnamefont{N.}~\bibnamefont{Cabibbo}},
  \bibinfo{author}{\bibfnamefont{L.}~\bibnamefont{Maiani}},
  \bibinfo{author}{\bibfnamefont{G.}~\bibnamefont{Parisi}}, \bibnamefont{and}
  \bibinfo{author}{\bibfnamefont{R.}~\bibnamefont{Petronzio}},
  \bibinfo{journal}{Nucl. Phys.} \textbf{\bibinfo{volume}{B158}},
  \bibinfo{pages}{295} (\bibinfo{year}{1979}).

\bibitem[{\citenamefont{Dashen and Neuberger}(1983)}]{Dashen:1983ts}
\bibinfo{author}{\bibfnamefont{R.~F.} \bibnamefont{Dashen}} \bibnamefont{and}
  \bibinfo{author}{\bibfnamefont{H.}~\bibnamefont{Neuberger}},
  \bibinfo{journal}{Phys. Rev. Lett.} \textbf{\bibinfo{volume}{50}},
  \bibinfo{pages}{1897} (\bibinfo{year}{1983}).

\bibitem[{\citenamefont{Callaway}(1984)}]{Callaway:1983zd}
\bibinfo{author}{\bibfnamefont{D.~J.~E.} \bibnamefont{Callaway}},
  \bibinfo{journal}{Nucl. Phys.} \textbf{\bibinfo{volume}{B233}},
  \bibinfo{pages}{189} (\bibinfo{year}{1984}).

\bibitem[{\citenamefont{Beg et~al.}(1984)\citenamefont{Beg, Panagiotakopoulos,
  and Sirlin}}]{Beg:1983tu}
\bibinfo{author}{\bibfnamefont{M.~A.~B.} \bibnamefont{Beg}},
  \bibinfo{author}{\bibfnamefont{C.}~\bibnamefont{Panagiotakopoulos}},
  \bibnamefont{and} \bibinfo{author}{\bibfnamefont{A.}~\bibnamefont{Sirlin}},
  \bibinfo{journal}{Phys. Rev. Lett.} \textbf{\bibinfo{volume}{52}},
  \bibinfo{pages}{883} (\bibinfo{year}{1984}).

\bibitem[{\citenamefont{Lindner}(1986)}]{Lindner:1985uk}
\bibinfo{author}{\bibfnamefont{M.}~\bibnamefont{Lindner}}, \bibinfo{journal}{Z.
  Phys.} \textbf{\bibinfo{volume}{C31}}, \bibinfo{pages}{295}
  (\bibinfo{year}{1986}).

\bibitem[{\citenamefont{Kuti et~al.}(1988)\citenamefont{Kuti, Lin, and
  Shen}}]{Kuti:1987nr}
\bibinfo{author}{\bibfnamefont{J.}~\bibnamefont{Kuti}},
  \bibinfo{author}{\bibfnamefont{L.}~\bibnamefont{Lin}}, \bibnamefont{and}
  \bibinfo{author}{\bibfnamefont{Y.}~\bibnamefont{Shen}},
  \bibinfo{journal}{Phys. Rev. Lett.} \textbf{\bibinfo{volume}{61}},
  \bibinfo{pages}{678} (\bibinfo{year}{1988}).

\bibitem[{\citenamefont{Hambye and Riesselmann}(1997)}]{Hambye:1996wb}
\bibinfo{author}{\bibfnamefont{T.}~\bibnamefont{Hambye}} \bibnamefont{and}
  \bibinfo{author}{\bibfnamefont{K.}~\bibnamefont{Riesselmann}},
  \bibinfo{journal}{Phys. Rev.} \textbf{\bibinfo{volume}{D55}},
  \bibinfo{pages}{7255} (\bibinfo{year}{1997}), \eprint{hep-ph/9610272}.

\bibitem[{\citenamefont{Gell-Mann and Low}(1954)}]{GellMann:1954fq}
\bibinfo{author}{\bibfnamefont{M.}~\bibnamefont{Gell-Mann}} \bibnamefont{and}
  \bibinfo{author}{\bibfnamefont{F.~E.} \bibnamefont{Low}},
  \bibinfo{journal}{Phys. Rev.} \textbf{\bibinfo{volume}{95}},
  \bibinfo{pages}{1300} (\bibinfo{year}{1954}).

\bibitem[{\citenamefont{Gockeler et~al.}(1998)\citenamefont{Gockeler, Horsley,
  Linke, Rakow, Schierholz, and Stuben}}]{Gockeler:1997dn}
\bibinfo{author}{\bibfnamefont{M.}~\bibnamefont{Gockeler}},
  \bibinfo{author}{\bibfnamefont{R.}~\bibnamefont{Horsley}},
  \bibinfo{author}{\bibfnamefont{V.}~\bibnamefont{Linke}},
  \bibinfo{author}{\bibfnamefont{P.~E.~L.} \bibnamefont{Rakow}},
  \bibinfo{author}{\bibfnamefont{G.}~\bibnamefont{Schierholz}},
  \bibnamefont{and} \bibinfo{author}{\bibfnamefont{H.}~\bibnamefont{Stuben}},
  \bibinfo{journal}{Phys. Rev. Lett.} \textbf{\bibinfo{volume}{80}},
  \bibinfo{pages}{4119} (\bibinfo{year}{1998}), \eprint{hep-th/9712244}.

\bibitem[{\citenamefont{Kim et~al.}(2001)\citenamefont{Kim, Kogut, and
  Lombardo}}]{Kim:2000rr}
\bibinfo{author}{\bibfnamefont{S.}~\bibnamefont{Kim}},
  \bibinfo{author}{\bibfnamefont{J.~B.} \bibnamefont{Kogut}}, \bibnamefont{and}
  \bibinfo{author}{\bibfnamefont{M.-P.} \bibnamefont{Lombardo}},
  \bibinfo{journal}{Phys. Lett.} \textbf{\bibinfo{volume}{B502}},
  \bibinfo{pages}{345} (\bibinfo{year}{2001}), \eprint{hep-lat/0009029}.

\bibitem[{\citenamefont{Gies and Jaeckel}(2004)}]{Gies:2004hy}
\bibinfo{author}{\bibfnamefont{H.}~\bibnamefont{Gies}} \bibnamefont{and}
  \bibinfo{author}{\bibfnamefont{J.}~\bibnamefont{Jaeckel}},
  \bibinfo{journal}{Phys. Rev. Lett.} \textbf{\bibinfo{volume}{93}},
  \bibinfo{pages}{110405} (\bibinfo{year}{2004}), \eprint{hep-ph/0405183}.

\bibitem[{\citenamefont{Freedman et~al.}(1982)\citenamefont{Freedman,
  Smolensky, and Weingarten}}]{Freedman:1981wr}
\bibinfo{author}{\bibfnamefont{B.}~\bibnamefont{Freedman}},
  \bibinfo{author}{\bibfnamefont{P.}~\bibnamefont{Smolensky}},
  \bibnamefont{and}
  \bibinfo{author}{\bibfnamefont{D.}~\bibnamefont{Weingarten}},
  \bibinfo{journal}{Phys. Lett.} \textbf{\bibinfo{volume}{113B}},
  \bibinfo{pages}{481} (\bibinfo{year}{1982}).

\bibitem[{\citenamefont{Aizenman}(1981)}]{Aizenman:1981du}
\bibinfo{author}{\bibfnamefont{M.}~\bibnamefont{Aizenman}},
  \bibinfo{journal}{Phys. Rev. Lett.} \textbf{\bibinfo{volume}{47}},
  \bibinfo{pages}{1} (\bibinfo{year}{1981}).

\bibitem[{\citenamefont{Frohlich}(1982)}]{Frohlich:1982tw}
\bibinfo{author}{\bibfnamefont{J.}~\bibnamefont{Frohlich}},
  \bibinfo{journal}{Nucl. Phys.} \textbf{\bibinfo{volume}{B200}},
  \bibinfo{pages}{281} (\bibinfo{year}{1982}).

\bibitem[{\citenamefont{Wetterich}(2019)}]{Wetterich:2019qzx}
\bibinfo{author}{\bibfnamefont{C.}~\bibnamefont{Wetterich}}
  (\bibinfo{year}{2019}), \eprint{1901.04741}.

\bibitem[{\citenamefont{Eichhorn}(2019{\natexlab{a}})}]{Eichhorn:2018yfc}
\bibinfo{author}{\bibfnamefont{A.}~\bibnamefont{Eichhorn}},
  \bibinfo{journal}{Front. Astron. Space Sci.} \textbf{\bibinfo{volume}{5}},
  \bibinfo{pages}{47} (\bibinfo{year}{2019}{\natexlab{a}}),
  \eprint{1810.07615}.

\bibitem[{\citenamefont{Gielen et~al.}(2018)\citenamefont{Gielen,
  de~Leon~Ardon, and Percacci}}]{Gielen:2018pvk}
\bibinfo{author}{\bibfnamefont{S.}~\bibnamefont{Gielen}},
  \bibinfo{author}{\bibfnamefont{R.}~\bibnamefont{de~Leon~Ardon}},
  \bibnamefont{and} \bibinfo{author}{\bibfnamefont{R.}~\bibnamefont{Percacci}},
  \bibinfo{journal}{Class. Quant. Grav.} \textbf{\bibinfo{volume}{35}},
  \bibinfo{pages}{195009} (\bibinfo{year}{2018}), \eprint{1805.11626}.

\bibitem[{\citenamefont{Reuter and Saueressig}(2002)}]{Reuter:2001ag}
\bibinfo{author}{\bibfnamefont{M.}~\bibnamefont{Reuter}} \bibnamefont{and}
  \bibinfo{author}{\bibfnamefont{F.}~\bibnamefont{Saueressig}},
  \bibinfo{journal}{Phys. Rev.} \textbf{\bibinfo{volume}{D65}},
  \bibinfo{pages}{065016} (\bibinfo{year}{2002}), \eprint{hep-th/0110054}.

\bibitem[{\citenamefont{Lauscher and
  Reuter}(2002{\natexlab{a}})}]{Lauscher:2001ya}
\bibinfo{author}{\bibfnamefont{O.}~\bibnamefont{Lauscher}} \bibnamefont{and}
  \bibinfo{author}{\bibfnamefont{M.}~\bibnamefont{Reuter}},
  \bibinfo{journal}{Phys. Rev.} \textbf{\bibinfo{volume}{D65}},
  \bibinfo{pages}{025013} (\bibinfo{year}{2002}{\natexlab{a}}),
  \eprint{hep-th/0108040}.

\bibitem[{\citenamefont{Litim}(2004)}]{Litim:2003vp}
\bibinfo{author}{\bibfnamefont{D.~F.} \bibnamefont{Litim}},
  \bibinfo{journal}{Phys. Rev. Lett.} \textbf{\bibinfo{volume}{92}},
  \bibinfo{pages}{201301} (\bibinfo{year}{2004}), \eprint{hep-th/0312114}.

\bibitem[{\citenamefont{Codello and Percacci}(2006)}]{Codello:2006in}
\bibinfo{author}{\bibfnamefont{A.}~\bibnamefont{Codello}} \bibnamefont{and}
  \bibinfo{author}{\bibfnamefont{R.}~\bibnamefont{Percacci}},
  \bibinfo{journal}{Phys. Rev. Lett.} \textbf{\bibinfo{volume}{97}},
  \bibinfo{pages}{221301} (\bibinfo{year}{2006}), \eprint{hep-th/0607128}.

\bibitem[{\citenamefont{Codello et~al.}(2009)\citenamefont{Codello, Percacci,
  and Rahmede}}]{Codello:2008vh}
\bibinfo{author}{\bibfnamefont{A.}~\bibnamefont{Codello}},
  \bibinfo{author}{\bibfnamefont{R.}~\bibnamefont{Percacci}}, \bibnamefont{and}
  \bibinfo{author}{\bibfnamefont{C.}~\bibnamefont{Rahmede}},
  \bibinfo{journal}{Annals Phys.} \textbf{\bibinfo{volume}{324}},
  \bibinfo{pages}{414} (\bibinfo{year}{2009}), \eprint{0805.2909}.

\bibitem[{\citenamefont{Benedetti et~al.}(2009)\citenamefont{Benedetti,
  Machado, and Saueressig}}]{Benedetti:2009rx}
\bibinfo{author}{\bibfnamefont{D.}~\bibnamefont{Benedetti}},
  \bibinfo{author}{\bibfnamefont{P.~F.} \bibnamefont{Machado}},
  \bibnamefont{and}
  \bibinfo{author}{\bibfnamefont{F.}~\bibnamefont{Saueressig}},
  \bibinfo{journal}{Mod. Phys. Lett.} \textbf{\bibinfo{volume}{A24}},
  \bibinfo{pages}{2233} (\bibinfo{year}{2009}), \eprint{0901.2984}.

\bibitem[{\citenamefont{Niedermaier}(2010)}]{Niedermaier:2010zz}
\bibinfo{author}{\bibfnamefont{M.}~\bibnamefont{Niedermaier}},
  \bibinfo{journal}{Nucl. Phys.} \textbf{\bibinfo{volume}{B833}},
  \bibinfo{pages}{226} (\bibinfo{year}{2010}).

\bibitem[{\citenamefont{Manrique
  et~al.}(2011{\natexlab{a}})\citenamefont{Manrique, Reuter, and
  Saueressig}}]{Manrique:2010am}
\bibinfo{author}{\bibfnamefont{E.}~\bibnamefont{Manrique}},
  \bibinfo{author}{\bibfnamefont{M.}~\bibnamefont{Reuter}}, \bibnamefont{and}
  \bibinfo{author}{\bibfnamefont{F.}~\bibnamefont{Saueressig}},
  \bibinfo{journal}{Annals Phys.} \textbf{\bibinfo{volume}{326}},
  \bibinfo{pages}{463} (\bibinfo{year}{2011}{\natexlab{a}}),
  \eprint{1006.0099}.

\bibitem[{\citenamefont{Manrique
  et~al.}(2011{\natexlab{b}})\citenamefont{Manrique, Rechenberger, and
  Saueressig}}]{Manrique:2011jc}
\bibinfo{author}{\bibfnamefont{E.}~\bibnamefont{Manrique}},
  \bibinfo{author}{\bibfnamefont{S.}~\bibnamefont{Rechenberger}},
  \bibnamefont{and}
  \bibinfo{author}{\bibfnamefont{F.}~\bibnamefont{Saueressig}},
  \bibinfo{journal}{Phys. Rev. Lett.} \textbf{\bibinfo{volume}{106}},
  \bibinfo{pages}{251302} (\bibinfo{year}{2011}{\natexlab{b}}),
  \eprint{1102.5012}.

\bibitem[{\citenamefont{Dietz and Morris}(2013)}]{Dietz:2012ic}
\bibinfo{author}{\bibfnamefont{J.~A.} \bibnamefont{Dietz}} \bibnamefont{and}
  \bibinfo{author}{\bibfnamefont{T.~R.} \bibnamefont{Morris}},
  \bibinfo{journal}{JHEP} \textbf{\bibinfo{volume}{01}}, \bibinfo{pages}{108}
  (\bibinfo{year}{2013}), \eprint{1211.0955}.

\bibitem[{\citenamefont{Donkin and Pawlowski}(2012)}]{Donkin:2012ud}
\bibinfo{author}{\bibfnamefont{I.}~\bibnamefont{Donkin}} \bibnamefont{and}
  \bibinfo{author}{\bibfnamefont{J.~M.} \bibnamefont{Pawlowski}}
  (\bibinfo{year}{2012}), \eprint{1203.4207}.

\bibitem[{\citenamefont{Codello et~al.}(2014)\citenamefont{Codello, D'Odorico,
  and Pagani}}]{Codello:2013fpa}
\bibinfo{author}{\bibfnamefont{A.}~\bibnamefont{Codello}},
  \bibinfo{author}{\bibfnamefont{G.}~\bibnamefont{D'Odorico}},
  \bibnamefont{and} \bibinfo{author}{\bibfnamefont{C.}~\bibnamefont{Pagani}},
  \bibinfo{journal}{Phys. Rev.} \textbf{\bibinfo{volume}{D89}},
  \bibinfo{pages}{081701} (\bibinfo{year}{2014}), \eprint{1304.4777}.

\bibitem[{\citenamefont{Falls et~al.}(2013)\citenamefont{Falls, Litim,
  Nikolakopoulos, and Rahmede}}]{Falls:2013bv}
\bibinfo{author}{\bibfnamefont{K.}~\bibnamefont{Falls}},
  \bibinfo{author}{\bibfnamefont{D.~F.} \bibnamefont{Litim}},
  \bibinfo{author}{\bibfnamefont{K.}~\bibnamefont{Nikolakopoulos}},
  \bibnamefont{and} \bibinfo{author}{\bibfnamefont{C.}~\bibnamefont{Rahmede}}
  (\bibinfo{year}{2013}), \eprint{1301.4191}.

\bibitem[{\citenamefont{Becker and Reuter}(2014)}]{Becker:2014qya}
\bibinfo{author}{\bibfnamefont{D.}~\bibnamefont{Becker}} \bibnamefont{and}
  \bibinfo{author}{\bibfnamefont{M.}~\bibnamefont{Reuter}},
  \bibinfo{journal}{Annals Phys.} \textbf{\bibinfo{volume}{350}},
  \bibinfo{pages}{225} (\bibinfo{year}{2014}), \eprint{1404.4537}.

\bibitem[{\citenamefont{Christiansen et~al.}(2016)\citenamefont{Christiansen,
  Knorr, Pawlowski, and Rodigast}}]{Christiansen:2014raa}
\bibinfo{author}{\bibfnamefont{N.}~\bibnamefont{Christiansen}},
  \bibinfo{author}{\bibfnamefont{B.}~\bibnamefont{Knorr}},
  \bibinfo{author}{\bibfnamefont{J.~M.} \bibnamefont{Pawlowski}},
  \bibnamefont{and} \bibinfo{author}{\bibfnamefont{A.}~\bibnamefont{Rodigast}},
  \bibinfo{journal}{Phys. Rev.} \textbf{\bibinfo{volume}{D93}},
  \bibinfo{pages}{044036} (\bibinfo{year}{2016}), \eprint{1403.1232}.

\bibitem[{\citenamefont{Demmel et~al.}(2015)\citenamefont{Demmel, Saueressig,
  and Zanusso}}]{Demmel:2015oqa}
\bibinfo{author}{\bibfnamefont{M.}~\bibnamefont{Demmel}},
  \bibinfo{author}{\bibfnamefont{F.}~\bibnamefont{Saueressig}},
  \bibnamefont{and} \bibinfo{author}{\bibfnamefont{O.}~\bibnamefont{Zanusso}},
  \bibinfo{journal}{JHEP} \textbf{\bibinfo{volume}{08}}, \bibinfo{pages}{113}
  (\bibinfo{year}{2015}), \eprint{1504.07656}.

\bibitem[{\citenamefont{Percacci and Vacca}(2015)}]{Percacci:2015wwa}
\bibinfo{author}{\bibfnamefont{R.}~\bibnamefont{Percacci}} \bibnamefont{and}
  \bibinfo{author}{\bibfnamefont{G.~P.} \bibnamefont{Vacca}},
  \bibinfo{journal}{Eur. Phys. J.} \textbf{\bibinfo{volume}{C75}},
  \bibinfo{pages}{188} (\bibinfo{year}{2015}), \eprint{1501.00888}.

\bibitem[{\citenamefont{Gies et~al.}(2016)\citenamefont{Gies, Knorr, Lippoldt,
  and Saueressig}}]{Gies:2016con}
\bibinfo{author}{\bibfnamefont{H.}~\bibnamefont{Gies}},
  \bibinfo{author}{\bibfnamefont{B.}~\bibnamefont{Knorr}},
  \bibinfo{author}{\bibfnamefont{S.}~\bibnamefont{Lippoldt}}, \bibnamefont{and}
  \bibinfo{author}{\bibfnamefont{F.}~\bibnamefont{Saueressig}},
  \bibinfo{journal}{Phys. Rev. Lett.} \textbf{\bibinfo{volume}{116}},
  \bibinfo{pages}{211302} (\bibinfo{year}{2016}), \eprint{1601.01800}.

\bibitem[{\citenamefont{Biemans
  et~al.}(2017{\natexlab{a}})\citenamefont{Biemans, Platania, and
  Saueressig}}]{Biemans:2016rvp}
\bibinfo{author}{\bibfnamefont{J.}~\bibnamefont{Biemans}},
  \bibinfo{author}{\bibfnamefont{A.}~\bibnamefont{Platania}}, \bibnamefont{and}
  \bibinfo{author}{\bibfnamefont{F.}~\bibnamefont{Saueressig}},
  \bibinfo{journal}{Phys. Rev.} \textbf{\bibinfo{volume}{D95}},
  \bibinfo{pages}{086013} (\bibinfo{year}{2017}{\natexlab{a}}),
  \eprint{1609.04813}.

\bibitem[{\citenamefont{Denz et~al.}(2018)\citenamefont{Denz, Pawlowski, and
  Reichert}}]{Denz:2016qks}
\bibinfo{author}{\bibfnamefont{T.}~\bibnamefont{Denz}},
  \bibinfo{author}{\bibfnamefont{J.~M.} \bibnamefont{Pawlowski}},
  \bibnamefont{and} \bibinfo{author}{\bibfnamefont{M.}~\bibnamefont{Reichert}},
  \bibinfo{journal}{Eur. Phys. J.} \textbf{\bibinfo{volume}{C78}},
  \bibinfo{pages}{336} (\bibinfo{year}{2018}), \eprint{1612.07315}.

\bibitem[{\citenamefont{Christiansen
  et~al.}(2018{\natexlab{a}})\citenamefont{Christiansen, Falls, Pawlowski, and
  Reichert}}]{Christiansen:2017bsy}
\bibinfo{author}{\bibfnamefont{N.}~\bibnamefont{Christiansen}},
  \bibinfo{author}{\bibfnamefont{K.}~\bibnamefont{Falls}},
  \bibinfo{author}{\bibfnamefont{J.~M.} \bibnamefont{Pawlowski}},
  \bibnamefont{and} \bibinfo{author}{\bibfnamefont{M.}~\bibnamefont{Reichert}},
  \bibinfo{journal}{Phys. Rev.} \textbf{\bibinfo{volume}{D97}},
  \bibinfo{pages}{046007} (\bibinfo{year}{2018}{\natexlab{a}}),
  \eprint{1711.09259}.

\bibitem[{\citenamefont{Knorr and Lippoldt}(2017)}]{Knorr:2017fus}
\bibinfo{author}{\bibfnamefont{B.}~\bibnamefont{Knorr}} \bibnamefont{and}
  \bibinfo{author}{\bibfnamefont{S.}~\bibnamefont{Lippoldt}},
  \bibinfo{journal}{Phys. Rev.} \textbf{\bibinfo{volume}{D96}},
  \bibinfo{pages}{065020} (\bibinfo{year}{2017}), \eprint{1707.01397}.

\bibitem[{\citenamefont{Knorr}(2018)}]{Knorr:2017mhu}
\bibinfo{author}{\bibfnamefont{B.}~\bibnamefont{Knorr}},
  \bibinfo{journal}{Class. Quant. Grav.} \textbf{\bibinfo{volume}{35}},
  \bibinfo{pages}{115005} (\bibinfo{year}{2018}), \eprint{1710.07055}.

\bibitem[{\citenamefont{Falls et~al.}(2018)\citenamefont{Falls, King, Litim,
  Nikolakopoulos, and Rahmede}}]{Falls:2017lst}
\bibinfo{author}{\bibfnamefont{K.}~\bibnamefont{Falls}},
  \bibinfo{author}{\bibfnamefont{C.~R.} \bibnamefont{King}},
  \bibinfo{author}{\bibfnamefont{D.~F.} \bibnamefont{Litim}},
  \bibinfo{author}{\bibfnamefont{K.}~\bibnamefont{Nikolakopoulos}},
  \bibnamefont{and} \bibinfo{author}{\bibfnamefont{C.}~\bibnamefont{Rahmede}},
  \bibinfo{journal}{Phys. Rev.} \textbf{\bibinfo{volume}{D97}},
  \bibinfo{pages}{086006} (\bibinfo{year}{2018}), \eprint{1801.00162}.

\bibitem[{\citenamefont{de~Alwis}(2018)}]{deAlwis:2017ysy}
\bibinfo{author}{\bibfnamefont{S.~P.} \bibnamefont{de~Alwis}},
  \bibinfo{journal}{JHEP} \textbf{\bibinfo{volume}{03}}, \bibinfo{pages}{118}
  (\bibinfo{year}{2018}), \eprint{1707.09298}.

\bibitem[{\citenamefont{Falls et~al.}(2019)\citenamefont{Falls, Litim, and
  Schröder}}]{Falls:2018ylp}
\bibinfo{author}{\bibfnamefont{K.~G.} \bibnamefont{Falls}},
  \bibinfo{author}{\bibfnamefont{D.~F.} \bibnamefont{Litim}}, \bibnamefont{and}
  \bibinfo{author}{\bibfnamefont{J.}~\bibnamefont{Schröder}},
  \bibinfo{journal}{Phys. Rev.} \textbf{\bibinfo{volume}{D99}},
  \bibinfo{pages}{126015} (\bibinfo{year}{2019}), \eprint{1810.08550}.

\bibitem[{\citenamefont{Pagani and Reuter}(2019)}]{Pagani:2019vfm}
\bibinfo{author}{\bibfnamefont{C.}~\bibnamefont{Pagani}} \bibnamefont{and}
  \bibinfo{author}{\bibfnamefont{M.}~\bibnamefont{Reuter}}
  (\bibinfo{year}{2019}), \eprint{1906.02507}.

\bibitem[{\citenamefont{Weinberg}(1979)}]{Hawking:1979ig}
\bibinfo{author}{\bibfnamefont{S.}~\bibnamefont{Weinberg}},
  \bibinfo{journal}{{Chap. 16 in General Relativity} ed. by Hawking, S.W. and
  Israel, W.}  (\bibinfo{year}{1979}).

\bibitem[{\citenamefont{Reuter}(1998)}]{Reuter:1996cp}
\bibinfo{author}{\bibfnamefont{M.}~\bibnamefont{Reuter}},
  \bibinfo{journal}{Phys. Rev.} \textbf{\bibinfo{volume}{D57}},
  \bibinfo{pages}{971} (\bibinfo{year}{1998}), \eprint{hep-th/9605030}.

\bibitem[{\citenamefont{Percacci}(2017)}]{Percacci:2017fkn}
\bibinfo{author}{\bibfnamefont{R.}~\bibnamefont{Percacci}},
  \emph{\bibinfo{title}{{An Introduction to Covariant Quantum Gravity and
  Asymptotic Safety}}}, vol.~\bibinfo{volume}{3} of \emph{\bibinfo{series}{100
  Years of General Relativity}} (\bibinfo{publisher}{World Scientific},
  \bibinfo{year}{2017}), ISBN \bibinfo{isbn}{9789813207172, 9789813207196,
  9789813207172, 9789813207196}.

\bibitem[{\citenamefont{Eichhorn}(2017)}]{Eichhorn:2017egq}
\bibinfo{author}{\bibfnamefont{A.}~\bibnamefont{Eichhorn}}, in
  \emph{\bibinfo{booktitle}{{Black Holes, Gravitational Waves and Spacetime
  Singularities Rome, Italy, May 9-12, 2017}}} (\bibinfo{year}{2017}),
  \eprint{1709.03696},
  \urlprefix\url{http://inspirehep.net/record/1623009/files/arXiv:1709.03696.pdf}.

\bibitem[{\citenamefont{Reuter and Saueressig}(2019)}]{Reuter:2019byg}
\bibinfo{author}{\bibfnamefont{M.}~\bibnamefont{Reuter}} \bibnamefont{and}
  \bibinfo{author}{\bibfnamefont{F.}~\bibnamefont{Saueressig}},
  \emph{\bibinfo{title}{{Quantum Gravity and the Functional Renormalization
  Group}}} (\bibinfo{publisher}{Cambridge University Press},
  \bibinfo{year}{2019}), ISBN \bibinfo{isbn}{9781107107328},
  \urlprefix\url{https://www.cambridge.org/academic/subjects/physics/theoretical-physics-and-mathematical-physics/quantum-gravity-and-functional-renormalization-group-road-towards-asymptotic-safety?format=HB&isbn=9781107107328}.

\bibitem[{\citenamefont{Shaposhnikov and
  Wetterich}(2010)}]{Shaposhnikov:2009pv}
\bibinfo{author}{\bibfnamefont{M.}~\bibnamefont{Shaposhnikov}}
  \bibnamefont{and}
  \bibinfo{author}{\bibfnamefont{C.}~\bibnamefont{Wetterich}},
  \bibinfo{journal}{Phys. Lett.} \textbf{\bibinfo{volume}{B683}},
  \bibinfo{pages}{196} (\bibinfo{year}{2010}), \eprint{0912.0208}.

\bibitem[{\citenamefont{Eichhorn and
  Held}(2018{\natexlab{a}})}]{Eichhorn:2017ylw}
\bibinfo{author}{\bibfnamefont{A.}~\bibnamefont{Eichhorn}} \bibnamefont{and}
  \bibinfo{author}{\bibfnamefont{A.}~\bibnamefont{Held}},
  \bibinfo{journal}{Phys. Lett.} \textbf{\bibinfo{volume}{B777}},
  \bibinfo{pages}{217} (\bibinfo{year}{2018}{\natexlab{a}}),
  \eprint{1707.01107}.

\bibitem[{\citenamefont{Eichhorn and Versteegen}(2018)}]{Eichhorn:2017lry}
\bibinfo{author}{\bibfnamefont{A.}~\bibnamefont{Eichhorn}} \bibnamefont{and}
  \bibinfo{author}{\bibfnamefont{F.}~\bibnamefont{Versteegen}},
  \bibinfo{journal}{JHEP} \textbf{\bibinfo{volume}{01}}, \bibinfo{pages}{030}
  (\bibinfo{year}{2018}), \eprint{1709.07252}.

\bibitem[{\citenamefont{Eichhorn and
  Held}(2018{\natexlab{b}})}]{Eichhorn:2018whv}
\bibinfo{author}{\bibfnamefont{A.}~\bibnamefont{Eichhorn}} \bibnamefont{and}
  \bibinfo{author}{\bibfnamefont{A.}~\bibnamefont{Held}},
  \bibinfo{journal}{Phys. Rev. Lett.} \textbf{\bibinfo{volume}{121}},
  \bibinfo{pages}{151302} (\bibinfo{year}{2018}{\natexlab{b}}),
  \eprint{1803.04027}.

\bibitem[{\citenamefont{Eichhorn and Schiffer}(2019)}]{Eichhorn:2019yzm}
\bibinfo{author}{\bibfnamefont{A.}~\bibnamefont{Eichhorn}} \bibnamefont{and}
  \bibinfo{author}{\bibfnamefont{M.}~\bibnamefont{Schiffer}},
  \bibinfo{journal}{Phys. Lett.} \textbf{\bibinfo{volume}{B793}},
  \bibinfo{pages}{383} (\bibinfo{year}{2019}), \eprint{1902.06479}.

\bibitem[{\citenamefont{Niedermaier}(2009)}]{Niedermaier:2009zz}
\bibinfo{author}{\bibfnamefont{M.~R.} \bibnamefont{Niedermaier}},
  \bibinfo{journal}{Phys. Rev. Lett.} \textbf{\bibinfo{volume}{103}},
  \bibinfo{pages}{101303} (\bibinfo{year}{2009}).

\bibitem[{\citenamefont{Falls et~al.}(2016)\citenamefont{Falls, Litim,
  Nikolakopoulos, and Rahmede}}]{Falls:2014tra}
\bibinfo{author}{\bibfnamefont{K.}~\bibnamefont{Falls}},
  \bibinfo{author}{\bibfnamefont{D.~F.} \bibnamefont{Litim}},
  \bibinfo{author}{\bibfnamefont{K.}~\bibnamefont{Nikolakopoulos}},
  \bibnamefont{and} \bibinfo{author}{\bibfnamefont{C.}~\bibnamefont{Rahmede}},
  \bibinfo{journal}{Phys. Rev.} \textbf{\bibinfo{volume}{D93}},
  \bibinfo{pages}{104022} (\bibinfo{year}{2016}), \eprint{1410.4815}.

\bibitem[{\citenamefont{Eichhorn
  et~al.}(2018{\natexlab{a}})\citenamefont{Eichhorn, Labus, Pawlowski, and
  Reichert}}]{Eichhorn:2018akn}
\bibinfo{author}{\bibfnamefont{A.}~\bibnamefont{Eichhorn}},
  \bibinfo{author}{\bibfnamefont{P.}~\bibnamefont{Labus}},
  \bibinfo{author}{\bibfnamefont{J.~M.} \bibnamefont{Pawlowski}},
  \bibnamefont{and} \bibinfo{author}{\bibfnamefont{M.}~\bibnamefont{Reichert}},
  \bibinfo{journal}{SciPost Phys.} \textbf{\bibinfo{volume}{5}},
  \bibinfo{pages}{031} (\bibinfo{year}{2018}{\natexlab{a}}),
  \eprint{1804.00012}.

\bibitem[{\citenamefont{Eichhorn
  et~al.}(2018{\natexlab{b}})\citenamefont{Eichhorn, Lippoldt, Pawlowski,
  Reichert, and Schiffer}}]{Eichhorn:2018ydy}
\bibinfo{author}{\bibfnamefont{A.}~\bibnamefont{Eichhorn}},
  \bibinfo{author}{\bibfnamefont{S.}~\bibnamefont{Lippoldt}},
  \bibinfo{author}{\bibfnamefont{J.~M.} \bibnamefont{Pawlowski}},
  \bibinfo{author}{\bibfnamefont{M.}~\bibnamefont{Reichert}}, \bibnamefont{and}
  \bibinfo{author}{\bibfnamefont{M.}~\bibnamefont{Schiffer}}
  (\bibinfo{year}{2018}{\natexlab{b}}), \eprint{1810.02828}.

\bibitem[{\citenamefont{Eichhorn
  et~al.}(2019{\natexlab{a}})\citenamefont{Eichhorn, Lippoldt, and
  Schiffer}}]{Eichhorn:2018nda}
\bibinfo{author}{\bibfnamefont{A.}~\bibnamefont{Eichhorn}},
  \bibinfo{author}{\bibfnamefont{S.}~\bibnamefont{Lippoldt}}, \bibnamefont{and}
  \bibinfo{author}{\bibfnamefont{M.}~\bibnamefont{Schiffer}},
  \bibinfo{journal}{Phys. Rev.} \textbf{\bibinfo{volume}{D99}},
  \bibinfo{pages}{086002} (\bibinfo{year}{2019}{\natexlab{a}}),
  \eprint{1812.08782}.

\bibitem[{\citenamefont{Becker et~al.}(2017)\citenamefont{Becker, Ripken, and
  Saueressig}}]{Becker:2017tcx}
\bibinfo{author}{\bibfnamefont{D.}~\bibnamefont{Becker}},
  \bibinfo{author}{\bibfnamefont{C.}~\bibnamefont{Ripken}}, \bibnamefont{and}
  \bibinfo{author}{\bibfnamefont{F.}~\bibnamefont{Saueressig}},
  \bibinfo{journal}{JHEP} \textbf{\bibinfo{volume}{12}}, \bibinfo{pages}{121}
  (\bibinfo{year}{2017}), \eprint{1709.09098}.

\bibitem[{\citenamefont{Adeifeoba et~al.}(2018)\citenamefont{Adeifeoba,
  Eichhorn, and Platania}}]{Adeifeoba:2018ydh}
\bibinfo{author}{\bibfnamefont{A.}~\bibnamefont{Adeifeoba}},
  \bibinfo{author}{\bibfnamefont{A.}~\bibnamefont{Eichhorn}}, \bibnamefont{and}
  \bibinfo{author}{\bibfnamefont{A.}~\bibnamefont{Platania}},
  \bibinfo{journal}{Class. Quant. Grav.} \textbf{\bibinfo{volume}{35}},
  \bibinfo{pages}{225007} (\bibinfo{year}{2018}), \eprint{1808.03472}.

\bibitem[{\citenamefont{Platania}(2019)}]{Platania:2019kyx}
\bibinfo{author}{\bibfnamefont{A.}~\bibnamefont{Platania}},
  \bibinfo{journal}{Eur. Phys. J.} \textbf{\bibinfo{volume}{C79}},
  \bibinfo{pages}{470} (\bibinfo{year}{2019}), \eprint{1903.10411}.

\bibitem[{\citenamefont{Bosma et~al.}(2019)\citenamefont{Bosma, Knorr, and
  Saueressig}}]{Bosma:2019aiu}
\bibinfo{author}{\bibfnamefont{L.}~\bibnamefont{Bosma}},
  \bibinfo{author}{\bibfnamefont{B.}~\bibnamefont{Knorr}}, \bibnamefont{and}
  \bibinfo{author}{\bibfnamefont{F.}~\bibnamefont{Saueressig}}
  (\bibinfo{year}{2019}), \eprint{1904.04845}.

\bibitem[{\citenamefont{Knorr et~al.}(2019)\citenamefont{Knorr, Ripken, and
  Saueressig}}]{Knorr:2019atm}
\bibinfo{author}{\bibfnamefont{B.}~\bibnamefont{Knorr}},
  \bibinfo{author}{\bibfnamefont{C.}~\bibnamefont{Ripken}}, \bibnamefont{and}
  \bibinfo{author}{\bibfnamefont{F.}~\bibnamefont{Saueressig}}
  (\bibinfo{year}{2019}), \eprint{1907.02903}.

\bibitem[{\citenamefont{Morris}(2016)}]{Morris:2016spn}
\bibinfo{author}{\bibfnamefont{T.~R.} \bibnamefont{Morris}},
  \bibinfo{journal}{JHEP} \textbf{\bibinfo{volume}{11}}, \bibinfo{pages}{160}
  (\bibinfo{year}{2016}), \eprint{1610.03081}.

\bibitem[{\citenamefont{Percacci and Vacca}(2017)}]{Percacci:2016arh}
\bibinfo{author}{\bibfnamefont{R.}~\bibnamefont{Percacci}} \bibnamefont{and}
  \bibinfo{author}{\bibfnamefont{G.~P.} \bibnamefont{Vacca}},
  \bibinfo{journal}{Eur. Phys. J.} \textbf{\bibinfo{volume}{C77}},
  \bibinfo{pages}{52} (\bibinfo{year}{2017}), \eprint{1611.07005}.

\bibitem[{\citenamefont{Ohta}(2017)}]{Ohta:2017dsq}
\bibinfo{author}{\bibfnamefont{N.}~\bibnamefont{Ohta}}, \bibinfo{journal}{PTEP}
  \textbf{\bibinfo{volume}{2017}}, \bibinfo{pages}{033E02}
  (\bibinfo{year}{2017}), \eprint{1701.01506}.

\bibitem[{\citenamefont{Eichhorn}(2019{\natexlab{b}})}]{Eichhorn:2019xav}
\bibinfo{author}{\bibfnamefont{A.}~\bibnamefont{Eichhorn}}, in
  \emph{\bibinfo{booktitle}{{9th International Conference: Spacetime - Matter -
  Quantum Mechanics: From discrete structures and dynamics to top-down
  causation (DICE2018) Castiglioncello , Tuscany , Italy, September 17-21,
  2018}}} (\bibinfo{year}{2019}{\natexlab{b}}), \eprint{1902.00391}.

\bibitem[{\citenamefont{Eichhorn
  et~al.}(2019{\natexlab{b}})\citenamefont{Eichhorn, Koslowski, and
  Pereira}}]{Eichhorn:2018phj}
\bibinfo{author}{\bibfnamefont{A.}~\bibnamefont{Eichhorn}},
  \bibinfo{author}{\bibfnamefont{T.}~\bibnamefont{Koslowski}},
  \bibnamefont{and} \bibinfo{author}{\bibfnamefont{A.~D.}
  \bibnamefont{Pereira}}, \bibinfo{journal}{Universe}
  \textbf{\bibinfo{volume}{5}}, \bibinfo{pages}{53}
  (\bibinfo{year}{2019}{\natexlab{b}}), \eprint{1811.12909}.

\bibitem[{\citenamefont{de~Alwis et~al.}(2019)\citenamefont{de~Alwis, Eichhorn,
  Held, Pawlowski, Schiffer, and Versteegen}}]{deAlwis:2019aud}
\bibinfo{author}{\bibfnamefont{S.}~\bibnamefont{de~Alwis}},
  \bibinfo{author}{\bibfnamefont{A.}~\bibnamefont{Eichhorn}},
  \bibinfo{author}{\bibfnamefont{A.}~\bibnamefont{Held}},
  \bibinfo{author}{\bibfnamefont{J.~M.} \bibnamefont{Pawlowski}},
  \bibinfo{author}{\bibfnamefont{M.}~\bibnamefont{Schiffer}}, \bibnamefont{and}
  \bibinfo{author}{\bibfnamefont{F.}~\bibnamefont{Versteegen}}
  (\bibinfo{year}{2019}), \eprint{1907.07894}.

\bibitem[{\citenamefont{Harst and Reuter}(2011)}]{Harst:2011zx}
\bibinfo{author}{\bibfnamefont{U.}~\bibnamefont{Harst}} \bibnamefont{and}
  \bibinfo{author}{\bibfnamefont{M.}~\bibnamefont{Reuter}},
  \bibinfo{journal}{JHEP} \textbf{\bibinfo{volume}{05}}, \bibinfo{pages}{119}
  (\bibinfo{year}{2011}), \eprint{1101.6007}.

\bibitem[{\citenamefont{Eichhorn and Gies}(2011)}]{Eichhorn:2011pc}
\bibinfo{author}{\bibfnamefont{A.}~\bibnamefont{Eichhorn}} \bibnamefont{and}
  \bibinfo{author}{\bibfnamefont{H.}~\bibnamefont{Gies}}, \bibinfo{journal}{New
  J. Phys.} \textbf{\bibinfo{volume}{13}}, \bibinfo{pages}{125012}
  (\bibinfo{year}{2011}), \eprint{1104.5366}.

\bibitem[{\citenamefont{Dona et~al.}(2014)\citenamefont{Dona, Eichhorn, and
  Percacci}}]{Dona:2013qba}
\bibinfo{author}{\bibfnamefont{P.}~\bibnamefont{Dona}},
  \bibinfo{author}{\bibfnamefont{A.}~\bibnamefont{Eichhorn}}, \bibnamefont{and}
  \bibinfo{author}{\bibfnamefont{R.}~\bibnamefont{Percacci}},
  \bibinfo{journal}{Phys. Rev.} \textbf{\bibinfo{volume}{D89}},
  \bibinfo{pages}{084035} (\bibinfo{year}{2014}), \eprint{1311.2898}.

\bibitem[{\citenamefont{Meibohm et~al.}(2016)\citenamefont{Meibohm, Pawlowski,
  and Reichert}}]{Meibohm:2015twa}
\bibinfo{author}{\bibfnamefont{J.}~\bibnamefont{Meibohm}},
  \bibinfo{author}{\bibfnamefont{J.~M.} \bibnamefont{Pawlowski}},
  \bibnamefont{and} \bibinfo{author}{\bibfnamefont{M.}~\bibnamefont{Reichert}},
  \bibinfo{journal}{Phys. Rev.} \textbf{\bibinfo{volume}{D93}},
  \bibinfo{pages}{084035} (\bibinfo{year}{2016}), \eprint{1510.07018}.

\bibitem[{\citenamefont{Oda and Yamada}(2016)}]{Oda:2015sma}
\bibinfo{author}{\bibfnamefont{K.-y.} \bibnamefont{Oda}} \bibnamefont{and}
  \bibinfo{author}{\bibfnamefont{M.}~\bibnamefont{Yamada}},
  \bibinfo{journal}{Class. Quant. Grav.} \textbf{\bibinfo{volume}{33}},
  \bibinfo{pages}{125011} (\bibinfo{year}{2016}), \eprint{1510.03734}.

\bibitem[{\citenamefont{Meibohm and Pawlowski}(2016)}]{Meibohm:2016mkp}
\bibinfo{author}{\bibfnamefont{J.}~\bibnamefont{Meibohm}} \bibnamefont{and}
  \bibinfo{author}{\bibfnamefont{J.~M.} \bibnamefont{Pawlowski}},
  \bibinfo{journal}{Eur. Phys. J.} \textbf{\bibinfo{volume}{C76}},
  \bibinfo{pages}{285} (\bibinfo{year}{2016}), \eprint{1601.04597}.

\bibitem[{\citenamefont{Biemans
  et~al.}(2017{\natexlab{b}})\citenamefont{Biemans, Platania, and
  Saueressig}}]{Biemans:2017zca}
\bibinfo{author}{\bibfnamefont{J.}~\bibnamefont{Biemans}},
  \bibinfo{author}{\bibfnamefont{A.}~\bibnamefont{Platania}}, \bibnamefont{and}
  \bibinfo{author}{\bibfnamefont{F.}~\bibnamefont{Saueressig}},
  \bibinfo{journal}{JHEP} \textbf{\bibinfo{volume}{05}}, \bibinfo{pages}{093}
  (\bibinfo{year}{2017}{\natexlab{b}}), \eprint{1702.06539}.

\bibitem[{\citenamefont{Eichhorn and Held}(2017)}]{Eichhorn:2017eht}
\bibinfo{author}{\bibfnamefont{A.}~\bibnamefont{Eichhorn}} \bibnamefont{and}
  \bibinfo{author}{\bibfnamefont{A.}~\bibnamefont{Held}},
  \bibinfo{journal}{Phys. Rev.} \textbf{\bibinfo{volume}{D96}},
  \bibinfo{pages}{086025} (\bibinfo{year}{2017}), \eprint{1705.02342}.

\bibitem[{\citenamefont{Eichhorn
  et~al.}(2018{\natexlab{c}})\citenamefont{Eichhorn, Held, and
  Wetterich}}]{Eichhorn:2017muy}
\bibinfo{author}{\bibfnamefont{A.}~\bibnamefont{Eichhorn}},
  \bibinfo{author}{\bibfnamefont{A.}~\bibnamefont{Held}}, \bibnamefont{and}
  \bibinfo{author}{\bibfnamefont{C.}~\bibnamefont{Wetterich}},
  \bibinfo{journal}{Phys. Lett.} \textbf{\bibinfo{volume}{B782}},
  \bibinfo{pages}{198} (\bibinfo{year}{2018}{\natexlab{c}}),
  \eprint{1711.02949}.

\bibitem[{\citenamefont{Eichhorn
  et~al.}(2018{\natexlab{d}})\citenamefont{Eichhorn, Hamada, Lumma, and
  Yamada}}]{Eichhorn:2017als}
\bibinfo{author}{\bibfnamefont{A.}~\bibnamefont{Eichhorn}},
  \bibinfo{author}{\bibfnamefont{Y.}~\bibnamefont{Hamada}},
  \bibinfo{author}{\bibfnamefont{J.}~\bibnamefont{Lumma}}, \bibnamefont{and}
  \bibinfo{author}{\bibfnamefont{M.}~\bibnamefont{Yamada}},
  \bibinfo{journal}{Phys. Rev.} \textbf{\bibinfo{volume}{D97}},
  \bibinfo{pages}{086004} (\bibinfo{year}{2018}{\natexlab{d}}),
  \eprint{1712.00319}.

\bibitem[{\citenamefont{Hamada and Yamada}(2017)}]{Hamada:2017rvn}
\bibinfo{author}{\bibfnamefont{Y.}~\bibnamefont{Hamada}} \bibnamefont{and}
  \bibinfo{author}{\bibfnamefont{M.}~\bibnamefont{Yamada}},
  \bibinfo{journal}{JHEP} \textbf{\bibinfo{volume}{08}}, \bibinfo{pages}{070}
  (\bibinfo{year}{2017}), \eprint{1703.09033}.

\bibitem[{\citenamefont{Christiansen
  et~al.}(2018{\natexlab{b}})\citenamefont{Christiansen, Litim, Pawlowski, and
  Reichert}}]{Christiansen:2017cxa}
\bibinfo{author}{\bibfnamefont{N.}~\bibnamefont{Christiansen}},
  \bibinfo{author}{\bibfnamefont{D.~F.} \bibnamefont{Litim}},
  \bibinfo{author}{\bibfnamefont{J.~M.} \bibnamefont{Pawlowski}},
  \bibnamefont{and} \bibinfo{author}{\bibfnamefont{M.}~\bibnamefont{Reichert}},
  \bibinfo{journal}{Phys. Rev.} \textbf{\bibinfo{volume}{D97}},
  \bibinfo{pages}{106012} (\bibinfo{year}{2018}{\natexlab{b}}),
  \eprint{1710.04669}.

\bibitem[{\citenamefont{Gies and Martini}(2018)}]{Gies:2018jnv}
\bibinfo{author}{\bibfnamefont{H.}~\bibnamefont{Gies}} \bibnamefont{and}
  \bibinfo{author}{\bibfnamefont{R.}~\bibnamefont{Martini}},
  \bibinfo{journal}{Phys. Rev.} \textbf{\bibinfo{volume}{D97}},
  \bibinfo{pages}{085017} (\bibinfo{year}{2018}), \eprint{1802.02865}.

\bibitem[{\citenamefont{Bonanno et~al.}(2018)\citenamefont{Bonanno, Platania,
  and Saueressig}}]{Bonanno:2018gck}
\bibinfo{author}{\bibfnamefont{A.}~\bibnamefont{Bonanno}},
  \bibinfo{author}{\bibfnamefont{A.}~\bibnamefont{Platania}}, \bibnamefont{and}
  \bibinfo{author}{\bibfnamefont{F.}~\bibnamefont{Saueressig}}
  (\bibinfo{year}{2018}), \eprint{1803.02355}.

\bibitem[{\citenamefont{Pawlowski et~al.}(2019)\citenamefont{Pawlowski,
  Reichert, Wetterich, and Yamada}}]{Pawlowski:2018ixd}
\bibinfo{author}{\bibfnamefont{J.~M.} \bibnamefont{Pawlowski}},
  \bibinfo{author}{\bibfnamefont{M.}~\bibnamefont{Reichert}},
  \bibinfo{author}{\bibfnamefont{C.}~\bibnamefont{Wetterich}},
  \bibnamefont{and} \bibinfo{author}{\bibfnamefont{M.}~\bibnamefont{Yamada}},
  \bibinfo{journal}{Phys. Rev.} \textbf{\bibinfo{volume}{D99}},
  \bibinfo{pages}{086010} (\bibinfo{year}{2019}), \eprint{1811.11706}.

\bibitem[{\citenamefont{De~Brito et~al.}(2019)\citenamefont{De~Brito, Hamada,
  Pereira, and Yamada}}]{deBrito:2019epw}
\bibinfo{author}{\bibfnamefont{G.~P.} \bibnamefont{De~Brito}},
  \bibinfo{author}{\bibfnamefont{Y.}~\bibnamefont{Hamada}},
  \bibinfo{author}{\bibfnamefont{A.~D.} \bibnamefont{Pereira}},
  \bibnamefont{and} \bibinfo{author}{\bibfnamefont{M.}~\bibnamefont{Yamada}}
  (\bibinfo{year}{2019}), \eprint{1905.11114}.

\bibitem[{\citenamefont{Wetterich and Yamada}(2019)}]{Wetterich:2019zdo}
\bibinfo{author}{\bibfnamefont{C.}~\bibnamefont{Wetterich}} \bibnamefont{and}
  \bibinfo{author}{\bibfnamefont{M.}~\bibnamefont{Yamada}}
  (\bibinfo{year}{2019}), \eprint{1906.01721}.

\bibitem[{\citenamefont{van~der Bij et~al.}(1982)\citenamefont{van~der Bij, van
  Dam, and Ng}}]{vanderBij:1981ym}
\bibinfo{author}{\bibfnamefont{J.~J.} \bibnamefont{van~der Bij}},
  \bibinfo{author}{\bibfnamefont{H.}~\bibnamefont{van Dam}}, \bibnamefont{and}
  \bibinfo{author}{\bibfnamefont{Y.~J.} \bibnamefont{Ng}},
  \bibinfo{journal}{Physica} \textbf{\bibinfo{volume}{116A}},
  \bibinfo{pages}{307} (\bibinfo{year}{1982}).

\bibitem[{\citenamefont{Herrero-Valea}(2018)}]{Herrero-Valea:2018ilg}
\bibinfo{author}{\bibfnamefont{M.}~\bibnamefont{Herrero-Valea}},
  \bibinfo{journal}{JHEP} \textbf{\bibinfo{volume}{12}}, \bibinfo{pages}{106}
  (\bibinfo{year}{2018}), \eprint{1806.01869}.

\bibitem[{\citenamefont{Hawking et~al.}(1976)\citenamefont{Hawking, King, and
  Mccarthy}}]{Hawking:1976fe}
\bibinfo{author}{\bibfnamefont{S.~W.} \bibnamefont{Hawking}},
  \bibinfo{author}{\bibfnamefont{A.~R.} \bibnamefont{King}}, \bibnamefont{and}
  \bibinfo{author}{\bibfnamefont{P.~J.} \bibnamefont{Mccarthy}},
  \bibinfo{journal}{J. Math. Phys.} \textbf{\bibinfo{volume}{17}},
  \bibinfo{pages}{174} (\bibinfo{year}{1976}).

\bibitem[{\citenamefont{Malament}(1977)}]{Malament:1977}
\bibinfo{author}{\bibfnamefont{D.~B.} \bibnamefont{Malament}},
  \bibinfo{journal}{J. Math. Phys.} \textbf{\bibinfo{volume}{18}},
  \bibinfo{pages}{1399} (\bibinfo{year}{1977}).

\bibitem[{\citenamefont{Feldbrugge et~al.}(2017)\citenamefont{Feldbrugge,
  Lehners, and Turok}}]{Feldbrugge:2017fcc}
\bibinfo{author}{\bibfnamefont{J.}~\bibnamefont{Feldbrugge}},
  \bibinfo{author}{\bibfnamefont{J.-L.} \bibnamefont{Lehners}},
  \bibnamefont{and} \bibinfo{author}{\bibfnamefont{N.}~\bibnamefont{Turok}},
  \bibinfo{journal}{Phys. Rev. Lett.} \textbf{\bibinfo{volume}{119}},
  \bibinfo{pages}{171301} (\bibinfo{year}{2017}), \eprint{1705.00192}.

\bibitem[{\citenamefont{Henneaux and Teitelboim}(1989)}]{Henneaux:1989zc}
\bibinfo{author}{\bibfnamefont{M.}~\bibnamefont{Henneaux}} \bibnamefont{and}
  \bibinfo{author}{\bibfnamefont{C.}~\bibnamefont{Teitelboim}},
  \bibinfo{journal}{Phys. Lett.} \textbf{\bibinfo{volume}{B222}},
  \bibinfo{pages}{195} (\bibinfo{year}{1989}).

\bibitem[{\citenamefont{Smolin}(2009)}]{Smolin:2009ti}
\bibinfo{author}{\bibfnamefont{L.}~\bibnamefont{Smolin}},
  \bibinfo{journal}{Phys. Rev.} \textbf{\bibinfo{volume}{D80}},
  \bibinfo{pages}{084003} (\bibinfo{year}{2009}), \eprint{0904.4841}.

\bibitem[{\citenamefont{Weinberg}(1989)}]{Weinberg:1988cp}
\bibinfo{author}{\bibfnamefont{S.}~\bibnamefont{Weinberg}},
  \bibinfo{journal}{Rev. Mod. Phys.} \textbf{\bibinfo{volume}{61}},
  \bibinfo{pages}{1} (\bibinfo{year}{1989}), \bibinfo{note}{[,569(1988)]}.

\bibitem[{\citenamefont{Finkelstein et~al.}(2001)\citenamefont{Finkelstein,
  Galiautdinov, and Baugh}}]{Finkelstein:2000pg}
\bibinfo{author}{\bibfnamefont{D.~R.} \bibnamefont{Finkelstein}},
  \bibinfo{author}{\bibfnamefont{A.~A.} \bibnamefont{Galiautdinov}},
  \bibnamefont{and} \bibinfo{author}{\bibfnamefont{J.~E.} \bibnamefont{Baugh}},
  \bibinfo{journal}{J. Math. Phys.} \textbf{\bibinfo{volume}{42}},
  \bibinfo{pages}{340} (\bibinfo{year}{2001}), \eprint{gr-qc/0009099}.

\bibitem[{\citenamefont{Alvarez et~al.}(2015)\citenamefont{Alvarez,
  Gonzalez-Martin, Herrero-Valea, and Martin}}]{Alvarez:2015sba}
\bibinfo{author}{\bibfnamefont{E.}~\bibnamefont{Alvarez}},
  \bibinfo{author}{\bibfnamefont{S.}~\bibnamefont{Gonzalez-Martin}},
  \bibinfo{author}{\bibfnamefont{M.}~\bibnamefont{Herrero-Valea}},
  \bibnamefont{and} \bibinfo{author}{\bibfnamefont{C.~P.}
  \bibnamefont{Martin}}, \bibinfo{journal}{JHEP} \textbf{\bibinfo{volume}{08}},
  \bibinfo{pages}{078} (\bibinfo{year}{2015}), \eprint{1505.01995}.

\bibitem[{\citenamefont{Torres}(2017)}]{Torres:2017ygl}
\bibinfo{author}{\bibfnamefont{R.}~\bibnamefont{Torres}},
  \bibinfo{journal}{Phys. Rev.} \textbf{\bibinfo{volume}{D95}},
  \bibinfo{pages}{124004} (\bibinfo{year}{2017}), \eprint{1703.09997}.

\bibitem[{\citenamefont{Ellis et~al.}(2011)\citenamefont{Ellis, van Elst,
  Murugan, and Uzan}}]{Ellis:2010uc}
\bibinfo{author}{\bibfnamefont{G.~F.~R.} \bibnamefont{Ellis}},
  \bibinfo{author}{\bibfnamefont{H.}~\bibnamefont{van Elst}},
  \bibinfo{author}{\bibfnamefont{J.}~\bibnamefont{Murugan}}, \bibnamefont{and}
  \bibinfo{author}{\bibfnamefont{J.-P.} \bibnamefont{Uzan}},
  \bibinfo{journal}{Class. Quant. Grav.} \textbf{\bibinfo{volume}{28}},
  \bibinfo{pages}{225007} (\bibinfo{year}{2011}), \eprint{1008.1196}.

\bibitem[{\citenamefont{Ellis}(2014)}]{Ellis:2013eqs}
\bibinfo{author}{\bibfnamefont{G.~F.~R.} \bibnamefont{Ellis}},
  \bibinfo{journal}{Gen. Rel. Grav.} \textbf{\bibinfo{volume}{46}},
  \bibinfo{pages}{1619} (\bibinfo{year}{2014}), \eprint{1306.3021}.

\bibitem[{\citenamefont{Eichhorn}(2015)}]{Eichhorn:2015bna}
\bibinfo{author}{\bibfnamefont{A.}~\bibnamefont{Eichhorn}},
  \bibinfo{journal}{JHEP} \textbf{\bibinfo{volume}{04}}, \bibinfo{pages}{096}
  (\bibinfo{year}{2015}), \eprint{1501.05848}.

\bibitem[{\citenamefont{Nojiri et~al.}(2016)\citenamefont{Nojiri, Odintsov, and
  Oikonomou}}]{Nojiri:2015sfd}
\bibinfo{author}{\bibfnamefont{S.}~\bibnamefont{Nojiri}},
  \bibinfo{author}{\bibfnamefont{S.~D.} \bibnamefont{Odintsov}},
  \bibnamefont{and} \bibinfo{author}{\bibfnamefont{V.~K.}
  \bibnamefont{Oikonomou}}, \bibinfo{journal}{JCAP}
  \textbf{\bibinfo{volume}{1605}}, \bibinfo{pages}{046} (\bibinfo{year}{2016}),
  \eprint{1512.07223}.

\bibitem[{\citenamefont{Saez-Gomez}(2016)}]{Saez-Gomez:2016gum}
\bibinfo{author}{\bibfnamefont{D.}~\bibnamefont{Saez-Gomez}},
  \bibinfo{journal}{Phys. Rev.} \textbf{\bibinfo{volume}{D93}},
  \bibinfo{pages}{124040} (\bibinfo{year}{2016}), \eprint{1602.04771}.

\bibitem[{\citenamefont{Alvarez}(2005)}]{Alvarez:2005iy}
\bibinfo{author}{\bibfnamefont{E.}~\bibnamefont{Alvarez}},
  \bibinfo{journal}{JHEP} \textbf{\bibinfo{volume}{03}}, \bibinfo{pages}{002}
  (\bibinfo{year}{2005}), \eprint{hep-th/0501146}.

\bibitem[{\citenamefont{de~Leon~Ardon et~al.}(2018)\citenamefont{de~Leon~Ardon,
  Ohta, and Percacci}}]{Ardon:2017atk}
\bibinfo{author}{\bibfnamefont{R.}~\bibnamefont{de~Leon~Ardon}},
  \bibinfo{author}{\bibfnamefont{N.}~\bibnamefont{Ohta}}, \bibnamefont{and}
  \bibinfo{author}{\bibfnamefont{R.}~\bibnamefont{Percacci}},
  \bibinfo{journal}{Phys. Rev.} \textbf{\bibinfo{volume}{D97}},
  \bibinfo{pages}{026007} (\bibinfo{year}{2018}), \eprint{1710.02457}.

\bibitem[{\citenamefont{Eichhorn}(2013{\natexlab{a}})}]{Eichhorn:2013xr}
\bibinfo{author}{\bibfnamefont{A.}~\bibnamefont{Eichhorn}},
  \bibinfo{journal}{Class. Quant. Grav.} \textbf{\bibinfo{volume}{30}},
  \bibinfo{pages}{115016} (\bibinfo{year}{2013}{\natexlab{a}}),
  \eprint{1301.0879}.

\bibitem[{\citenamefont{Benedetti}(2016)}]{Benedetti:2015zsw}
\bibinfo{author}{\bibfnamefont{D.}~\bibnamefont{Benedetti}},
  \bibinfo{journal}{Gen. Rel. Grav.} \textbf{\bibinfo{volume}{48}},
  \bibinfo{pages}{68} (\bibinfo{year}{2016}), \eprint{1511.06560}.

\bibitem[{\citenamefont{Scholz}(2018)}]{Scholz:2017pfo}
\bibinfo{author}{\bibfnamefont{E.}~\bibnamefont{Scholz}},
  \bibinfo{journal}{Einstein Stud.} \textbf{\bibinfo{volume}{14}},
  \bibinfo{pages}{261} (\bibinfo{year}{2018}), \eprint{1703.03187}.

\bibitem[{\citenamefont{Percacci}(2011)}]{Percacci:2011uf}
\bibinfo{author}{\bibfnamefont{R.}~\bibnamefont{Percacci}},
  \bibinfo{journal}{New J. Phys.} \textbf{\bibinfo{volume}{13}},
  \bibinfo{pages}{125013} (\bibinfo{year}{2011}), \eprint{1110.6758}.

\bibitem[{\citenamefont{Codello et~al.}(2013)\citenamefont{Codello, D'Odorico,
  Pagani, and Percacci}}]{Codello:2012sn}
\bibinfo{author}{\bibfnamefont{A.}~\bibnamefont{Codello}},
  \bibinfo{author}{\bibfnamefont{G.}~\bibnamefont{D'Odorico}},
  \bibinfo{author}{\bibfnamefont{C.}~\bibnamefont{Pagani}}, \bibnamefont{and}
  \bibinfo{author}{\bibfnamefont{R.}~\bibnamefont{Percacci}},
  \bibinfo{journal}{Class. Quant. Grav.} \textbf{\bibinfo{volume}{30}},
  \bibinfo{pages}{115015} (\bibinfo{year}{2013}), \eprint{1210.3284}.

\bibitem[{\citenamefont{Pagani and Percacci}(2014)}]{Pagani:2013fca}
\bibinfo{author}{\bibfnamefont{C.}~\bibnamefont{Pagani}} \bibnamefont{and}
  \bibinfo{author}{\bibfnamefont{R.}~\bibnamefont{Percacci}},
  \bibinfo{journal}{Class. Quant. Grav.} \textbf{\bibinfo{volume}{31}},
  \bibinfo{pages}{115005} (\bibinfo{year}{2014}), \eprint{1312.7767}.

\bibitem[{\citenamefont{Shaposhnikov and
  Shkerin}(2018{\natexlab{a}})}]{Shaposhnikov:2018jag}
\bibinfo{author}{\bibfnamefont{M.}~\bibnamefont{Shaposhnikov}}
  \bibnamefont{and} \bibinfo{author}{\bibfnamefont{A.}~\bibnamefont{Shkerin}},
  \bibinfo{journal}{JHEP} \textbf{\bibinfo{volume}{10}}, \bibinfo{pages}{024}
  (\bibinfo{year}{2018}{\natexlab{a}}), \eprint{1804.06376}.

\bibitem[{\citenamefont{Shaposhnikov and
  Shkerin}(2018{\natexlab{b}})}]{Shaposhnikov:2018xkv}
\bibinfo{author}{\bibfnamefont{M.}~\bibnamefont{Shaposhnikov}}
  \bibnamefont{and} \bibinfo{author}{\bibfnamefont{A.}~\bibnamefont{Shkerin}},
  \bibinfo{journal}{Phys. Lett.} \textbf{\bibinfo{volume}{B783}},
  \bibinfo{pages}{253} (\bibinfo{year}{2018}{\natexlab{b}}),
  \eprint{1803.08907}.

\bibitem[{\citenamefont{Mooij et~al.}(2019)\citenamefont{Mooij, Shaposhnikov,
  and Voumard}}]{Mooij:2018hew}
\bibinfo{author}{\bibfnamefont{S.}~\bibnamefont{Mooij}},
  \bibinfo{author}{\bibfnamefont{M.}~\bibnamefont{Shaposhnikov}},
  \bibnamefont{and} \bibinfo{author}{\bibfnamefont{T.}~\bibnamefont{Voumard}},
  \bibinfo{journal}{Phys. Rev.} \textbf{\bibinfo{volume}{D99}},
  \bibinfo{pages}{085013} (\bibinfo{year}{2019}), \eprint{1812.07946}.

\bibitem[{\citenamefont{Shaposhnikov and Shimada}(2019)}]{Shaposhnikov:2018nnm}
\bibinfo{author}{\bibfnamefont{M.}~\bibnamefont{Shaposhnikov}}
  \bibnamefont{and} \bibinfo{author}{\bibfnamefont{K.}~\bibnamefont{Shimada}},
  \bibinfo{journal}{Phys. Rev.} \textbf{\bibinfo{volume}{D99}},
  \bibinfo{pages}{103528} (\bibinfo{year}{2019}), \eprint{1812.08706}.

\bibitem[{\citenamefont{'t~Hooft}(2011{\natexlab{a}})}]{tHooft:2009wdx}
\bibinfo{author}{\bibfnamefont{G.}~\bibnamefont{'t~Hooft}},
  \bibinfo{journal}{Subnucl. Ser.} \textbf{\bibinfo{volume}{47}},
  \bibinfo{pages}{251} (\bibinfo{year}{2011}{\natexlab{a}}),
  \eprint{0909.3426}.

\bibitem[{\citenamefont{'t~Hooft}(2011{\natexlab{b}})}]{tHooft:2011aa}
\bibinfo{author}{\bibfnamefont{G.}~\bibnamefont{'t~Hooft}},
  \bibinfo{journal}{Found. Phys.} \textbf{\bibinfo{volume}{41}},
  \bibinfo{pages}{1829} (\bibinfo{year}{2011}{\natexlab{b}}),
  \eprint{1104.4543}.

\bibitem[{\citenamefont{'t~Hooft}(2014)}]{Hooft:2014daa}
\bibinfo{author}{\bibfnamefont{G.}~\bibnamefont{'t~Hooft}}
  (\bibinfo{year}{2014}), \eprint{1410.6675}.

\bibitem[{\citenamefont{Smolin}(1979)}]{Smolin:1979uz}
\bibinfo{author}{\bibfnamefont{L.}~\bibnamefont{Smolin}},
  \bibinfo{journal}{Nucl. Phys.} \textbf{\bibinfo{volume}{B160}},
  \bibinfo{pages}{253} (\bibinfo{year}{1979}).

\bibitem[{\citenamefont{Cheng}(1988)}]{Cheng:1988zx}
\bibinfo{author}{\bibfnamefont{H.}~\bibnamefont{Cheng}},
  \bibinfo{journal}{Phys. Rev. Lett.} \textbf{\bibinfo{volume}{61}},
  \bibinfo{pages}{2182} (\bibinfo{year}{1988}).

\bibitem[{\citenamefont{Ghilencea}(2019)}]{Ghilencea:2018dqd}
\bibinfo{author}{\bibfnamefont{D.~M.} \bibnamefont{Ghilencea}},
  \bibinfo{journal}{JHEP} \textbf{\bibinfo{volume}{03}}, \bibinfo{pages}{049}
  (\bibinfo{year}{2019}), \eprint{1812.08613}.

\bibitem[{\citenamefont{Ferreira et~al.}(2018)\citenamefont{Ferreira, Hill, and
  Ross}}]{Ferreira:2018itt}
\bibinfo{author}{\bibfnamefont{P.~G.} \bibnamefont{Ferreira}},
  \bibinfo{author}{\bibfnamefont{C.~T.} \bibnamefont{Hill}}, \bibnamefont{and}
  \bibinfo{author}{\bibfnamefont{G.~G.} \bibnamefont{Ross}},
  \bibinfo{journal}{Phys. Rev.} \textbf{\bibinfo{volume}{D98}},
  \bibinfo{pages}{116012} (\bibinfo{year}{2018}), \eprint{1801.07676}.

\bibitem[{\citenamefont{Stelle}(1977)}]{Stelle:1976gc}
\bibinfo{author}{\bibfnamefont{K.~S.} \bibnamefont{Stelle}},
  \bibinfo{journal}{Phys. Rev.} \textbf{\bibinfo{volume}{D16}},
  \bibinfo{pages}{953} (\bibinfo{year}{1977}).

\bibitem[{\citenamefont{Fradkin and Tseytlin}(1982)}]{Fradkin:1981iu}
\bibinfo{author}{\bibfnamefont{E.~S.} \bibnamefont{Fradkin}} \bibnamefont{and}
  \bibinfo{author}{\bibfnamefont{A.~A.} \bibnamefont{Tseytlin}},
  \bibinfo{journal}{Nucl. Phys.} \textbf{\bibinfo{volume}{B201}},
  \bibinfo{pages}{469} (\bibinfo{year}{1982}).

\bibitem[{\citenamefont{Lee and van Nieuwenhuizen}(1982)}]{Lee:1982cp}
\bibinfo{author}{\bibfnamefont{S.~C.} \bibnamefont{Lee}} \bibnamefont{and}
  \bibinfo{author}{\bibfnamefont{P.}~\bibnamefont{van Nieuwenhuizen}},
  \bibinfo{journal}{Phys. Rev.} \textbf{\bibinfo{volume}{D26}},
  \bibinfo{pages}{934} (\bibinfo{year}{1982}).

\bibitem[{\citenamefont{Riegert}(1984)}]{Riegert:1984hf}
\bibinfo{author}{\bibfnamefont{R.~J.} \bibnamefont{Riegert}},
  \bibinfo{journal}{Phys. Lett.} \textbf{\bibinfo{volume}{A105}},
  \bibinfo{pages}{110} (\bibinfo{year}{1984}).

\bibitem[{\citenamefont{Holdom and Ren}(2016{\natexlab{a}})}]{Holdom:2016xfn}
\bibinfo{author}{\bibfnamefont{B.}~\bibnamefont{Holdom}} \bibnamefont{and}
  \bibinfo{author}{\bibfnamefont{J.}~\bibnamefont{Ren}}, \bibinfo{journal}{Int.
  J. Mod. Phys.} \textbf{\bibinfo{volume}{D25}}, \bibinfo{pages}{1643004}
  (\bibinfo{year}{2016}{\natexlab{a}}), \eprint{1605.05006}.

\bibitem[{\citenamefont{Donoghue}(2017)}]{Donoghue:2017fvm}
\bibinfo{author}{\bibfnamefont{J.~F.} \bibnamefont{Donoghue}},
  \bibinfo{journal}{Phys. Rev.} \textbf{\bibinfo{volume}{D96}},
  \bibinfo{pages}{044007} (\bibinfo{year}{2017}), \eprint{1704.01533}.

\bibitem[{\citenamefont{Anselmi and
  Piva}(2018{\natexlab{a}})}]{Anselmi:2018ibi}
\bibinfo{author}{\bibfnamefont{D.}~\bibnamefont{Anselmi}} \bibnamefont{and}
  \bibinfo{author}{\bibfnamefont{M.}~\bibnamefont{Piva}},
  \bibinfo{journal}{JHEP} \textbf{\bibinfo{volume}{05}}, \bibinfo{pages}{027}
  (\bibinfo{year}{2018}{\natexlab{a}}), \eprint{1803.07777}.

\bibitem[{\citenamefont{Mannheim and O'Brien}(2012)}]{Mannheim:2010xw}
\bibinfo{author}{\bibfnamefont{P.~D.} \bibnamefont{Mannheim}} \bibnamefont{and}
  \bibinfo{author}{\bibfnamefont{J.~G.} \bibnamefont{O'Brien}},
  \bibinfo{journal}{Phys. Rev.} \textbf{\bibinfo{volume}{D85}},
  \bibinfo{pages}{124020} (\bibinfo{year}{2012}), \eprint{1011.3495}.

\bibitem[{\citenamefont{Bars et~al.}(2014)\citenamefont{Bars, Steinhardt, and
  Turok}}]{Bars:2013yba}
\bibinfo{author}{\bibfnamefont{I.}~\bibnamefont{Bars}},
  \bibinfo{author}{\bibfnamefont{P.}~\bibnamefont{Steinhardt}},
  \bibnamefont{and} \bibinfo{author}{\bibfnamefont{N.}~\bibnamefont{Turok}},
  \bibinfo{journal}{Phys. Rev.} \textbf{\bibinfo{volume}{D89}},
  \bibinfo{pages}{043515} (\bibinfo{year}{2014}), \eprint{1307.1848}.

\bibitem[{\citenamefont{de~Cesare et~al.}(2017)\citenamefont{de~Cesare, Moffat,
  and Sakellariadou}}]{deCesare:2016mml}
\bibinfo{author}{\bibfnamefont{M.}~\bibnamefont{de~Cesare}},
  \bibinfo{author}{\bibfnamefont{J.~W.} \bibnamefont{Moffat}},
  \bibnamefont{and}
  \bibinfo{author}{\bibfnamefont{M.}~\bibnamefont{Sakellariadou}},
  \bibinfo{journal}{Eur. Phys. J.} \textbf{\bibinfo{volume}{C77}},
  \bibinfo{pages}{605} (\bibinfo{year}{2017}), \eprint{1612.08066}.

\bibitem[{\citenamefont{Oda}(2018)}]{Oda:2018zth}
\bibinfo{author}{\bibfnamefont{I.}~\bibnamefont{Oda}}, \bibinfo{journal}{Eur.
  Phys. J.} \textbf{\bibinfo{volume}{C78}}, \bibinfo{pages}{798}
  (\bibinfo{year}{2018}), \eprint{1806.03420}.

\bibitem[{\citenamefont{Pereira}(2019)}]{Pereira:2019dbn}
\bibinfo{author}{\bibfnamefont{A.~D.} \bibnamefont{Pereira}}, in
  \emph{\bibinfo{booktitle}{{Progress and Visions in Quantum Theory in View of
  Gravity: Bridging foundations of physics and mathematics Leipzig, Germany,
  October 1-5, 2018}}} (\bibinfo{year}{2019}), \eprint{1904.07042}.

\bibitem[{\citenamefont{Kawai et~al.}(1993)\citenamefont{Kawai, Kitazawa, and
  Ninomiya}}]{Kawai:1992np}
\bibinfo{author}{\bibfnamefont{H.}~\bibnamefont{Kawai}},
  \bibinfo{author}{\bibfnamefont{Y.}~\bibnamefont{Kitazawa}}, \bibnamefont{and}
  \bibinfo{author}{\bibfnamefont{M.}~\bibnamefont{Ninomiya}},
  \bibinfo{journal}{Nucl. Phys.} \textbf{\bibinfo{volume}{B393}},
  \bibinfo{pages}{280} (\bibinfo{year}{1993}), \eprint{hep-th/9206081}.

\bibitem[{\citenamefont{Nink}(2015)}]{Nink:2014yya}
\bibinfo{author}{\bibfnamefont{A.}~\bibnamefont{Nink}}, \bibinfo{journal}{Phys.
  Rev.} \textbf{\bibinfo{volume}{D91}}, \bibinfo{pages}{044030}
  (\bibinfo{year}{2015}), \eprint{1410.7816}.

\bibitem[{\citenamefont{Gies et~al.}(2015)\citenamefont{Gies, Knorr, and
  Lippoldt}}]{Gies:2015tca}
\bibinfo{author}{\bibfnamefont{H.}~\bibnamefont{Gies}},
  \bibinfo{author}{\bibfnamefont{B.}~\bibnamefont{Knorr}}, \bibnamefont{and}
  \bibinfo{author}{\bibfnamefont{S.}~\bibnamefont{Lippoldt}},
  \bibinfo{journal}{Phys. Rev.} \textbf{\bibinfo{volume}{D92}},
  \bibinfo{pages}{084020} (\bibinfo{year}{2015}), \eprint{1507.08859}.

\bibitem[{\citenamefont{Ohta et~al.}(2016)\citenamefont{Ohta, Percacci, and
  Pereira}}]{Ohta:2016npm}
\bibinfo{author}{\bibfnamefont{N.}~\bibnamefont{Ohta}},
  \bibinfo{author}{\bibfnamefont{R.}~\bibnamefont{Percacci}}, \bibnamefont{and}
  \bibinfo{author}{\bibfnamefont{A.~D.} \bibnamefont{Pereira}},
  \bibinfo{journal}{JHEP} \textbf{\bibinfo{volume}{06}}, \bibinfo{pages}{115}
  (\bibinfo{year}{2016}), \eprint{1605.00454}.

\bibitem[{\citenamefont{Ohta et~al.}(2017)\citenamefont{Ohta, Percacci, and
  Pereira}}]{Ohta:2016jvw}
\bibinfo{author}{\bibfnamefont{N.}~\bibnamefont{Ohta}},
  \bibinfo{author}{\bibfnamefont{R.}~\bibnamefont{Percacci}}, \bibnamefont{and}
  \bibinfo{author}{\bibfnamefont{A.~D.} \bibnamefont{Pereira}},
  \bibinfo{journal}{Eur. Phys. J.} \textbf{\bibinfo{volume}{C77}},
  \bibinfo{pages}{611} (\bibinfo{year}{2017}), \eprint{1610.07991}.

\bibitem[{\citenamefont{De~Brito et~al.}(2018)\citenamefont{De~Brito, Ohta,
  Pereira, Tomaz, and Yamada}}]{deBrito:2018jxt}
\bibinfo{author}{\bibfnamefont{G.~P.} \bibnamefont{De~Brito}},
  \bibinfo{author}{\bibfnamefont{N.}~\bibnamefont{Ohta}},
  \bibinfo{author}{\bibfnamefont{A.~D.} \bibnamefont{Pereira}},
  \bibinfo{author}{\bibfnamefont{A.~A.} \bibnamefont{Tomaz}}, \bibnamefont{and}
  \bibinfo{author}{\bibfnamefont{M.}~\bibnamefont{Yamada}},
  \bibinfo{journal}{Phys. Rev.} \textbf{\bibinfo{volume}{D98}},
  \bibinfo{pages}{026027} (\bibinfo{year}{2018}), \eprint{1805.09656}.

\bibitem[{\citenamefont{Reuter and Wetterich}(1997)}]{Reuter:1996eg}
\bibinfo{author}{\bibfnamefont{M.}~\bibnamefont{Reuter}} \bibnamefont{and}
  \bibinfo{author}{\bibfnamefont{C.}~\bibnamefont{Wetterich}},
  \bibinfo{journal}{Nucl. Phys.} \textbf{\bibinfo{volume}{B506}},
  \bibinfo{pages}{483} (\bibinfo{year}{1997}), \eprint{hep-th/9605039}.

\bibitem[{\citenamefont{Litim}(2001)}]{Litim:2001up}
\bibinfo{author}{\bibfnamefont{D.~F.} \bibnamefont{Litim}},
  \bibinfo{journal}{Phys. Rev.} \textbf{\bibinfo{volume}{D64}},
  \bibinfo{pages}{105007} (\bibinfo{year}{2001}), \eprint{hep-th/0103195}.

\bibitem[{\citenamefont{Litim}(2000)}]{Litim:2000ci}
\bibinfo{author}{\bibfnamefont{D.~F.} \bibnamefont{Litim}},
  \bibinfo{journal}{Phys. Lett.} \textbf{\bibinfo{volume}{B486}},
  \bibinfo{pages}{92} (\bibinfo{year}{2000}), \eprint{hep-th/0005245}.

\bibitem[{\citenamefont{Lauscher and
  Reuter}(2002{\natexlab{b}})}]{Lauscher:2002sq}
\bibinfo{author}{\bibfnamefont{O.}~\bibnamefont{Lauscher}} \bibnamefont{and}
  \bibinfo{author}{\bibfnamefont{M.}~\bibnamefont{Reuter}},
  \bibinfo{journal}{Phys. Rev.} \textbf{\bibinfo{volume}{D66}},
  \bibinfo{pages}{025026} (\bibinfo{year}{2002}{\natexlab{b}}),
  \eprint{hep-th/0205062}.

\bibitem[{\citenamefont{Groh and Saueressig}(2010)}]{Groh:2010ta}
\bibinfo{author}{\bibfnamefont{K.}~\bibnamefont{Groh}} \bibnamefont{and}
  \bibinfo{author}{\bibfnamefont{F.}~\bibnamefont{Saueressig}},
  \bibinfo{journal}{J. Phys.} \textbf{\bibinfo{volume}{A43}},
  \bibinfo{pages}{365403} (\bibinfo{year}{2010}), \eprint{1001.5032}.

\bibitem[{\citenamefont{Eichhorn and Gies}(2010)}]{Eichhorn:2010tb}
\bibinfo{author}{\bibfnamefont{A.}~\bibnamefont{Eichhorn}} \bibnamefont{and}
  \bibinfo{author}{\bibfnamefont{H.}~\bibnamefont{Gies}},
  \bibinfo{journal}{Phys. Rev.} \textbf{\bibinfo{volume}{D81}},
  \bibinfo{pages}{104010} (\bibinfo{year}{2010}), \eprint{1001.5033}.

\bibitem[{\citenamefont{Wetterich}(1993)}]{Wetterich:1992yh}
\bibinfo{author}{\bibfnamefont{C.}~\bibnamefont{Wetterich}},
  \bibinfo{journal}{Phys. Lett.} \textbf{\bibinfo{volume}{B301}},
  \bibinfo{pages}{90} (\bibinfo{year}{1993}).

\bibitem[{\citenamefont{Morris}(1994)}]{Morris:1993qb}
\bibinfo{author}{\bibfnamefont{T.~R.} \bibnamefont{Morris}},
  \bibinfo{journal}{Int. J. Mod. Phys.} \textbf{\bibinfo{volume}{A9}},
  \bibinfo{pages}{2411} (\bibinfo{year}{1994}), \eprint{hep-ph/9308265}.

\bibitem[{\citenamefont{Ellwanger}(1994)}]{Ellwanger:1993mw}
\bibinfo{author}{\bibfnamefont{U.}~\bibnamefont{Ellwanger}},
  \bibinfo{journal}{Z. Phys.} \textbf{\bibinfo{volume}{C62}},
  \bibinfo{pages}{503} (\bibinfo{year}{1994}), \bibinfo{note}{[,206(1993)]},
  \eprint{hep-ph/9308260}.

\bibitem[{\citenamefont{Balog et~al.}(2019)\citenamefont{Balog, Chate,
  Delamotte, Marohnic, and Wschebor}}]{Balog:2019rrg}
\bibinfo{author}{\bibfnamefont{I.}~\bibnamefont{Balog}},
  \bibinfo{author}{\bibfnamefont{H.}~\bibnamefont{Chate}},
  \bibinfo{author}{\bibfnamefont{B.}~\bibnamefont{Delamotte}},
  \bibinfo{author}{\bibfnamefont{M.}~\bibnamefont{Marohnic}}, \bibnamefont{and}
  \bibinfo{author}{\bibfnamefont{N.}~\bibnamefont{Wschebor}}
  (\bibinfo{year}{2019}), \eprint{1907.01829}.

\bibitem[{\citenamefont{Visser}(2017)}]{Visser:2017atf}
\bibinfo{author}{\bibfnamefont{M.}~\bibnamefont{Visser}}
  (\bibinfo{year}{2017}), \eprint{1702.05572}.

\bibitem[{\citenamefont{Baldazzi et~al.}(2019)\citenamefont{Baldazzi, Percacci,
  and Skrinjar}}]{Baldazzi:2018mtl}
\bibinfo{author}{\bibfnamefont{A.}~\bibnamefont{Baldazzi}},
  \bibinfo{author}{\bibfnamefont{R.}~\bibnamefont{Percacci}}, \bibnamefont{and}
  \bibinfo{author}{\bibfnamefont{V.}~\bibnamefont{Skrinjar}},
  \bibinfo{journal}{Class. Quant. Grav.} \textbf{\bibinfo{volume}{36}},
  \bibinfo{pages}{105008} (\bibinfo{year}{2019}), \eprint{1811.03369}.

\bibitem[{\citenamefont{Percacci}(2018)}]{Percacci:2017fsy}
\bibinfo{author}{\bibfnamefont{R.}~\bibnamefont{Percacci}},
  \bibinfo{journal}{Found. Phys.} \textbf{\bibinfo{volume}{48}},
  \bibinfo{pages}{1364} (\bibinfo{year}{2018}), \eprint{1712.09903}.

\bibitem[{\citenamefont{Eichhorn}(2012)}]{Eichhorn:2012va}
\bibinfo{author}{\bibfnamefont{A.}~\bibnamefont{Eichhorn}},
  \bibinfo{journal}{Phys. Rev.} \textbf{\bibinfo{volume}{D86}},
  \bibinfo{pages}{105021} (\bibinfo{year}{2012}), \eprint{1204.0965}.

\bibitem[{\citenamefont{Eichhorn et~al.}(2016)\citenamefont{Eichhorn, Held, and
  Pawlowski}}]{Eichhorn:2016esv}
\bibinfo{author}{\bibfnamefont{A.}~\bibnamefont{Eichhorn}},
  \bibinfo{author}{\bibfnamefont{A.}~\bibnamefont{Held}}, \bibnamefont{and}
  \bibinfo{author}{\bibfnamefont{J.~M.} \bibnamefont{Pawlowski}},
  \bibinfo{journal}{Phys. Rev.} \textbf{\bibinfo{volume}{D94}},
  \bibinfo{pages}{104027} (\bibinfo{year}{2016}), \eprint{1604.02041}.

\bibitem[{\citenamefont{Eichhorn
  et~al.}(2018{\natexlab{e}})\citenamefont{Eichhorn, Lippoldt, and
  Skrinjar}}]{Eichhorn:2017sok}
\bibinfo{author}{\bibfnamefont{A.}~\bibnamefont{Eichhorn}},
  \bibinfo{author}{\bibfnamefont{S.}~\bibnamefont{Lippoldt}}, \bibnamefont{and}
  \bibinfo{author}{\bibfnamefont{V.}~\bibnamefont{Skrinjar}},
  \bibinfo{journal}{Phys. Rev.} \textbf{\bibinfo{volume}{D97}},
  \bibinfo{pages}{026002} (\bibinfo{year}{2018}{\natexlab{e}}),
  \eprint{1710.03005}.

\bibitem[{\citenamefont{Christiansen and
  Eichhorn}(2017)}]{Christiansen:2017gtg}
\bibinfo{author}{\bibfnamefont{N.}~\bibnamefont{Christiansen}}
  \bibnamefont{and} \bibinfo{author}{\bibfnamefont{A.}~\bibnamefont{Eichhorn}},
  \bibinfo{journal}{Phys. Lett.} \textbf{\bibinfo{volume}{B770}},
  \bibinfo{pages}{154} (\bibinfo{year}{2017}), \eprint{1702.07724}.

\bibitem[{\citenamefont{Upadhyay et~al.}(2017)\citenamefont{Upadhyay, Oksanen,
  and Bufalo}}]{Upadhyay:2015fna}
\bibinfo{author}{\bibfnamefont{S.}~\bibnamefont{Upadhyay}},
  \bibinfo{author}{\bibfnamefont{M.}~\bibnamefont{Oksanen}}, \bibnamefont{and}
  \bibinfo{author}{\bibfnamefont{R.}~\bibnamefont{Bufalo}},
  \bibinfo{journal}{Braz. J. Phys.} \textbf{\bibinfo{volume}{47}},
  \bibinfo{pages}{350} (\bibinfo{year}{2017}), \eprint{1510.00188}.

\bibitem[{\citenamefont{York}(1973)}]{York:1973ia}
\bibinfo{author}{\bibfnamefont{J.}~\bibnamefont{York},
  \bibfnamefont{James~W.}}, \bibinfo{journal}{J.Math.Phys.}
  \textbf{\bibinfo{volume}{14}}, \bibinfo{pages}{456} (\bibinfo{year}{1973}).

\bibitem[{\citenamefont{Dona et~al.}(2016)\citenamefont{Dona, Eichhorn, Labus,
  and Percacci}}]{Dona:2015tnf}
\bibinfo{author}{\bibfnamefont{P.}~\bibnamefont{Dona}},
  \bibinfo{author}{\bibfnamefont{A.}~\bibnamefont{Eichhorn}},
  \bibinfo{author}{\bibfnamefont{P.}~\bibnamefont{Labus}}, \bibnamefont{and}
  \bibinfo{author}{\bibfnamefont{R.}~\bibnamefont{Percacci}},
  \bibinfo{journal}{Phys. Rev.} \textbf{\bibinfo{volume}{D93}},
  \bibinfo{pages}{044049} (\bibinfo{year}{2016}), \bibinfo{note}{[Erratum:
  Phys. Rev.D93,no.12,129904(2016)]}, \eprint{1512.01589}.

\bibitem[{\citenamefont{Holdom and Ren}(2016{\natexlab{b}})}]{Holdom:2015kbf}
\bibinfo{author}{\bibfnamefont{B.}~\bibnamefont{Holdom}} \bibnamefont{and}
  \bibinfo{author}{\bibfnamefont{J.}~\bibnamefont{Ren}},
  \bibinfo{journal}{Phys. Rev.} \textbf{\bibinfo{volume}{D93}},
  \bibinfo{pages}{124030} (\bibinfo{year}{2016}{\natexlab{b}}),
  \eprint{1512.05305}.

\bibitem[{\citenamefont{Donoghue and Menezes}(2019)}]{Donoghue:2018lmc}
\bibinfo{author}{\bibfnamefont{J.~F.} \bibnamefont{Donoghue}} \bibnamefont{and}
  \bibinfo{author}{\bibfnamefont{G.}~\bibnamefont{Menezes}},
  \bibinfo{journal}{Phys. Rev.} \textbf{\bibinfo{volume}{D99}},
  \bibinfo{pages}{065017} (\bibinfo{year}{2019}), \eprint{1812.03603}.

\bibitem[{\citenamefont{Anselmi and
  Piva}(2018{\natexlab{b}})}]{Anselmi:2018tmf}
\bibinfo{author}{\bibfnamefont{D.}~\bibnamefont{Anselmi}} \bibnamefont{and}
  \bibinfo{author}{\bibfnamefont{M.}~\bibnamefont{Piva}},
  \bibinfo{journal}{JHEP} \textbf{\bibinfo{volume}{11}}, \bibinfo{pages}{021}
  (\bibinfo{year}{2018}{\natexlab{b}}), \eprint{1806.03605}.

\bibitem[{\citenamefont{Woodard}(2015)}]{Woodard:2015zca}
\bibinfo{author}{\bibfnamefont{R.~P.} \bibnamefont{Woodard}},
  \bibinfo{journal}{Scholarpedia} \textbf{\bibinfo{volume}{10}},
  \bibinfo{pages}{32243} (\bibinfo{year}{2015}), \eprint{1506.02210}.

\bibitem[{\citenamefont{Calmet et~al.}(2017)\citenamefont{Calmet, Capozziello,
  and Pryer}}]{Calmet:2017rxl}
\bibinfo{author}{\bibfnamefont{X.}~\bibnamefont{Calmet}},
  \bibinfo{author}{\bibfnamefont{S.}~\bibnamefont{Capozziello}},
  \bibnamefont{and} \bibinfo{author}{\bibfnamefont{D.}~\bibnamefont{Pryer}},
  \bibinfo{journal}{Eur. Phys. J.} \textbf{\bibinfo{volume}{C77}},
  \bibinfo{pages}{589} (\bibinfo{year}{2017}), \eprint{1708.08253}.

\bibitem[{\citenamefont{Chen et~al.}(2019)\citenamefont{Chen, Qin, Tan, and
  Shao}}]{Chen:2019stu}
\bibinfo{author}{\bibfnamefont{Y.-F.} \bibnamefont{Chen}},
  \bibinfo{author}{\bibfnamefont{C.-G.} \bibnamefont{Qin}},
  \bibinfo{author}{\bibfnamefont{Y.-J.} \bibnamefont{Tan}}, \bibnamefont{and}
  \bibinfo{author}{\bibfnamefont{C.-G.} \bibnamefont{Shao}},
  \bibinfo{journal}{Phys. Rev.} \textbf{\bibinfo{volume}{D99}},
  \bibinfo{pages}{104008} (\bibinfo{year}{2019}).

\bibitem[{\citenamefont{Kim et~al.}(2019)\citenamefont{Kim, Kobakhidze, and
  Picker}}]{Kim:2019sqk}
\bibinfo{author}{\bibfnamefont{Y.}~\bibnamefont{Kim}},
  \bibinfo{author}{\bibfnamefont{A.}~\bibnamefont{Kobakhidze}},
  \bibnamefont{and} \bibinfo{author}{\bibfnamefont{Z.~S.~C.}
  \bibnamefont{Picker}} (\bibinfo{year}{2019}), \eprint{1906.12034}.

\bibitem[{\citenamefont{Ohta and Percacci}(2016)}]{Ohta:2015zwa}
\bibinfo{author}{\bibfnamefont{N.}~\bibnamefont{Ohta}} \bibnamefont{and}
  \bibinfo{author}{\bibfnamefont{R.}~\bibnamefont{Percacci}},
  \bibinfo{journal}{Class. Quant. Grav.} \textbf{\bibinfo{volume}{33}},
  \bibinfo{pages}{035001} (\bibinfo{year}{2016}), \eprint{1506.05526}.

\bibitem[{\citenamefont{Pawlowski}(2007)}]{Pawlowski:2005xe}
\bibinfo{author}{\bibfnamefont{J.~M.} \bibnamefont{Pawlowski}},
  \bibinfo{journal}{Annals Phys.} \textbf{\bibinfo{volume}{322}},
  \bibinfo{pages}{2831} (\bibinfo{year}{2007}), \eprint{hep-th/0512261}.

\bibitem[{\citenamefont{Gies}(2012)}]{Gies:2006wv}
\bibinfo{author}{\bibfnamefont{H.}~\bibnamefont{Gies}},
  \bibinfo{journal}{Lect.Notes Phys.} \textbf{\bibinfo{volume}{852}},
  \bibinfo{pages}{287} (\bibinfo{year}{2012}), \eprint{hep-ph/0611146}.

\bibitem[{\citenamefont{Buttazzo et~al.}(2013)\citenamefont{Buttazzo, Degrassi,
  Giardino, Giudice, Sala, Salvio, and Strumia}}]{Buttazzo:2013uya}
\bibinfo{author}{\bibfnamefont{D.}~\bibnamefont{Buttazzo}},
  \bibinfo{author}{\bibfnamefont{G.}~\bibnamefont{Degrassi}},
  \bibinfo{author}{\bibfnamefont{P.~P.} \bibnamefont{Giardino}},
  \bibinfo{author}{\bibfnamefont{G.~F.} \bibnamefont{Giudice}},
  \bibinfo{author}{\bibfnamefont{F.}~\bibnamefont{Sala}},
  \bibinfo{author}{\bibfnamefont{A.}~\bibnamefont{Salvio}}, \bibnamefont{and}
  \bibinfo{author}{\bibfnamefont{A.}~\bibnamefont{Strumia}},
  \bibinfo{journal}{JHEP} \textbf{\bibinfo{volume}{12}}, \bibinfo{pages}{089}
  (\bibinfo{year}{2013}), \eprint{1307.3536}.

\bibitem[{\citenamefont{Bezrukov and Shaposhnikov}(2015)}]{Bezrukov:2014ina}
\bibinfo{author}{\bibfnamefont{F.}~\bibnamefont{Bezrukov}} \bibnamefont{and}
  \bibinfo{author}{\bibfnamefont{M.}~\bibnamefont{Shaposhnikov}},
  \bibinfo{journal}{J. Exp. Theor. Phys.} \textbf{\bibinfo{volume}{120}},
  \bibinfo{pages}{335} (\bibinfo{year}{2015}), \bibinfo{note}{[Zh. Eksp. Teor.
  Fiz.147,389(2015)]}, \eprint{1411.1923}.

\bibitem[{\citenamefont{Narain and Percacci}(2010)}]{Narain:2009fy}
\bibinfo{author}{\bibfnamefont{G.}~\bibnamefont{Narain}} \bibnamefont{and}
  \bibinfo{author}{\bibfnamefont{R.}~\bibnamefont{Percacci}},
  \bibinfo{journal}{Class. Quant. Grav.} \textbf{\bibinfo{volume}{27}},
  \bibinfo{pages}{075001} (\bibinfo{year}{2010}), \eprint{0911.0386}.

\bibitem[{\citenamefont{Gonzalez-Martin and
  Martin}(2018{\natexlab{a}})}]{Gonzalez-Martin:2018dmy}
\bibinfo{author}{\bibfnamefont{S.}~\bibnamefont{Gonzalez-Martin}}
  \bibnamefont{and} \bibinfo{author}{\bibfnamefont{C.~P.}
  \bibnamefont{Martin}}, \bibinfo{journal}{Eur. Phys. J.}
  \textbf{\bibinfo{volume}{C78}}, \bibinfo{pages}{236}
  (\bibinfo{year}{2018}{\natexlab{a}}), \eprint{1802.03755}.

\bibitem[{\citenamefont{Folkerts et~al.}(2012)\citenamefont{Folkerts, Litim,
  and Pawlowski}}]{Folkerts:2011jz}
\bibinfo{author}{\bibfnamefont{S.}~\bibnamefont{Folkerts}},
  \bibinfo{author}{\bibfnamefont{D.~F.} \bibnamefont{Litim}}, \bibnamefont{and}
  \bibinfo{author}{\bibfnamefont{J.~M.} \bibnamefont{Pawlowski}},
  \bibinfo{journal}{Phys. Lett.} \textbf{\bibinfo{volume}{B709}},
  \bibinfo{pages}{234} (\bibinfo{year}{2012}), \eprint{1101.5552}.

\bibitem[{\citenamefont{Cheng et~al.}(1974)\citenamefont{Cheng, Eichten, and
  Li}}]{Cheng:1973nv}
\bibinfo{author}{\bibfnamefont{T.~P.} \bibnamefont{Cheng}},
  \bibinfo{author}{\bibfnamefont{E.}~\bibnamefont{Eichten}}, \bibnamefont{and}
  \bibinfo{author}{\bibfnamefont{L.-F.} \bibnamefont{Li}},
  \bibinfo{journal}{Phys. Rev.} \textbf{\bibinfo{volume}{D9}},
  \bibinfo{pages}{2259} (\bibinfo{year}{1974}).

\bibitem[{\citenamefont{Gonzalez-Martin and
  Martin}(2017)}]{Gonzalez-Martin:2017bvw}
\bibinfo{author}{\bibfnamefont{S.}~\bibnamefont{Gonzalez-Martin}}
  \bibnamefont{and} \bibinfo{author}{\bibfnamefont{C.~P.}
  \bibnamefont{Martin}}, \bibinfo{journal}{Phys. Lett.}
  \textbf{\bibinfo{volume}{B773}}, \bibinfo{pages}{585} (\bibinfo{year}{2017}),
  \eprint{1707.06667}.

\bibitem[{\citenamefont{Gonzalez-Martin and
  Martin}(2018{\natexlab{b}})}]{Gonzalez-Martin:2017fwz}
\bibinfo{author}{\bibfnamefont{S.}~\bibnamefont{Gonzalez-Martin}}
  \bibnamefont{and} \bibinfo{author}{\bibfnamefont{C.~P.}
  \bibnamefont{Martin}}, \bibinfo{journal}{JCAP}
  \textbf{\bibinfo{volume}{1801}}, \bibinfo{pages}{028}
  (\bibinfo{year}{2018}{\natexlab{b}}), \eprint{1711.08009}.

\bibitem[{\citenamefont{Bezrukov et~al.}(2012)\citenamefont{Bezrukov, Kalmykov,
  Kniehl, and Shaposhnikov}}]{Bezrukov:2012sa}
\bibinfo{author}{\bibfnamefont{F.}~\bibnamefont{Bezrukov}},
  \bibinfo{author}{\bibfnamefont{M.~{\relax Yu}.} \bibnamefont{Kalmykov}},
  \bibinfo{author}{\bibfnamefont{B.~A.} \bibnamefont{Kniehl}},
  \bibnamefont{and}
  \bibinfo{author}{\bibfnamefont{M.}~\bibnamefont{Shaposhnikov}},
  \bibinfo{journal}{JHEP} \textbf{\bibinfo{volume}{10}}, \bibinfo{pages}{140}
  (\bibinfo{year}{2012}), \bibinfo{note}{[,275(2012)]}, \eprint{1205.2893}.

\bibitem[{\citenamefont{Eichhorn et~al.}()\citenamefont{Eichhorn, Held, and
  Wetterich}}]{Held:2019}
\bibinfo{author}{\bibfnamefont{A.}~\bibnamefont{Eichhorn}},
  \bibinfo{author}{\bibfnamefont{A.}~\bibnamefont{Held}}, \bibnamefont{and}
  \bibinfo{author}{\bibfnamefont{C.}~\bibnamefont{Wetterich}} (????).

\bibitem[{\citenamefont{Eichhorn}(2013{\natexlab{b}})}]{Eichhorn:2013ug}
\bibinfo{author}{\bibfnamefont{A.}~\bibnamefont{Eichhorn}},
  \bibinfo{journal}{Phys. Rev.} \textbf{\bibinfo{volume}{D87}},
  \bibinfo{pages}{124016} (\bibinfo{year}{2013}{\natexlab{b}}),
  \eprint{1301.0632}.

\bibitem[{\citenamefont{de~Berredo-Peixoto and
  Shapiro}(2004)}]{deBerredoPeixoto:2003pj}
\bibinfo{author}{\bibfnamefont{G.}~\bibnamefont{de~Berredo-Peixoto}}
  \bibnamefont{and} \bibinfo{author}{\bibfnamefont{I.~L.}
  \bibnamefont{Shapiro}}, \bibinfo{journal}{Phys. Rev.}
  \textbf{\bibinfo{volume}{D70}}, \bibinfo{pages}{044024}
  (\bibinfo{year}{2004}), \eprint{hep-th/0307030}.

\bibitem[{\citenamefont{Ohta and Percacci}(2014)}]{Ohta:2013uca}
\bibinfo{author}{\bibfnamefont{N.}~\bibnamefont{Ohta}} \bibnamefont{and}
  \bibinfo{author}{\bibfnamefont{R.}~\bibnamefont{Percacci}},
  \bibinfo{journal}{Class. Quant. Grav.} \textbf{\bibinfo{volume}{31}},
  \bibinfo{pages}{015024} (\bibinfo{year}{2014}), \eprint{1308.3398}.

\bibitem[{\citenamefont{Narain and
  Anishetty}(2013{\natexlab{a}})}]{Narain:2012te}
\bibinfo{author}{\bibfnamefont{G.}~\bibnamefont{Narain}} \bibnamefont{and}
  \bibinfo{author}{\bibfnamefont{R.}~\bibnamefont{Anishetty}},
  \bibinfo{journal}{JHEP} \textbf{\bibinfo{volume}{07}}, \bibinfo{pages}{106}
  (\bibinfo{year}{2013}{\natexlab{a}}), \eprint{1211.5040}.

\bibitem[{\citenamefont{Narain and
  Anishetty}(2013{\natexlab{b}})}]{Narain:2013eea}
\bibinfo{author}{\bibfnamefont{G.}~\bibnamefont{Narain}} \bibnamefont{and}
  \bibinfo{author}{\bibfnamefont{R.}~\bibnamefont{Anishetty}},
  \bibinfo{journal}{JHEP} \textbf{\bibinfo{volume}{10}}, \bibinfo{pages}{203}
  (\bibinfo{year}{2013}{\natexlab{b}}), \eprint{1309.0473}.

\bibitem[{\citenamefont{Pietrykowski}(2007)}]{Pietrykowski:2006xy}
\bibinfo{author}{\bibfnamefont{A.~R.} \bibnamefont{Pietrykowski}},
  \bibinfo{journal}{Phys. Rev. Lett.} \textbf{\bibinfo{volume}{98}},
  \bibinfo{pages}{061801} (\bibinfo{year}{2007}), \eprint{hep-th/0606208}.

\bibitem[{\citenamefont{Toms}(2007)}]{Toms:2007sk}
\bibinfo{author}{\bibfnamefont{D.~J.} \bibnamefont{Toms}},
  \bibinfo{journal}{Phys. Rev.} \textbf{\bibinfo{volume}{D76}},
  \bibinfo{pages}{045015} (\bibinfo{year}{2007}), \eprint{0708.2990}.

\bibitem[{\citenamefont{Ebert et~al.}(2008)\citenamefont{Ebert, Plefka, and
  Rodigast}}]{Ebert:2007gf}
\bibinfo{author}{\bibfnamefont{D.}~\bibnamefont{Ebert}},
  \bibinfo{author}{\bibfnamefont{J.}~\bibnamefont{Plefka}}, \bibnamefont{and}
  \bibinfo{author}{\bibfnamefont{A.}~\bibnamefont{Rodigast}},
  \bibinfo{journal}{Phys. Lett.} \textbf{\bibinfo{volume}{B660}},
  \bibinfo{pages}{579} (\bibinfo{year}{2008}), \eprint{0710.1002}.

\bibitem[{\citenamefont{Felipe et~al.}(2011)\citenamefont{Felipe, Brito,
  Sampaio, and Nemes}}]{Felipe:2011rs}
\bibinfo{author}{\bibfnamefont{J.~C.~C.} \bibnamefont{Felipe}},
  \bibinfo{author}{\bibfnamefont{L.~C.~T.} \bibnamefont{Brito}},
  \bibinfo{author}{\bibfnamefont{M.}~\bibnamefont{Sampaio}}, \bibnamefont{and}
  \bibinfo{author}{\bibfnamefont{M.~C.} \bibnamefont{Nemes}},
  \bibinfo{journal}{Phys. Lett.} \textbf{\bibinfo{volume}{B700}},
  \bibinfo{pages}{86} (\bibinfo{year}{2011}), \eprint{1103.5824}.

\bibitem[{\citenamefont{Ellis and Mavromatos}(2012)}]{Ellis:2010rw}
\bibinfo{author}{\bibfnamefont{J.}~\bibnamefont{Ellis}} \bibnamefont{and}
  \bibinfo{author}{\bibfnamefont{N.~E.} \bibnamefont{Mavromatos}},
  \bibinfo{journal}{Phys. Lett.} \textbf{\bibinfo{volume}{B711}},
  \bibinfo{pages}{139} (\bibinfo{year}{2012}), \eprint{1012.4353}.

\bibitem[{\citenamefont{Anber et~al.}(2011)\citenamefont{Anber, Donoghue, and
  El-Houssieny}}]{Anber:2010uj}
\bibinfo{author}{\bibfnamefont{M.~M.} \bibnamefont{Anber}},
  \bibinfo{author}{\bibfnamefont{J.~F.} \bibnamefont{Donoghue}},
  \bibnamefont{and}
  \bibinfo{author}{\bibfnamefont{M.}~\bibnamefont{El-Houssieny}},
  \bibinfo{journal}{Phys. Rev.} \textbf{\bibinfo{volume}{D83}},
  \bibinfo{pages}{124003} (\bibinfo{year}{2011}), \eprint{1011.3229}.

\bibitem[{\citenamefont{Toms}(2011)}]{Toms:2011zza}
\bibinfo{author}{\bibfnamefont{D.~J.} \bibnamefont{Toms}},
  \bibinfo{journal}{Phys. Rev.} \textbf{\bibinfo{volume}{D84}},
  \bibinfo{pages}{084016} (\bibinfo{year}{2011}).

\bibitem[{\citenamefont{Nielsen}(2012)}]{Nielsen:2012fm}
\bibinfo{author}{\bibfnamefont{N.~K.} \bibnamefont{Nielsen}},
  \bibinfo{journal}{Annals Phys.} \textbf{\bibinfo{volume}{327}},
  \bibinfo{pages}{861} (\bibinfo{year}{2012}), \eprint{1109.2699}.

\bibitem[{\citenamefont{Pietrykowski}(2013)}]{Pietrykowski:2012nc}
\bibinfo{author}{\bibfnamefont{A.~R.} \bibnamefont{Pietrykowski}},
  \bibinfo{journal}{Phys. Rev.} \textbf{\bibinfo{volume}{D87}},
  \bibinfo{pages}{024026} (\bibinfo{year}{2013}), \eprint{1210.0507}.

\bibitem[{\citenamefont{Weinberg}(2013)}]{Weinberg:1996kr}
\bibinfo{author}{\bibfnamefont{S.}~\bibnamefont{Weinberg}},
  \emph{\bibinfo{title}{{The quantum theory of fields. Vol. 2: Modern
  applications}}} (\bibinfo{publisher}{Cambridge University Press},
  \bibinfo{year}{2013}), ISBN \bibinfo{isbn}{9781139632478, 9780521670548,
  9780521550024}.

\bibitem[{\citenamefont{Eichhorn et~al.}(2015)\citenamefont{Eichhorn, Gies,
  Jaeckel, Plehn, Scherer, and Sondenheimer}}]{Eichhorn:2015kea}
\bibinfo{author}{\bibfnamefont{A.}~\bibnamefont{Eichhorn}},
  \bibinfo{author}{\bibfnamefont{H.}~\bibnamefont{Gies}},
  \bibinfo{author}{\bibfnamefont{J.}~\bibnamefont{Jaeckel}},
  \bibinfo{author}{\bibfnamefont{T.}~\bibnamefont{Plehn}},
  \bibinfo{author}{\bibfnamefont{M.~M.} \bibnamefont{Scherer}},
  \bibnamefont{and}
  \bibinfo{author}{\bibfnamefont{R.}~\bibnamefont{Sondenheimer}},
  \bibinfo{journal}{JHEP} \textbf{\bibinfo{volume}{04}}, \bibinfo{pages}{022}
  (\bibinfo{year}{2015}), \eprint{1501.02812}.

\bibitem[{\citenamefont{Salvio and Strumia}(2018)}]{Salvio:2017qkx}
\bibinfo{author}{\bibfnamefont{A.}~\bibnamefont{Salvio}} \bibnamefont{and}
  \bibinfo{author}{\bibfnamefont{A.}~\bibnamefont{Strumia}},
  \bibinfo{journal}{Eur. Phys. J.} \textbf{\bibinfo{volume}{C78}},
  \bibinfo{pages}{124} (\bibinfo{year}{2018}), \eprint{1705.03896}.

\bibitem[{\citenamefont{Woodard}(1984)}]{Woodard:1984sj}
\bibinfo{author}{\bibfnamefont{R.~P.} \bibnamefont{Woodard}},
  \bibinfo{journal}{Phys. Lett.} \textbf{\bibinfo{volume}{148B}},
  \bibinfo{pages}{440} (\bibinfo{year}{1984}).

\bibitem[{\citenamefont{van Nieuwenhuizen}(1981)}]{vanNieuwenhuizen:1981uf}
\bibinfo{author}{\bibfnamefont{P.}~\bibnamefont{van Nieuwenhuizen}},
  \bibinfo{journal}{Phys. Rev.} \textbf{\bibinfo{volume}{D24}},
  \bibinfo{pages}{3315} (\bibinfo{year}{1981}).

\bibitem[{\citenamefont{Gies and Lippoldt}(2014)}]{Gies:2013noa}
\bibinfo{author}{\bibfnamefont{H.}~\bibnamefont{Gies}} \bibnamefont{and}
  \bibinfo{author}{\bibfnamefont{S.}~\bibnamefont{Lippoldt}},
  \bibinfo{journal}{Phys. Rev.} \textbf{\bibinfo{volume}{D89}},
  \bibinfo{pages}{064040} (\bibinfo{year}{2014}), \eprint{1310.2509}.

\bibitem[{\citenamefont{Gies and Lippoldt}(2015)}]{Gies:2015cka}
\bibinfo{author}{\bibfnamefont{H.}~\bibnamefont{Gies}} \bibnamefont{and}
  \bibinfo{author}{\bibfnamefont{S.}~\bibnamefont{Lippoldt}},
  \bibinfo{journal}{Phys. Lett.} \textbf{\bibinfo{volume}{B743}},
  \bibinfo{pages}{415} (\bibinfo{year}{2015}), \eprint{1502.00918}.

\bibitem[{\citenamefont{Lippoldt}(2015)}]{Lippoldt:2015cea}
\bibinfo{author}{\bibfnamefont{S.}~\bibnamefont{Lippoldt}},
  \bibinfo{journal}{Phys. Rev.} \textbf{\bibinfo{volume}{D91}},
  \bibinfo{pages}{104006} (\bibinfo{year}{2015}), \eprint{1502.05607}.

\bibitem[{\citenamefont{Ellwanger et~al.}(1996)\citenamefont{Ellwanger, Hirsch,
  and Weber}}]{Ellwanger:1995qf}
\bibinfo{author}{\bibfnamefont{U.}~\bibnamefont{Ellwanger}},
  \bibinfo{author}{\bibfnamefont{M.}~\bibnamefont{Hirsch}}, \bibnamefont{and}
  \bibinfo{author}{\bibfnamefont{A.}~\bibnamefont{Weber}},
  \bibinfo{journal}{Z.Phys.} \textbf{\bibinfo{volume}{C69}},
  \bibinfo{pages}{687} (\bibinfo{year}{1996}), \eprint{hep-th/9506019}.

\bibitem[{\citenamefont{Pawlowski}(2003)}]{Pawlowski:2003sk}
\bibinfo{author}{\bibfnamefont{J.~M.} \bibnamefont{Pawlowski}}
  (\bibinfo{year}{2003}), \eprint{hep-th/0310018}.

\bibitem[{\citenamefont{Manrique and Reuter}(2010)}]{Manrique:2009uh}
\bibinfo{author}{\bibfnamefont{E.}~\bibnamefont{Manrique}} \bibnamefont{and}
  \bibinfo{author}{\bibfnamefont{M.}~\bibnamefont{Reuter}},
  \bibinfo{journal}{Annals Phys.} \textbf{\bibinfo{volume}{325}},
  \bibinfo{pages}{785} (\bibinfo{year}{2010}), \eprint{0907.2617}.

\end{thebibliography}

\end{document}